\documentclass[
reprint, 
secnumarabic,
amssymb, 
nobibnotes,
aps, 
prd,
superscriptaddress,
nofootinbib,
]{revtex4-1}
\usepackage{algorithm}

\usepackage[ddmmyyyy]{datetime} 
\usepackage{amsmath} 
\usepackage[dvipsnames]{xcolor}
\usepackage{cancel} 
\usepackage{graphicx}
\usepackage{comment} 
\usepackage{soul,ulem} 
\usepackage{eqnarray}
\soulregister\cite7
\soulregister\ref7 
\soulregister\eqref7 
\usepackage{bbding}
\usepackage{pifont}
\usepackage{wasysym}
\usepackage{amssymb}
\usepackage{textcomp} 
\usepackage{marginnote} 
\usepackage[colorinlistoftodos
, textsize=tiny]{todonotes} 
\usepackage{textcomp} 
\usepackage{lineno} 
\usepackage{hyperref}
\usepackage{cleveref}
\hypersetup{
    colorlinks=true,
    citecolor=[rgb]{0.85,0.17,0.17},
    linkcolor=blue,
    filecolor=magenta,      
    urlcolor=cyan,
    urlbordercolor=red
}

\newcommand*\araa{ARA\&A}

\newcommand{\GNeff}{\ensuremath{G_{\rm N}^{\rm eff}}}
\newcommand{\gNtilde}{\ensuremath{\tilde{\gamma}_{\rm N}}}
\newcommand{\gN}{\ensuremath{{\gamma}_{\rm N}}}
\newcommand{\GN}{\ensuremath{G_{\rm N}}}
\newcommand{\M}{\ensuremath{M_{500}}}
\newcommand{\R}{\ensuremath{R_{500}}}

\newcommand{\ksM}{\text{km/s Mpc$^{-1} $}}
\newcommand{\Psz}{\ensuremath{P_{\rm SZ}}}
\newcommand{\Px}{\ensuremath{P_{\rm X}}}
\newcommand{\Tx}{\ensuremath{T_{\rm X}}}
\newcommand{\aH}{\ensuremath{\alpha_{\rm H}}}
\newcommand{\bone}{\ensuremath{\beta_{1}}}
\defcitealias{Ettori:2018tus}{E19}

\definecolor{mypurple}{RGB}{135,10,190} 

\begin{document}

\title{ Testing generalized scalar-tensor theories of gravity with clusters of galaxies }

\author{Balakrishna S. Haridasu}
 \email{sandeep.haridasu@sissa.it}
 \affiliation{SISSA-International School for Advanced Studies, Via Bonomea 265, 34136 Trieste, Italy}
 \affiliation{INFN, Sezione di Trieste, Via Valerio 2, I-34127 Trieste, Italy}
 \affiliation{IFPU, Institute for Fundamental Physics of the Universe, via Beirut 2, 34151 Trieste, Italy}

\author{Purnendu Karmakar}
 \email{purnendu.karmakar@pd.infn.it}
 \affiliation{Dipartimento di Fisica, Sapienza Universit\'a di Roma, P.le Aldo Moro 2, 00185, Roma, Italy}
 
  \author{Marco De Petris}
 \email{marco.depetris@roma1.infn.it}
 \affiliation{Dipartimento di Fisica, Sapienza Universit\'a di Roma, P.le Aldo Moro 2, 00185, Roma, Italy}
  \affiliation{INAF - Osservatorio Astronomico di Roma, Via Frascati 33, 00040, Monteporzio Catone, Roma, Italy}
 \affiliation{INFN- Sezione di Roma 1, P.le Aldo Moro 2, 00185, Roma, Italy}
 
 \author{Vincenzo F. Cardone}
 \email{vincenzo.cardone@inaf.it}
 \affiliation{INAF - Osservatorio Astronomico di Roma, Via Frascati 33, 00040, Monteporzio Catone, Roma, Italy}
 \affiliation{INFN- Sezione di Roma 1, P.le Aldo Moro 2, 00185, Roma, Italy}
 
 \author{Roberto Maoli}
 \email{roberto.maoli@roma1.infn.it}
 \affiliation{Dipartimento di Fisica, Sapienza Universit\'a di Roma, P.le Aldo Moro 2, 00185, Roma, Italy}
 \affiliation{INAF - Osservatorio Astronomico di Roma, Via Frascati 33, 00040, Monteporzio Catone, Roma, Italy}

\begin{abstract}

We test the generalized scalar-tensor theory in static systems, namely galaxy clusters. The Degenerate higher-order scalar-tensor (DHOST) theory modifies the Newtonian potential through effective Newtonian constant and $\Xi_1$ parameter in the small scale, which modifies the hydrostatic equilibrium. We utilize the well-compiled X-COP catalog consisting of 12 clusters with Intra Cluster Medium (ICM) pressure profile by Sunyaev-Zel'dovich effect data and temperature profile by X-ray data for each cluster. We perform a fully Bayesian analysis modeling Navarro–Frenk–White (NFW) for the mass profile and the simplified Vikhlinin model for the electron density. Carefully selecting suitable clusters to present our results, we find a mild to moderate, i.e, $\sim 2\sigma$ significance for a deviation from the standard scenario in 4 of the clusters. However, in terms of Bayesian evidence, we find either equivalent or mild preference for GR. We estimate a joint constraint of $\Xi_1 = -0.030 \pm 0.043$ using 8 clusters, for a modification from a $\Lambda$CDM scenario. This limit is in very good agreement with theoretical ones and an order of magnitude more stringent than the previous constraint obtained using clusters. We also quote a more conservative limit of $\Xi_1 = -0.061 \pm 0.074$. Finally, we comment on the tentative redshift dependence  ($\Xi_1(z)$), finding a mild preference ($ \lesssim 2\sigma$) for the same.

\end{abstract}

\date{\today\ at \currenttime\ of Roma} 
\maketitle


\section{Introduction}
\label{sec:Introduction}

To date, the majority of cosmological observations \cite{Adam:2015rua, Ade:2015xua} related to the space-time of the Universe can be explained by General Relativity (GR), particularly in the smaller scales.
To explain the observed late time acceleration \cite{Riess:1998cb, Perlmutter:1998np} with metric based GR framework, one can: i) introduce a cosmological constant \cite{Weinberg:1988cp, Bull:2015stt}, ii) introduce dark energy (DE) such as a dynamical scalar field (e.g., quintessence \cite{Caldwell:1997ii}, k essence \cite{Chiba:1999ka,ArmendarizPicon:2000ah})  or (iii) modify the gravitational coupling differing from GR at cosmological distance, known as the modified gravity (MG) theory \cite{Tsujikawa:2010zza,Nojiri:2010wj,Clifton:2011jh,Joyce:2016vqv}.

To test a large number of MG models, a generalized platform is necessary.
The degenerate higher-order scalar-tensor (DHOST) theory is claimed to be the most general class of a scalar-tensor theory, where a propagating scalar degree of freedom (DOF) is added on the top of two tensor DOF of metric based GR given under the general-covariance (see Appendix~\ref{app:dhost} for more details) \cite{Langlois:2015cwa,Crisostomi:2016czh,Achour:2016rkg,Motohashi:2016ftl,BenAchour:2016fzp} \cite{Kobayashi:2019hrl}. The DHOST theory includes many popular modified gravity models such as Brans-Dicke theory \cite{Brans:1961sx}, $f(R)$ gravity \cite{DeFelice:2010aj,Nojiri:2010wj}, covariant Galileon \cite{Deffayet:2009wt,Neveu:2013mfa,Neveu:2016gxp}, Horndeski \cite{Horndeski:1974wa,Kobayashi:2011nu}, transforming gravity \cite{Zumalacarregui:2013pma}, and GLPV theory \cite{Gleyzes:2014dya}. A large class of DHOST theories produces gravitational waves with velocity ($c_g$) equal to the velocity of light ($c$) in agreement with recent observations. These are referred to as viable DHOST$c_g^2=c^2$ theories in this article. 

In order to explain the small scale observations, the additional degrees of freedom (DOF) should be screened through a non-linear mechanism, and GR should be recovered on the small scale, known as the so-called screening mechanism. The Vainshtein screening \cite{Vainshtein:1972sx} is useful for the higher-order MG theories \cite{Babichev:2013usa,Brax:2004qh}. In Vainshtein screening, the gravitational potentials for the DHOST$c_g=c$ are modified inside the matter sources  \cite{Crisostomi:2017lbg,Langlois:2017dyl,Bartolo:2017ibw,Dima:2017pwp,Hirano:2019scf,Crisostomi:2019yfo}. Horndeski theory, a subset of DHOST theory, recovers GR in the small scale approximation \cite{Kimura:2011dc,Narikawa:2013pjr,Koyama:2013paa,Kase:2013uja}. However, Vainshtein screening for GLPV theory, another subset of DHOST theory, breaks down inside the matter \cite{Kobayashi:2014ida,Koyama:2015oma, Saito:2015fza,Sakstein:2015zoa, Sakstein:2015aac,Jain:2015edg,Babichev:2016jom}. DHOST theory has not been studied extensively on the galaxy cluster scale, other than a few bounds on $\Xi_1$ \cite{Sakstein:2016ggl, Salzano:2017qac}. However, as shown in \cite{Saito:2015fza}, one must always have  $\Xi_1> -1/6$ to guarantee a stable static solution for stars, and the upper bound  $\Xi_1< 7\times 10^{-3}$ to fulfill the consistency of the minimum mass for hydrogen burning in stars with the lowest mass red dwarf~\cite{Sakstein:2015zoa, Sakstein:2015aac}. Which will be very useful in order to compare with the constraints that can be obtained from the galaxy cluster scales. 

Given the task at hand, we utilize the XMM-Newton Cluster Outskirts Project (X-COP) data products \citep{Eckert:2016bfe, Ettori:2018tus, Eckert:2018mlz, Ghirardini:2018byi}, which consists of 12 clusters with well-observed X-ray emission and high signal to noise ratio in the Planck Sunyaev-Zel'dovich (SZ) survey \citep{Planck:2015lwi}, essentially providing both ICM temperature and pressure data over the large radial range of $0.2 \,{\rm Mpc} \lesssim r \lesssim 2 \, {\rm Mpc}$. Such a compilation is in fact very helpful to assess the mass profiles out to a large radial range and, as in the current work, subsequently test modifications to the gravitational potential. 

A similar approach relying on the same galaxy cluster phenomenology has been utilized in several previous analyses \citep{Sakstein:2016ggl} (see in particular \citep{Terukina:2013eqa, Wilcox:2015kna} for tests of Chameleon screening). In \citep{Sakstein:2016ggl} utilize both X-ray and weak lensing profiles of 58 stacked clusters, obtaining constraints on the modifications to gravitational potential which point towards consistency with GR. On the contrary, in this work for the very first time (as far as we are aware) we  exploit individual clusters with well-observed ICM to finally obtain a joint constraint, while simultaneously assessing a possible redshift dependence. More recently, \citep{Pizzuti:2020tdl} have presented generalized frameworks for testing these scenarios, also including the beyond Horndeski and DHOST scenarios, essentially constraining the Vainshtein screening. 

The paper is organized as follows. We provide a brief description of the DHOST theory in \Cref{sec:dhost}, while the modeling of the hydrostatic equilibrium in the case of a cluster of galaxies is summarized in \Cref{sec:pressure_profile}. The datasets utilized and method is described in \Cref{sec:data}. Finally we present and discuss our results in \Cref{sec:constraints}. In the current analysis, we assume $H_0 = 70 $ \ksM  and $\Omega_{\rm m } = 0.3$, when computing the critical density $\rho_{\rm c} (z) =   3H^2(z)/ 8\pi \GNeff$\footnote{For the case of GR $\GNeff \to \GN$, please see \Cref{sec:dhost} for details.}, where $H^2(z)/H^2_0 = \Omega_{\rm m}(1+z)^3 + (1 - \Omega_{\rm m})$. 

\section{Perturbed gravitational forces of DHOST in small spatial scale}
\label{sec:dhost}

Gravity is well tested at small scales. We expect to recover GR within the Vainstein radius (including inside the object), and full extended theory outside. However, for the DHOST theory, GR is not recovered everywhere inside the Vainstein radius, i.e., Vainstein screening breaks down inside the object. The gravitational potentials for the DHOST theory in the galaxy scales modifies to 
\cite{1711.07403, Crisostomi:2017lbg, Bartolo:2017ibw, Dima:2017pwp}

\begin{eqnarray}\label{exp:grav_force:phi}
\frac{d\Phi (r)}{dr}&=& \frac{G_{\rm N}^{\rm eff}{\cal M}(r)}{r^2}+\Xi_1 G_{\rm N}^{\rm eff}{\cal M}''(r)\,,
\\
\frac{d\Psi (r)}{dr}&=& \frac{G_{\rm N}^{\rm eff}{\cal M}(r)}{r^2}+ \Xi_2 \frac{G_{\rm N}^{\rm eff}{\cal M}'(r)}{r}+\Xi_3\, G_{\rm N}^{\rm eff}{\cal M}''(r)\,, \label{exp:grav_force:psi}
\end{eqnarray}
with the \textit{modified} or \textit{effective Newton's constant} $G_{\rm N}^{\rm eff}$ defined by the expression
\begin{eqnarray} \label{exp:geff}
\frac{1}{8\pi G_{\rm N}^{\rm eff}} &=& {2 F-2{F_X} X- \frac{3}{2}A_3 X^2}\equiv 2F (1+\Xi_0)\,,
\end{eqnarray}
and the dimensionless coefficients
\begin{equation} \label{exp:xi}
\left.\begin{aligned}
\Xi_1 &= -\frac{(4{F_X} -X A_3)^2}{16F A_3}\,, \\
\Xi_2 &= -\frac{2{X F_X}}{F}\,, \\ 
\Xi_3 &= \frac{16{F_X}^2-A_3^2 X^2}{16A_3 F}\,,\\
\Xi_0 &= -\frac{F_X X}{F}- \frac{3}{4} \frac{A_3 X^2}{F}\,.
\end{aligned} \right.
\end{equation}

From \eqref{exp:xi}, one can derive a consistency relation, $\Xi_2 = (2/\Xi_1) \left(\Xi_3^2 - \Xi_1^2\right)$, which allows to reduce to $(\Xi_1, \Xi_3)$ the number of parameters regulating the deviations from $\Lambda$CDM in the perturbations regime \cite{Langlois:2017dyl}. 

One can recover GR by setting $F= 1 / 2\kappa$ with $\kappa =8\pi \GN c^{-4}$, and $P, Q, A_3=0$ (see \Cref{app:dhost}), which would lead to $\Xi_{0,1,2,3}=0$.

In terms of the cosmological effective parameters (EFT parameters) \cite{Langlois:2017mxy}, 

\begin{equation} \label{exp:xi:EFT}
\left.\begin{aligned}
\Xi_1 &=-\frac{(\alpha_{\text{H}}+\beta_1)^2}{2(\alpha_{\text{H}}+2\beta_1)}\,,\quad
\Xi_2 =\alpha_{\text{H}}\,,\quad \\ 
\Xi_3 &=-\frac{\beta_1(\alpha_{\text{H}}+\beta_1)}{2(\alpha_{\text{H}}+2\beta_1)}\,,\quad 
\Xi_0 = -\alpha_{\text{H}} - 3\beta_1\,.
\end{aligned} \right.
\end{equation}

Therefore, 
$\GNeff \equiv \GNeff (\aH, \beta_1)$ and $\Xi_\mu \equiv \Xi_\mu(\aH, \beta_1)$  and can be explicitly written as, 

\begin{equation}
\left.\begin{aligned}
\GNeff &= [16 \pi F (1 + \Xi_0)]^{-1} \\
                &= [8 \pi M_{\star}^2 (1 + \Xi_0)]^{-1} \\
                &= \gN \GN/(1 - \alpha_{\rm H} - 3 \beta_1) \,,
\end{aligned} \right.
\label{eq: gneff}
\end{equation}
where the effective Planck mass $M_{\star} = M_{\rm Pl}/\gN$. Hereafter, we write $\GNeff = \gNtilde \GN$ for brevity. Note that it is  always possible to convert the parameterization from the effective ones to the physical ones after the analysis is performed. We elaborate on this in \Cref{sec:physical_parameters}. 

As the current modification is a linear order perturbation to the metric, the EFT parameters depend only on the background. Thereby, the $G_{\rm N}^{\rm eff} (t)$ and $\Xi_i(t)$ are solely functions of time.

\section{ICM Pressure profile of galaxy cluster}
\label{sec:pressure_profile}
As per the standard model, the galaxy clusters consists of Dark Matter (DM) ($\sim$ 85\% of the total mass), the hot ionized hydrogen and helium gas called intracluster medium (ICM) ($\sim$12\%), and visible stars and galaxies ($\sim$3\%).  The galaxy clusters are expected to retain all the ICM and stars accreted since the formation epoch in the deep gravitational well created by DM. On one hand, the hot ICM is visible in the X-ray band through thermal bremsstrahlung and line emission. On the other hand, the ICM can be detected at millimetre wavelengths through the distortion of the cosmic microwave background (CMB) induced by inverse Compton scattering, called the SZ effect. Since ICM traces the main baryonic component, X-ray and SZ observations are crucial for testing gravity in the clusters and filaments, which is however a challenging task.

Among the possible approaches to infer the total mass, one possible method is to assume that the ionized gas is in \textit{hydrostatic equilibrium} with the gravitational potential, $\Phi$,  (mostly created by DM \citep{Kravtsov12}). In the hydrostatic equilibrium, 
\begin{equation}\label{eq:hydro_equil}
\frac{1}{\rho_{\rm gas}(r)}\frac{{\rm d} P_{\rm gas} (r)}{{\rm d} r}=-\frac{{\rm d}\Phi (r)}{{\rm d} r} \,. 
\end{equation}

where we have implicitly assumed spherical symmetry. $P_{\rm gas} (r)$ is the gas pressure radial profile while the gas density is: 
\begin{eqnarray} 
 \rho_{\rm gas} (r) &=& n_{\rm gas}(r)\, \mu m_p \,, \\
 &\sim & 1.8 n_e(r)\, \mu m_p \,,
 \label{exp:rho_gas}
\end{eqnarray} 
where $n_{\rm gas} (r)$ is the gas density, the sum of the electron and proton number densities, i.e., $n_{\rm gas} (r) = n_e (r) + n_p (r)$, $\mu$ is the mean molecular weight in a.m.u. ($\mu = 0.61$ is the corresponding mean molecular weight in atomic mass unit), and $m_p$ is the proton mass ($\sim$  1 atomic mass unit, $m_u$) therefore $n_p\approx 0.8 n_e$ which gives $n_{\rm gas} (r) \approx 1.826\, n_e (r) $.

Among the possible models for  $n_e(r)$, such as the $\beta$ profile, or double $\beta$ profile, or the more recent Vikhlinin profile, we choose the latter \cite{Vikhlinin:2005mp}. We use the simplified Vikhlinin\footnote{Here we have utilized the Vikhlinin profile with only 6 parameters, neglecting a second part of the parametric model which adds 3 more parameters. We comment on the implications for this at in \Cref{sec:Mass_comparison}.} parametric model setting,  

\begin{equation}
\frac{n_e(r)}{n_0} = \frac{(r/r_c)^{-\alpha/2} [1 + (r/r_s)^{\gamma}]^{-\varepsilon/(2 \gamma)}}
{[1 + (r/r_c)^2]^{(3/2) \beta - \alpha/4}}
\label{eq: vikgas}
\end{equation}
where $n_{0}$ is the cluster central density, $r_s$ is the transition radius at which the logarithmic slope changes, and $r_c$ is the core radius. The $\beta$ and $\epsilon$ parameters give the inner and outer profile slope, respectively. The $\gamma$ parameter gives the width of the transition in the profile. We fix $\gamma = 3$ as suggested in \cite{Vikhlinin:2005mp,astro-ph/0504098}, so the electron density parameter space is given by ${\Theta_{e}} = \{n_0, \alpha, \beta, \varepsilon, r_{\rm c}, r_{\rm s}\}$. Moreover, we limit $\epsilon< 5$ to avoid nonphysical sharp features in the electron density profile, which we however find to not make a major difference on the mass estimates once rest of the parameters are marginalized upon. 

The right side of \Cref{eq:hydro_equil} is governed by the gravitational potential, which depends on the assumed theory of gravitation. In the standard GR scenario we have,
\begin{eqnarray}\label{exp:grav_force:phi:xray}
\frac{d\Phi (r)}{dr}&=& \frac{G_{\rm N}}{r^2}{M}_{\rm HSE}(r). 
\end{eqnarray}

For the DHOST theory elaborated in \Cref{exp:grav_force:phi}, the gravitational potential can be written as,

\begin{equation}
\frac{d\Phi(r)}{dr} = \frac{G_{\rm N}^{\rm eff} M_{\rm HSE}(r)}{r^2} + \Xi_{1} G_{\rm N}^{\rm eff} \frac{d^2M_{\rm HSE}(r)}{dr^2}.
\label{eq: dphidr}
\end{equation}
In \Cref{eq: dphidr}, $G_{\rm N}^{\rm eff}$ is the effective gravitational constant which is related to the Newton one $\GN$ in the DHOST model through \Cref{eq: gneff}. One could ideally write $\GNeff = \gNtilde \GN$, where $\gNtilde$ is a redshift dependent function. However, it is evident that $\gNtilde$ can not be constrained by the data on the pressure profile alone since it only enters as a multiplicative term fully degenerate with the total mass. Also, for viable DHOST theories, it is expected that $\GNeff$ does not deviate from $\GN$ so we perform the analysis fixing $\gNtilde =1$, and then $\gNtilde \neq 1$. 

Under the assumption that the cluster mass is dominated by the dark matter component, we model the mass density using the NFW profile \cite{Navarro:1995iw} given as,
\begin{equation}
\rho(r) = \frac{\rho_s}{(r/r_s) (1 + r/r_s)^2}
\label{eq: nfwrho}
\end{equation}
where $\rho_s$ is a characteristic density, and $r_s$ the radius where the logarithmic slope $s = d\ln{\rho}/d\ln{r}$ takes the isothermal value $s = -2$. Now the mass profile can be straightforwardly obtained and it is conveniently rewritten as 

\begin{equation}
M(<r) = M_{500} \frac{\ln{(1 + c_{500} x)} - c_{500} x/(1 + c_{500} x)}{\ln{(1 + c_{500})} - c_{500}/(1 + c_{500})}
\label{eq: massnfw}
\end{equation}
with $x = r/R_{500}$, $c_{500} = \R/r_s$\footnote{{$\R$ in the DHOST case already takes into account the variation due to modification to gravity on the background through $\rho_{\rm c}(z)$, hence having $\R^{3} \sim \gNtilde\M$.}} the halo concentration, and

\begin{equation}
\M = 500\, \frac{4}{3} \pi  \rho_{\rm c}(z) \R^{3} \ .
\label{eq: mdeltadef}
\end{equation}

From \Cref{eq: massnfw}, we obtain, 

\begin{equation}
M''(r)= \frac{M_{500}}{R_{500}^2} \frac{c_{500}^2 (1 - c_{500} x) (1 + c_{500} x)^{-3}}{\ln{(1 + c_{500})} - c_{500}/(1 + c_{500})} \ ,
\label{eq: d2mdr2nfw}
\end{equation}

which can be plugged into \Cref{eq: dphidr} together with \Cref{eq: massnfw} to get the modified gravitational potential. Now, the parameters for the mass profile are $\Theta_{\rm M} = \{\M, c_{500}\}$. One can get the SZ pressure profile for the galaxy cluster by integrating  \eqref{eq:hydro_equil} and using \eqref{exp:grav_force:phi}, 

\begin{widetext}
\begin{equation}
P^{\rm th}(r) = P^{\rm th}(0) - 1.8 \mu m_{\rm p}\int_0^r n_{\rm e}(\tilde r)\left[\frac{G_{\rm N}^{\rm eff} {M}_{\rm HSE}(\tilde r)}{\tilde r^2}+ \Xi_1 G_{\rm N}^{\rm eff}{ M''}_{\rm HSE}(\tilde r) \right]{\rm d} \tilde r
\label{eq:pressure:profile:sz}
\end{equation}
\end{widetext}
where, $\GNeff = \gNtilde\times\GN$ and $M_{\rm HSE}$ are modeled using the NFW profile \Cref{eq: massnfw}. {Note that here $P^{\rm th}(0)$ is not a free parameter and can be reconstructed form the integration assuming $P^{\rm th}(r\to \infty) \sim 0$.} Effectively, the DHOST modification to the standard GR case, is dictated by the combination of two parameters $\{\Xi_1, \gNtilde\}$, which we hereafter denote as $\Theta_{\rm DHOST}$. The formalism adopted here is in fact termed as \textit{Backward} method, where a given parametric profile is assumed for the mass model and the pressure is obtained by the integration of the same. In contrast, assuming a parametric profile for the pressure is termed as \textit{Forward} method, in which case assessing the modifications to gravitational potential is not possible.

\section{Data and Likelihood}
\label{sec:data}

We utilize the collection of clusters within the XMM-Newton Cluster Outskirts Project (X-COP)\footnote{The datasets are publicly available at the following link: \href{https://dominiqueeckert.wixsite.com/xcop/about-x-cop}{https://dominiqueeckert.wixsite.com/xcop/about-x-cop}} catalog \citep{Eckert:2016bfe} with joint X-ray and millimeter observations. The compilation consists of 12 massive ($10^{14} \, M_{\odot} \lesssim \M \lesssim 10^{15} \, M_{\odot} $) galaxy clusters in the redshift range $0.04 < z < 0.1$, selected for high signal-to-noise ratios (${\rm S/N}>12$) in the \textit{Planck} SZ survey \cite{Ade:2013skr}. The physical observables of interest are, $i)$ Electron density, $ii)$ Temperature of gas, in the X-ray and the $iii)$ Pressure of the gas observed in the SZ. These observables indeed present a suitable scenario to test the formalism described in \Cref{sec:pressure_profile} and the modifications of gravity on cluster scales.

\begin{figure}
\includegraphics[scale=0.5]{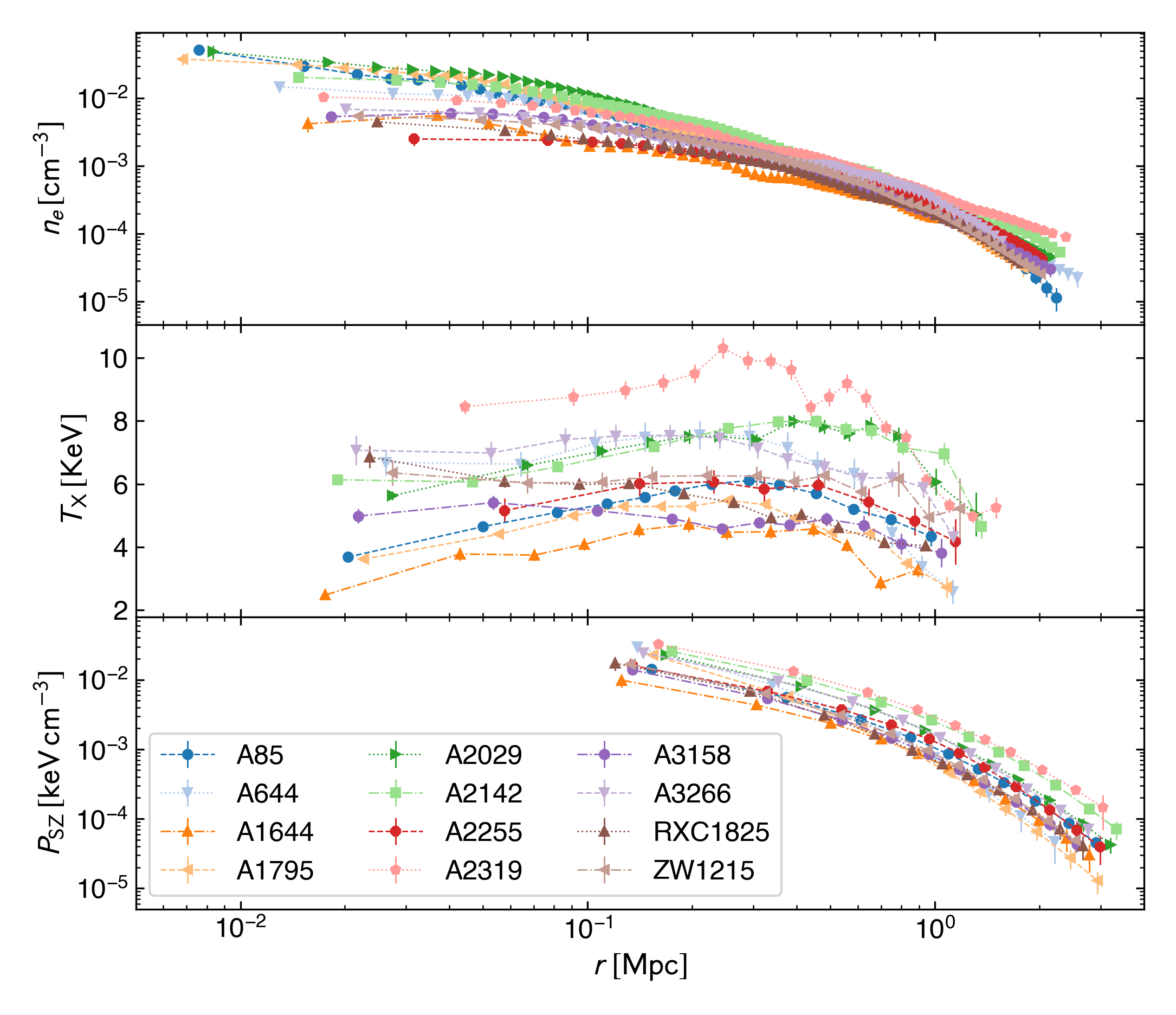}
 \caption{\textit{Top}: Electron density radial profiles obtained using the L1 regularization method \citep{Ghirardini:2018byi}. \textit{Middle}: Temperature of the X-ray emitting gas. \textit{Bottom}: The SZ electron pressure profile. This figure is comparable to the Fig.1 of \citet{Ettori:2018tus}, except that we have removed the self-similar scaling, as reported therein. Also note that they show the reconstructed electron density profiles. }
 \label{fig:All_data}
\end{figure}

In \Cref{fig:All_data}, we present the collection of the 12 clusters data currently analyzed. Firstly, we re-scale the data for the self-similar normalization (i.e, the data were provided as a function of $R/\R$, for each cluster), which they were provided with and show the radial profiles. 

In order to constrain the DHOST modification parameters ($\Theta_{\rm DHOST}$), the halo mass profile parameters ($\Theta_{M}$), and the electron density parameters (${\Theta}_e$), we perform a joint fit to the measured $n_e(r)$, $P_{\rm X}(r)$ and $P_{\rm SZ}(r)$ data from the X-COP sample. The total number of parameters is then,
\begin{equation}
{\Theta} = \Theta_e \cup {\Theta}_{\rm M} \cup \Theta_{\rm DHOST}.
\label{eq:theta}
\end{equation}
accounting for a total of 10 parameters in the DHOST scenarios and 8 within GR case, having $\Theta_{\rm DHOST} \equiv \{0.0, 1.0\}$. Similar to \cite{Ettori:2018tus, Ghirardini:2017apw}, we define the likelihood function for the observed data denoted as ($\rm obs$) and for the model ($\rm mod$) as in the following way,

\begin{eqnarray}
-2 \ln{{\cal{L}}}  &=&  ({\bf P}^{\rm obs}_{\rm SZ} - {\bf P}^{\rm mod}_{\rm SZ})\Sigma_{\rm TOT}^{-1} ({\bf P}^{\rm obs}_{\rm SZ} - {\bf P}^{\rm mod}_{\rm SZ})^{T}{+ \ln\mid \Sigma_{\rm TOT}\mid}  \nonumber\\
&+& \sum_{j=1}^{N_{P_{\rm X}}} \left[ \frac{(P_{{\rm X},i}^{\rm obs} - P_{{\rm X},i}^{\rm mod})^2}{\sigma_{P_{{\rm X},i}}^2 + \sigma_{P,{\rm int}}^2} + \ln({\sigma_{P_{{\rm X},i}}^2 + \sigma_{P,{\rm int}}^2}) \right ] \nonumber\\
&+&\sum_{i=1}^{N_{\rm n_{e}}} \left[ \frac{(n_{e,i}^{\rm obs} - n_{e,i}^{\rm mod})^2}{\sigma_{n_{e,i}}^2} \right]
\label{eq:chisqeff}
\end{eqnarray}

where the first term accounts for the co-varying SZ pressure data, with the covariance matrix $\Sigma_{\rm TOT} = \Sigma_{\rm P} +\Sigma_{{\rm P, int}}$\footnote{$\Sigma_{{\rm P, int}}$ is a diagonal matrix of $\sigma_{P,{\rm int}}^2$, accounting for an additional intrinsic scatter for the $\Px$ and $\Psz$ data points.}, $\Sigma_{\rm P}$ being the covariance matrix of the $\Psz$ data itself. The second and the third term account for the X-ray pressure and the electron density data, respectively. Note that following \citep{Ghirardini:2017apw, Ettori:2018tus}, we include an additional intrinsic scatter to the measurement uncertainty of the pressure and the temperature datasets. Our Likelihood is practically identical to the one implemented in \cite{Ettori:2018tus}, except that we have utilized the publicly available electron density data instead of the emissivity $(\epsilon)$ data utilized therein. The electron density data taken into account in our analysis is obtained through the L1 regularization method \citep{Ghirardini:2018byi}, in contrast to the smooth reconstructed electron density profiles utilized in \cite{Ettori:2018tus}. Their reconstructed profiles were obtained through the multiscale fitting \cite{Eckert:2015rlr} procedure to the emissivity data. The multiscale fitting of the emissivity data essentially provides smooth reconstructed electron density with minimal features and in this way the reconstructed electron density profiles show limited features resulting in a better estimate of the hydrostatic mass, having better control over the gradient. The L1 regularization data we utilize here shows more pronounced features\footnote{The gas clumping in the electron density data should be accounted for while obtaining the smoothed reconstructed profiles. However, the possible bias due to the clumping is mild \cite{Eckert:2013faa}. } and represents the electron density observations  better to study the shape of the profile \citep{Ghirardini:2018byi}. However, we proceed to utilize the data here, expecting a mild to moderate deviation from the hydrostatic masses reported in \citep{Ettori:2018tus}. This is partly due to the fact that, in \citep{Ghirardini:2018byi}, it has been shown that the two methods agree incredibly well and show an utmost scatter of the order $\sim 5\%$ at a given radius, at least for the clusters at hand in the current compilation. And that using the simplified Vikhlinin profile (\Cref{eq: vikgas}) provides smooth reconstructions for the electron density and varies even more minimally with respect to the multiscale fitting reconstruction. However, the L1 based electron density data also have mild correlations among the data points\footnote{ Dominique Eckert and Stefano Ettori in private communication.}, which we have not taken into account here and could potentially have an impact on the final mass estimates. In any case, as the current analysis is performed to assess the deviations from GR to the DHOST scenarios, we expect the use of $n_e$ data as here, instead of the $\epsilon$ data, to equivalently affect the GR and the DHOST scenarios. This we indeed verify a posteriori, finding minimal correlations between the $\Theta_{\rm M}$ and $\Theta_{e}$ parameters. 

We also utilize the likelihood using the $\Tx$ data to perform validations, which is the direct observable while $\Px$ is obtained by $\Tx$ data and the measured electron density. In this case, the second term in the Equation (\ref{eq:chisqeff}) is replaced with the $\Tx$ data and the rest remains unchanged. As shown in \citep{Ghirardini:2017apw}, a small intrinsic scatter ($\sigma_{\rm int}$) is modeled up on $\log(P)$ such that the intrinsic error on the pressure can be written as $\sigma_{P, {\rm int}} = P\, \sinh({\sigma_{\rm int}})$. In our analysis we utilize the $\sinh({\sigma_{\rm int}})$ as a free parameter\footnote{It is usually convenient to sample on the $\sinh({\sigma_{\rm int}})$ as a free parameter, to avoid loss of probability for the posterior of parameter $\sigma_{\rm int} \xrightarrow{} 0$ \citep{Hogg:2017akh}. } and the intrinsic scatter is propagated to the temperature profile as $\sigma_{T, {\rm int}} = T\, \sinh({\sigma_{\rm int}})$. Therefore, the intrinsic scatter is included to the parameter array $\Theta$ in \Cref{eq:theta}, as an additional parameter.

Finally, the posterior defined in Equation (\ref{eq:chisqeff}), is utilized to perform a Bayesian analysis through MCMC sampling. We use the \texttt{emcee}\footnote{\href{http://dfm.io/emcee/current/}{http://dfm.io/emcee/current/}} package \citep{Foreman-Mackey13, Hogg:2017akh}, which implements an affine-invariant ensemble sampler. We utilise publicly available \texttt{ChainConsumer}\footnote{\href{https://github.com/Samreay/ChainConsumer/tree/Final-Paper}{https://github.com/Samreay/ChainConsumer/tree/Final-Paper}.} package \citep{Hinton16},  to perform analysis of the chains and plot the contours. To asses the evidence for the DHOST modification we compute the Bayesian evidence \cite{Trotta:2008qt, Trotta:2017wnx, Heavens:2017hkr}, through the \texttt{MCEvidence} package \cite{Heavens:2017afc}\footnote{We utilized the \texttt{MCEvidence} package publicly available at \href{https://github.com/yabebalFantaye/MCEvidence}{https://github.com/yabebalFantaye/MCEvidence}.}. We impose uniform flat priors on all the parameters, in particular we impose $\Xi_1 \in \{-2.0, 2.0\}$ and $\gNtilde \in \{0.001, 2.0\}$, for the DHOST parameters. We also separately test the effects of different initialization of the mass profile parameters for the walkers in the affine-invariant procedure, finding no implications for the posteriors. Note that the information on the prior volume is important in assessing the Bayesian evidence. Through the Bayes' rule the posterior distribution w.r.t. the parameters $(\Theta)$ of the given model $\mathcal{M}(\Theta)$ and the observations $\mathcal{D}$, can be written as,

\begin{equation}
    \label{eqn:BayesRule}
    p(\Theta|\mathcal{D},\mathcal{M}) = \dfrac{p(\mathcal{D}|\Theta,\mathcal{M})\pi(\Theta|\mathcal{M})}{p(\mathcal{D}|\mathcal{M})},
\end{equation}
where $\pi(\Theta|\mathcal{M})$ is the prior volume and $p(\mathcal{D}|\mathcal{M})$ is the Bayesian `evidence' $(\cal{B})$. A comparison of the evidence can be utilized to assess the preference of a given model $\mathcal{M}_{1}(\Theta_{1})$ over the base model, which in our case are the DHOST and GR, respectively. As is the usual practice we contrast the Bayesian evidence using the Jeffrey's scale \citep{Jeffreys:1939xee}, where $\Delta\log(\cal{B}) $ $ < 2.5 $ and $\Delta\log(\cal{B}) $ $> 5$, imply a weak and strong preference for the extended model against the base model, respectively.

\section{Results and Discussion}

\begin{figure}[!t]
\centering
\includegraphics[scale=0.5]{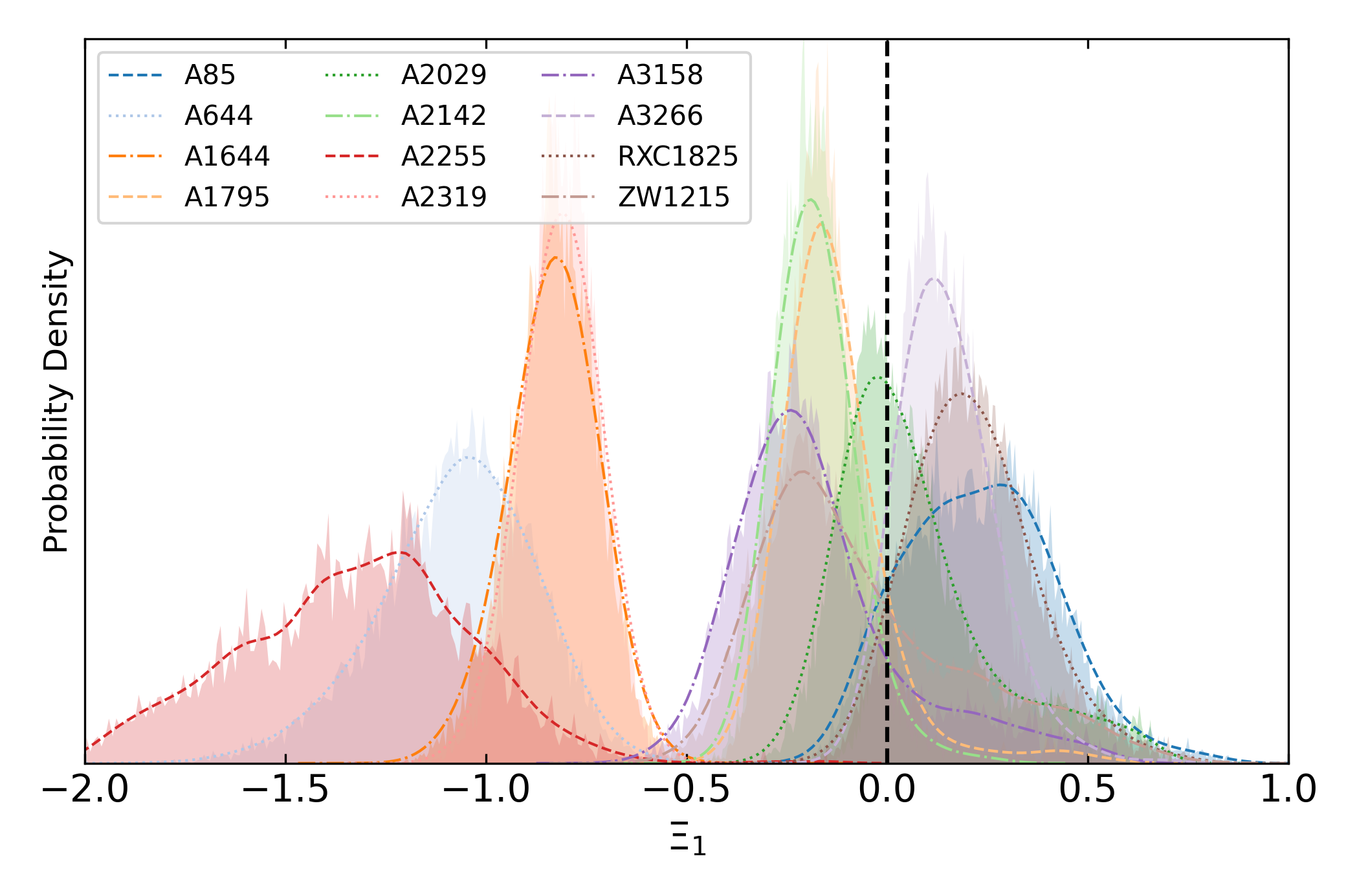}
 \caption{Probability density for the distribution of the parameter $\Xi_1$. We perform a simple Gaussian kernel density estimation to obtain the smooth profiles, which are over-plotted on the distributions. The vertical dashed line marks the GR ($\Xi_1 = 0$) case.}
 \label{fig:XI}
\end{figure}

We begin by performing the analysis for the standard GR case, aiming to reproduce the results of \citetalias{Ettori:2018tus}. The constraints obtained in our analysis are shown in the first three columns of \Cref{tab:Constraints_PP}. We find a good agreement with the constraints reported in \citetalias{Ettori:2018tus}, however, note that here we have reported asymmetric uncertainties accounting for the maximum likelihood, in contrast to the symmetric dispersion between the $84^{\rm th}$ and $16^{\rm th}$ percentiles presented therein. In general, we find a good agreement within the order of $\lesssim 1.5\sigma$, for the mass parameter $M_{500}$, also with very similar uncertainty estimates. We stress that a difference of the order $\sim 1.5\sigma$ is yet within the usual expectations of a bias using different methods or datasets of estimating $M_{500}$. The difference between the analysis in \citetalias{Ettori:2018tus} and here is within the modeling of the electron density data, where they utilize the smooth reconstructed profiles from the multiscale fitting, whereas we utilize the publicly available L1 regularization based data points (see \citep{Ameglio:2008ip, Croston:2006nq}). The presence of clumpiness, if assumed to be smooth essentially yields an underestimate of the total mass \citep{Mathiesen:1999pn, Ettori:2013tka}, and should indicate a systematic bias in our mass estimates. However, as can be seen in \Cref{fig:mass_comp}, our mass estimates do not show such a systematic behavior allowing us to validate the formalism implemented here. Therefore, owing to the agreement with the analysis in \citetalias{Ettori:2018tus} and no systematic behavior in mass constraints we find our results to be consistent with those presented in \citetalias{Ettori:2018tus} and proceed with further analysis of comparing the GR and modified gravity scenarios, where any difference from the earlier results will be equivalently present in both cases and is not expected to bias our inferences for the DHOST scenario. We present a more detailed comparison of masses both in the GR and the DHOST cases in \Cref{sec:Mass_comparison}.  


{\renewcommand{\arraystretch}{1.25}%
    \setlength{\tabcolsep}{4pt}%
\begin{table*}[!ht]
    \centering
    \caption{Constraints obtained for the parameters of the mass profiles in both the GR and the DHOST scenarios. We show the maximum likelihood statistics as the $68\%$ C.L. limits. Here we have utilized the $\Psz+\Px$ datasets. In the last column, we show the Bayesian evidence for the DHOST modification w.r.t GR, where a negative value indicates that GR is preferred. The four non-NFW clusters are denoted with the superscript $^\dagger$ and additionally, the cluster A2319 carries $^\ddagger$, representing the large non-thermal effects reported in \citep{Eckert:2018mlz}. }
    \label{tab:Constraints_PP}
    \begin{tabular}{cccc|ccccc|c}
    \hline
        Cluster& \multicolumn{3}{c}{GR} & \multicolumn{5}{c}{DHOST} &  \\ 
        & $c_{500}$ & $M_{500}\,$ & $R_{500}\, $ & $c_{500}$ & $M_{500}\, $ & $R_{500}\,$ & $\Xi_1$ & $\tilde{\gamma}_{\rm N}\times M_{500}\, $ & $\Delta\log(\cal{B})$ \\ 
        & & $ [10^{14}\, M_{\odot}]$ & $[{\rm Mpc}]$ & & $[10^{14}\, M_{\odot}]$ & $[{\rm Mpc}]$ & & $[10^{14}\, M_{\odot}]$ \\
        \hline
        \hline
        
        \begin{tabular}{@{}c@{}}A85 \\ (z=0.0555)\end{tabular} & $2.05^{+0.09}_{-0.06}$ & $6.14^{+0.14}_{-0.21}$ & $ 1.270^{+0.010}_{-0.015}$ & $1.16^{+0.57}_{-0.32}$ & $4.10^{+3.43}_{-0.63}$ & $1.292^{+0.017}_{-0.030}$ & $0.30^{+0.11}_{-0.27}$ &  $6.46^{+0.26}_{-0.45}$ & $-3.7$\\
        
        \begin{tabular}{@{}c@{}}A644$^\dagger$ \\ (z=0.0704)\end{tabular} & $4.22^{+0.31}_{-0.17}$ & $4.93^{+0.25}_{-0.18}$ & $1.175^{+0.020}_{-0.015}$ & $6.76^{+0.28}_{-0.33}$ & $3.80^{+5.13}_{-0.71}$ & $0.980^{+0.028}_{-0.030}$ & $-1.04^{+0.18}_{-0.19}$ & $2.85^{+0.24}_{-0.26}$ & $46.5$\\ 
        
        \begin{tabular}{@{}c@{}}A1644$^\dagger$ \\ (z=0.0473)\end{tabular} & $0.95^{+0.10}_{-0.10}$ & $3.00^{+0.17}_{-0.15}$ & $1.003^{+0.019}_{-0.017}$ & $2.89^{+0.23}_{-0.20}$ & $2.03^{+3.77}_{-0.67}$ & $0.844^{+0.020}_{-0.027}$ & $-0.837^{+0.119}_{-0.090}$  & $1.78^{+0.13}_{-0.16}$ & $39.9$\\ 
        
        \begin{tabular}{@{}c@{}}A1759 \\ (z=0.0622)\end{tabular} & $3.08^{+0.15}_{-0.10}$ & $4.59^{+0.18}_{-0.12}$ & $1.150^{+0.015}_{-0.010}$ & $3.71^{+0.47}_{-0.35}$ & $2.88^{+2.69}_{-0.81}$ & $1.101^{+0.032}_{-0.035}$ & $-0.169^{+0.111}_{-0.090}$ &  $4.03^{+0.34}_{-0.39}$ & $-2.9$\\ 
        
        \begin{tabular}{@{}c@{}}A2029 \\ (z=0.0773)\end{tabular} & $3.14^{+0.12}_{-0.17}$ & $7.84^{+0.33}_{-0.26}$ & $1.369^{+0.019}_{-0.015}$ & $3.31^{+0.49}_{-0.84}$ & $6.0^{+4.9}_{-1.6}$ & $1.352^{+0.089}_{-0.016}$ & $-0.04^{+0.19}_{-0.12}$ & $8.04^{+0.96}_{-0.88}$ & $-3.0$ \\ 
        
        \begin{tabular}{@{}c@{}}A2142 \\ (z=0.0909)\end{tabular} & $2.25^{+0.10}_{-0.12}$ & $8.30^{+0.33}_{-0.28}$ & $1.389^{+0.017}_{-0.017}$ & $2.86^{+0.33}_{-0.29}$ & $4.55^{+6.06}_{-0.92}$ & $1.326^{+0.040}_{-0.024}$ & $-0.203^{+0.101}_{-0.079}$ & $7.21^{+0.65}_{-0.40}$ & $0.4$\\ 
        
        \begin{tabular}{@{}c@{}}A2255$^\dagger$ \\ (z=0.0809)\end{tabular} & $0.68^{+0.13}_{-0.10}$ & $5.02^{+0.31}_{-0.26}$ & $1.180^{+0.023}_{-0.021}$ & $2.44^{+0.15}_{-0.21}$ & $5.2^{+2.4}_{-2.1}$ & $0.953^{+0.046}_{-0.043}$ & $-1.1^{+0.26}_{-0.32}$ &  $2.66^{+0.38}_{-0.36}$ & $28.2$ \\ 
        
        \begin{tabular}{@{}c@{}}A2319$^{\dagger,\, \ddagger}$ \\ (z=0.0557)\end{tabular} & $3.40^{+0.13}_{-0.09}$ & $7.15^{+0.16}_{-0.09}$ & $1.336^{+0.010}_{-0.006}$ & $5.14^{+0.13}_{-0.16}$ & $8.83^{+3.63}_{-2.26}$ & $1.151^{+0.020}_{-0.016}$ & $-0.827^{+0.108}_{-0.076}$ &  $4.57^{+0.23}_{-0.19}$ & $109.6$\\ 
        
        \begin{tabular}{@{}c@{}}A3158 \\ (z=0.0597)\end{tabular} & $1.81\pm 0.12$ & $4.21^{+0.19}_{-0.14}$ & $1.119^{+0.016}_{-0.012}$ &
        $2.62^{+0.38}_{-0.56}$ & $2.46^{+2.90}_{-0.62}$ & $1.054^{+0.057}_{-0.029}$ & $-0.23^{+0.15}_{-0.18}$ &  $3.51^{+0.57}_{-0.31}$ & $-0.4$\\ 
        
        \begin{tabular}{@{}c@{}}{A3266} \\ (z=0.0589)\end{tabular} & $0.93^{+0.10}_{-0.10}$ & $9.90^{+0.57}_{-0.59}$ & $1.489^{+0.027}_{-0.030}$ & $0.71^{+0.17}_{-0.20}$ & $6.3^{+4.0}_{-1.8}$ & $1.455^{+0.045}_{-0.055}$ & $0.100^{+0.137}_{-0.079}$ & $9.23^{+0.87}_{-0.99}$ & $-1.3$ \\ 
        
        \begin{tabular}{@{}c@{}}RXC1825 \\ (z=0.0650)\end{tabular} & $2.16^{+0.15}_{-0.12}$ & $4.11^{+0.15}_{-0.13}$ & $1.108^{+0.013}_{-0.012}$ & $1.54^{+0.44}_{-0.43}$ & $2.68^{+2.49}_{-0.46}$ & $1.130^{+0.016}_{-0.018}$ & $0.17^{+0.17}_{-0.13}$  & $4.37^{+0.17}_{-0.22}$ & $-0.3$ \\ 
        
        \begin{tabular}{@{}c@{}}ZW1215 \\ (z=0.0766)\end{tabular} & $1.32^{+0.11}_{-0.14}$ & $7.82^{+0.51}_{-0.50}$ & $1.368^{+0.029}_{-0.029}$ & $1.98^{+0.51}_{-0.84}$ & $7.23^{+4.31}_{-2.85}$ & $1.331^{+0.041}_{-0.076}$ & $-0.21^{+0.27}_{-0.18}$ &  $7.14^{+0.65}_{-1.18}$ & $-4.5$ \\ 
        \hline
    \end{tabular}
    \end{table*}
}

\subsection{Constraints on the DHOST modification}

The pressure profile for the DHOST gravity is estimated following \Cref{eq:pressure:profile:sz}, which now accounts for the modified background evolution and the small scale clustering effects through the effective parameters $\gNtilde$ and $\Xi_1$, respectively. We show the constraints on the mass and the DHOST parameters in \Cref{tab:Constraints_PP}, under the column titled DHOST. To begin with, we notice that the clusters A644, A1644, A2319 and A2255 show large negative values for $\Xi_1$ parameter, significantly different from the GR ($\Xi_1 = 0$) expectation. However, it has already been shown in \citetalias{Ettori:2018tus} that these same four clusters present a mild preference for mass profiles other than the NFW one, with a mild-to-moderate evidence for either the King approximation to Isothermal sphere (A2225, A2319), the Burkert (A644), or the Hernquist (A1644) profiles. Moreover, these 4 clusters are also outliers with respect to the $c_{200}-M_{200}$ scaling relation. Therefore, we do not immediately take our constraints for these clusters as evidence for modification of gravity, but warn the reader that any inferences should be made with caution. Hereafter we refer to these four clusters as `non-NFW' clusters for the ease of discussion and represent them with a superscript $^{\dagger}$ in the tables. This clearly indicates a degenerate scenario between an assumption of the mass profile in the GR case and a modification to the gravity itself, such as the DHOST theory in the current work.

\begin{figure}[!t]
\centering
\includegraphics[scale=0.47]{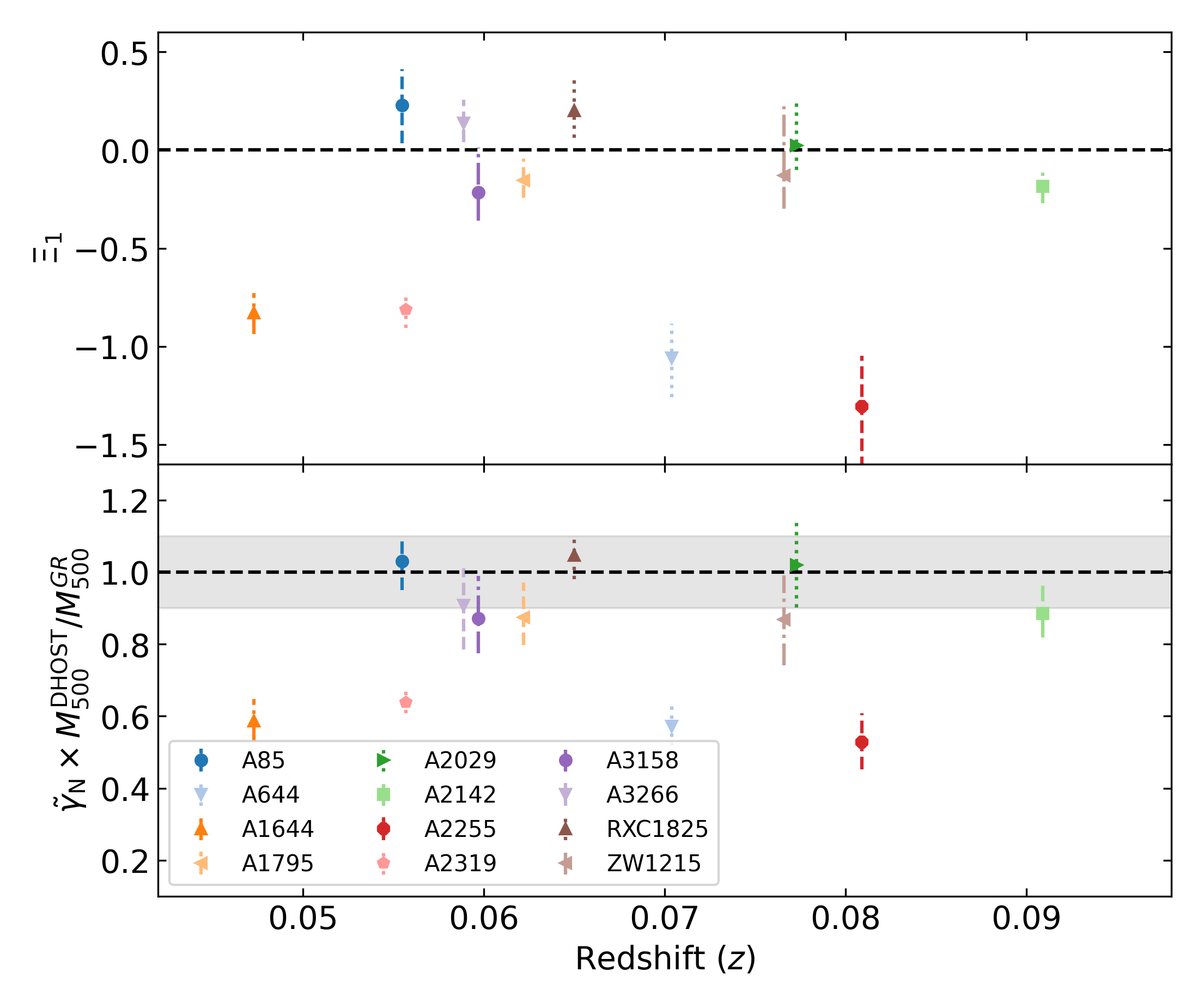}
 \caption{Constraints on $\Xi_1$ and the ratios of $\gNtilde\times M_{500}^{\rm DHOST}$ and $M_{500}^{\rm GR}$, plotted against the redshift of the cluster. The horizontal dashed lines in the top and bottom panels mark the GR case with $\Xi_1 = 0$ and $\gNtilde = 1$, respectively. The shaded gray region represents a $10\%$ deviation from GR for the mass parameter. }
 \label{fig:Xi_gamma_redshift}
\end{figure}

{In \Cref{fig:XI} we show the posterior probability density for the parameter $\Xi_1$ obtained for each of the clusters. It is evident that the four non-NFW clusters have posteriors notably far from the $\Xi_1 =0$, with a $>95\%$ C.L upper limit of $\Xi_1<-0.5$. On the contrary, the other 8 clusters are scattered around $\Xi_1 = 0$ showing no immediate preference for either a negative or positive values for this parameter. Note that these results are obtained under the prior assumption that NFW mass profile provides the best description of data in the GR scenario, as shown in \citetalias{Ettori:2018tus}. Assuming an incorrect or less-preferred mass profile lowers the quality of fit within the GR case and could be partially compensated by the additional DHOST degree of freedom showing a mislead preference for modifications to gravity. In this context, we find it convenient that the validation for an assumption of the NFW mass profile has already been performed in \citetalias{Ettori:2018tus}. It is very much possible that modeling the four non-NFW clusters with a different profile would push the constraints on $\Xi_1$ back towards the GR value. We do not perform this analysis here since the large majority of the clusters we consider is well fitted by the NFW profile. At a face-value the significance at which the non-NFW mass profiles are preferred for these 4 clusters is much less than the evidence for a DHOST modification we obtain here through the Bayesian evidence (see Fig 2. of \citetalias{Ettori:2018tus}). However, we prefer to be conservative, and do not draw any definitive conclusion from these considerations.}

As described earlier, within the DHOST scenario, $\M$ and $\gNtilde$ are degenerate parameters so that they are both not well constrained when letting them free to independently vary in the the MCMC analysis because of their expected strong correlation. Therefore, we also present in \Cref{tab:Constraints_PP} the derived constraints on $\gNtilde \times \M$, which encompasses the total deviation from the GR case and will be equal to the mass estimates in DHOST when $\gNtilde=1$ is assumed. As expected, one can immediately notice that $\gNtilde \times \M$ is better constrained than the $\M$ alone. In the bottom panel of \Cref{fig:Xi_gamma_redshift}, we show the redshift variation of the same, normalized\footnote{The normalization is performed by separately drawing $\sim 10000$ random samples from the MCMC chains of DHOST and the GR analyses. Therefore, this is indicative of the total variation in $\gNtilde\times\M$ from GR when marginalizing on $\Xi_1$, which is $0$ within GR.} to $\M$ in the GR case. As in the case of $\Xi_1$, the four non-NFW clusters once again show a large deviation from $\gNtilde\times \M^{\rm DHOST}/\M^{\rm GR} \sim 1$, which indicates either a very low total mass in comparison to GR or that $\gNtilde << 1$\footnote{In principle, the value of $\gN$ is expected to be $\sim 1$, unless $1-\aH - 3\bone <<1$, not implying a drastic change in $\GNeff$, which will be an outlier behavior. }. The remaining 8 clusters, do not immediately show any discernible trend in redsfhit for the estimates of both $\Xi_1$ and $\gNtilde\times\M^{\rm DHOST}/\M^{\rm GR}$, however, see \Cref{sec:XI_evolution} for an elaborate discussion. A particular trend of these parameters in redshift could provide strong implications for the redshift evolution of the physical parameters in the DHOST scenario. 

\begin{figure}[!t]
\includegraphics[scale=0.55]{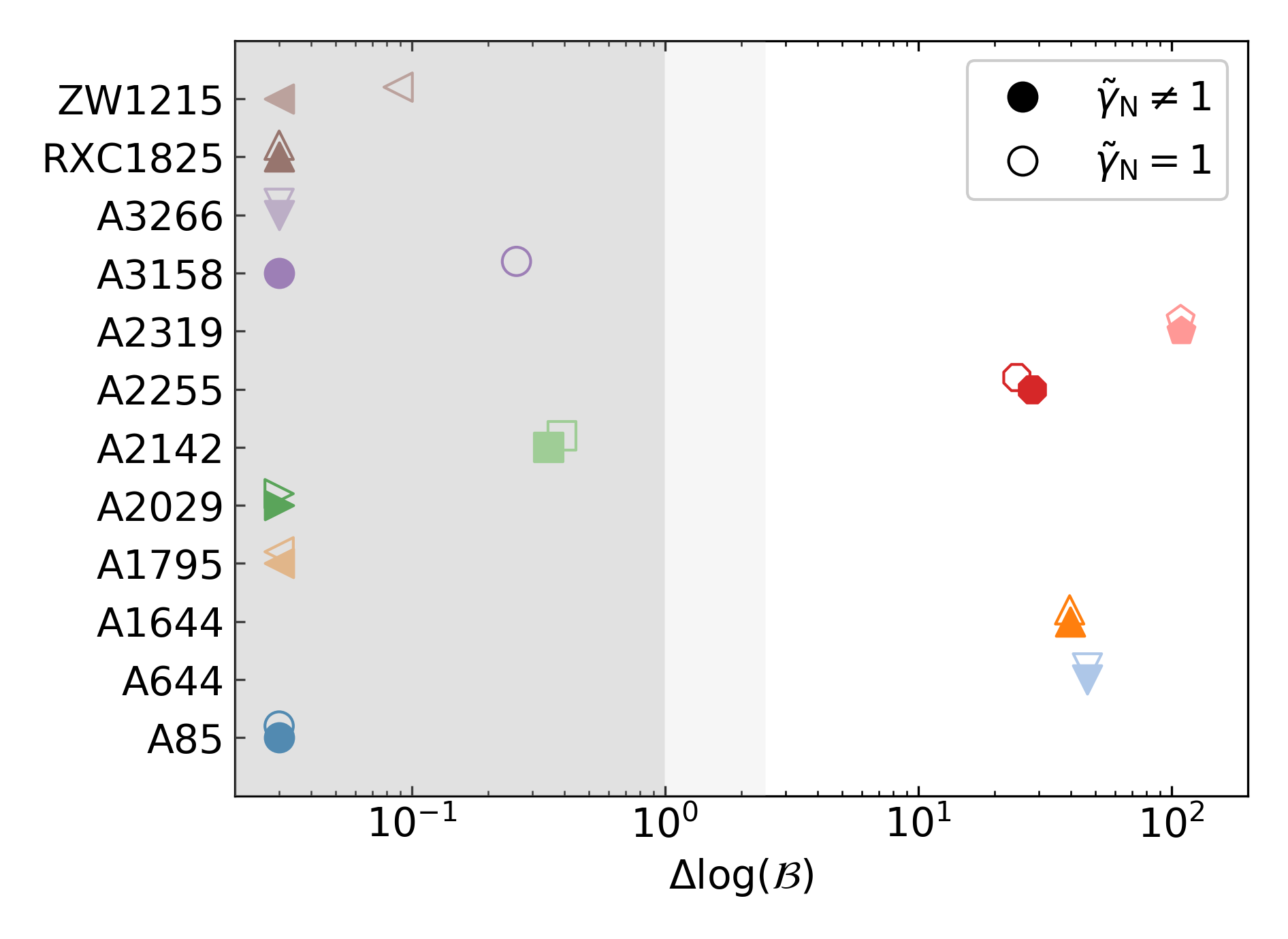}
 \caption{We show the Bayesian evidence estimates for the DHOST model w.r.t GR. Here, DHOST analyses with $\gNtilde \neq 1$ and $\gNtilde = 1$ are indicated with filled and open markers, respectively. To be contrasted using the Jeffery's scale. }
 \label{fig:Bayesian_evidence}
\end{figure}

\begin{figure*}[!t]
\centering
\includegraphics[scale=0.35]{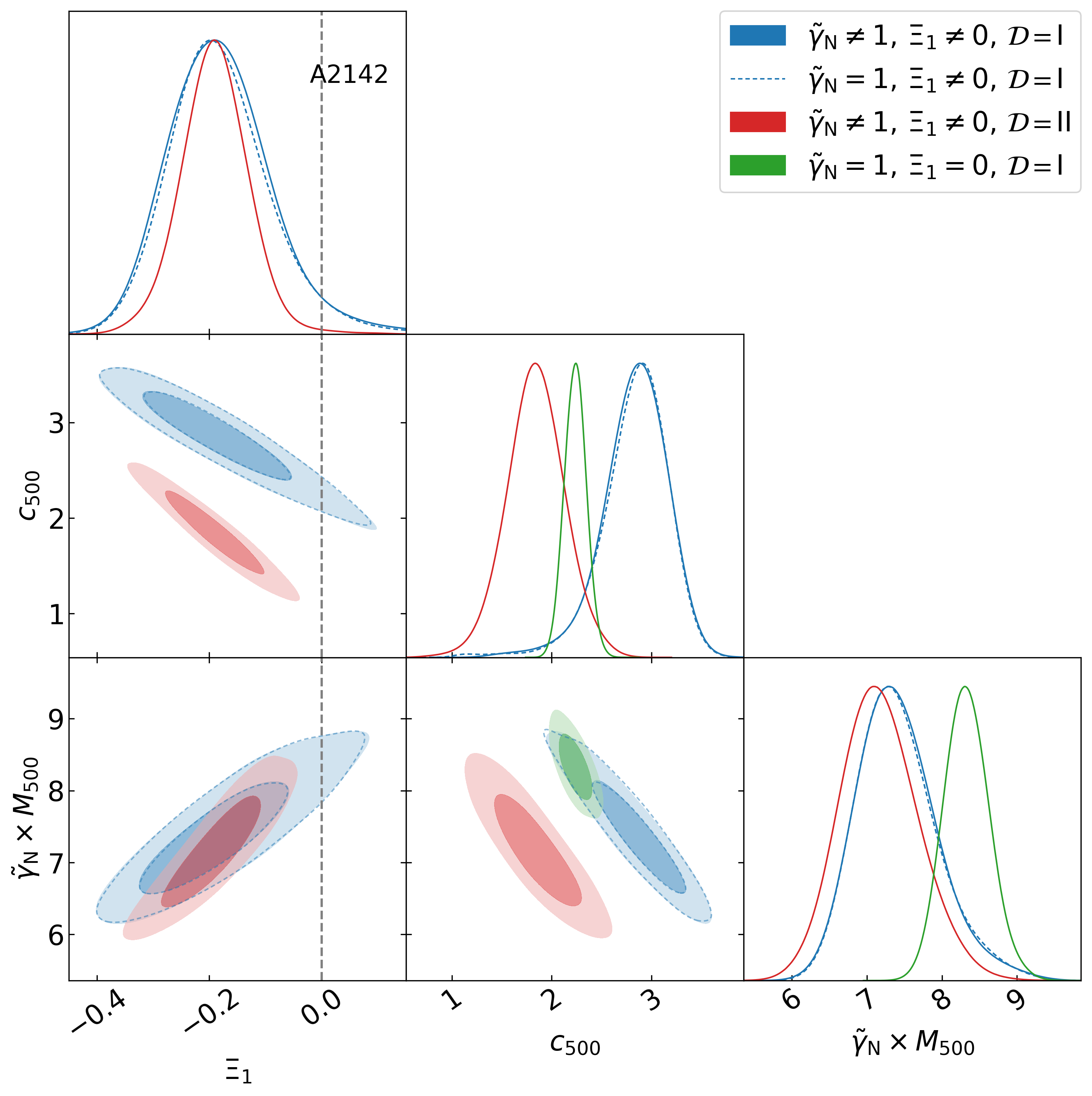}
\includegraphics[scale=0.35]{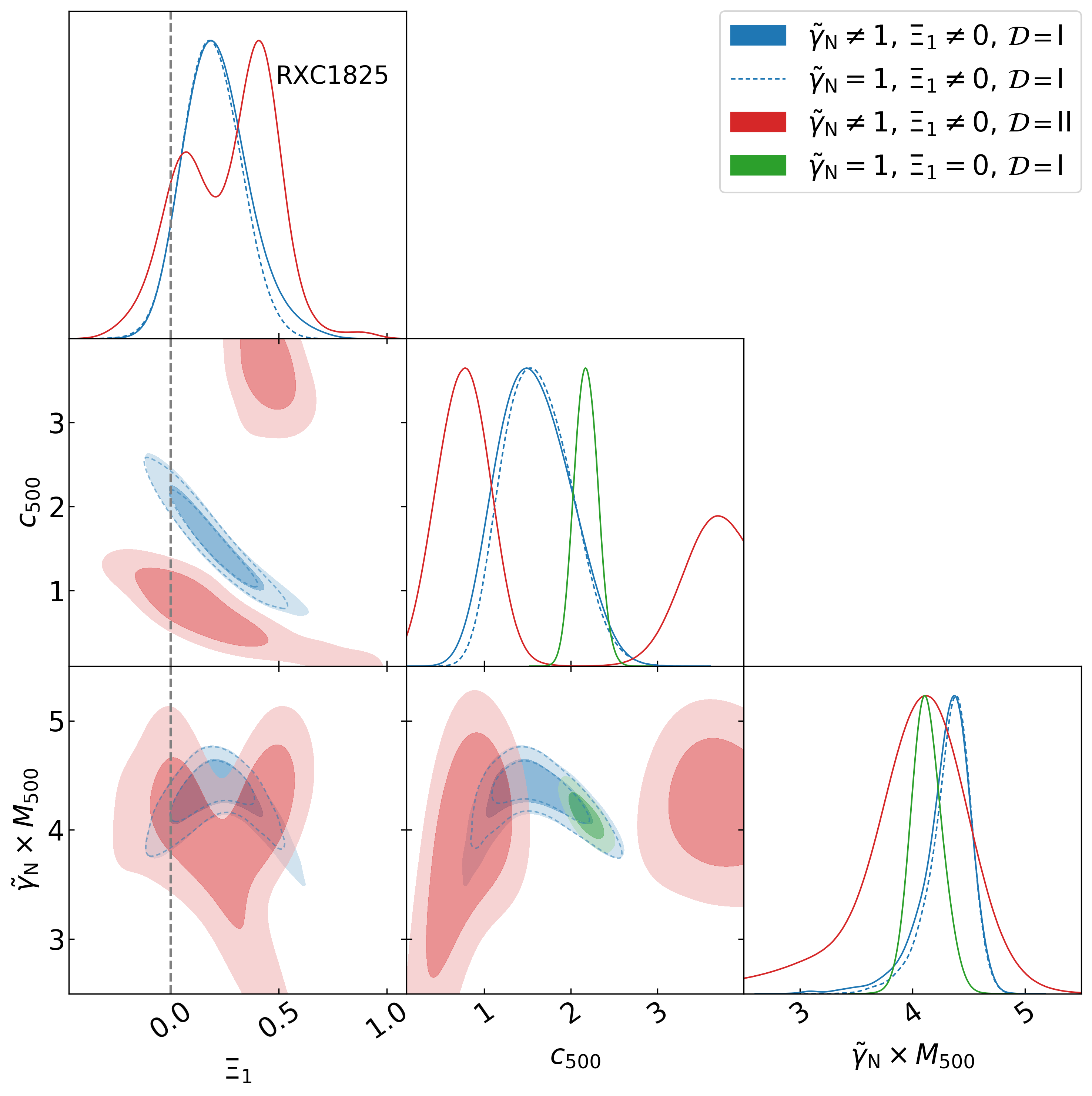}
\caption{ We show the $68\%$ and $95\%$ C.L. contours for the concentration, mass and DHOST parameters for the clusters A2142 (\textit{Left}) and RXC1825 (\textit{Right}). Here $\cal{D} = \rm{I}$ stands for the dataset ($\cal{D}$) utilized being $\Px+\Psz$ and $\cal{D} = \rm{II}$, shows the constraints obtained using only the $\Psz$ data. The dashed line marks the GR scenario of $\Xi_1 = 0$. }
\label{fig:Contours}
\end{figure*}

To assess the statistical evidence for/against the DHOST models, we compute the Bayesian evidence ($\Delta\log{\cal B}$) with respect to the GR scenario. These are reported in the last column of \Cref{tab:Constraints_PP} clearly showing that GR is mostly the preferred theory of gravity although DHOST is not  strongly disfavored. As for the constraints on $\Xi_1$, we still find that the situation is reversed for the four non-NFW clusters. Note that the Bayesian evidence is affected by the prior volume assumed on the parameters $\Xi_1, \, \gNtilde, \, \M$ for the DHOST model. We, therefore, perform a second MCMC analysis by fixing $\gNtilde = 1$ and sampling only over $\Xi_1, \, \M$ as free parameters. Firstly, we notice that the posteriors for $\Xi_1$ remain equivalent to the ones obtained with $\gNtilde \neq 1$ for all the 12 clusters (see e.g., \Cref{fig:Contours}), indicating no correlation between the two effective parameters of the DHOST model. We show the Bayesian evidence for the DHOST analyses with $\gNtilde \neq 1$ and the latter with $\gNtilde = 1$ in \Cref{fig:Bayesian_evidence}, indicated with filled and open markers, respectively. While we see a small change/preference for the $\gNtilde = 1$ w.r.t to the DHOST analysis with $\gNtilde \neq 1$, GR yet remains the preferred model of gravity. Having established that there is no statistical preference for the DHOST modification, from the individual X-COP clusters utilized in this work, we proceed to discuss the nature of constraints for the DHOST modifications and later a possible redshift dependence. 

In \Cref{fig:Contours}, we show the confidence regions for two clusters A2142 and RXC1825, for the concentration and the DHOST parameters. Clearly, we have very distinct distributions for the two clusters with RXC1825 showing more degenerate boomerang-like contours in the parameter space of $c_{500}\, \rm{vs.}\, \gNtilde \times \M$. We specifically choose these two clusters A2142 and RXC1825, to represent $\Xi_1<0$ and $\Xi_1>0$, cases (see also \Cref{sec:Mass_comparison}), respectively. As can be clearly noticed in \Cref{fig:Contours}, the negative (positive) values of $\Xi_1$ indicate strengthening (weakening) of gravity, which thereby implies a lower (higher) value of $\M^{\gNtilde = 1}$, with respect to the GR case. In this context, we show the contours for the analysis with $\gNtilde = 1$, essentially to validate that the constrains on the $\Xi_1$ parameter remain unaltered, in comparison to the $\gNtilde\neq1$ case. For comparison, we also show the constraints for the DHOST modification obtained using only the $\Psz$ data. Through which we find that cluster A2142 has more Gaussian-like constrains for the $\M$, $\Xi_1$ parameters and hence is a simpler case for the discussion of extended gravity theories\footnote{Note that in \citep{Ettori:2016kll}, a study of an emergent gravity model was conducted for A2142 and A2319 clusters, as they are massive clusters and makes them more suitable to study the modifications to gravity. Which, however, was later extended to all the 12 clusters in X-COP in \citetalias{Ettori:2018tus}.}, while there is a shift in the constraints only for the $c_{500}$ parameter. Note also that the constraints on $\gNtilde\times \M$ remain more consistent for all contrasted combinations, for both the clusters. This we notice to be the case for all the clusters except the 4 non-NFW clusters. 

Using the $\Psz$ data alone, the RXC1825 cluster shows a multi-modal behavior for the constrains on $c_{500}$ and consequently mild, yet a similar double peak posterior for the $\Xi_1$ parameter. The posterior for the mass however remains very much in agreement with the $\Psz+\Px$ analysis, also for the GR case. This is also equivalently shown in \Cref{fig:XI_PSZ} of \Cref{sec:SZ_alone}, where we show the distributions for $\Xi_1$ obtained for all the clusters using $\Psz$ data alone. While posterior distribution on the $\Xi_1$ appears as a double peak, it should be noted that the two peaks come from very distinct distributions for the concentration parameter. The inclusion of $\Px$ data reduces the posteriors to the aforementioned boomerang-like distributions, which highlights a change in the correlation of the mass parameter with the $c_{500}$ and $\Xi_1$ parameters. Also emphasizing the advantage of performing a joint analysis of $\Px+\Psz$ in our analysis. Interestingly, a similar effect was observed for the degeneracy between the NFW parameters and the DHOST parameter $\Xi_1$ in a simulated data based analysis in \citep{Pizzuti:2020tdl}\footnote{The analysis therein is equivalent to the case of $\gNtilde =1$, performed here.}, however for much larger values of $\Xi_1$. In fact, this is the reason for extended posteriors for $\Xi_1$ towards the positive end of the distributions, as shown in \Cref{fig:XI}.

Also, it is interesting to note that the degeneracy manifests primarily between the $c_{500}$ parameter and the DHOST parameter, rather than with the $M_{500}$ parameter which remains consistent among different data combinations within the DHOST analysis. In other words, $M_{500}$ and hence $R_{500}$ is constrained in a more consistent way rather than the shape of the mass profile, as it might be expected for the change in gravitational potential and as an implication of a screening mechanism at play.

\begin{figure}[!t]
\centering
\includegraphics[scale=0.38]{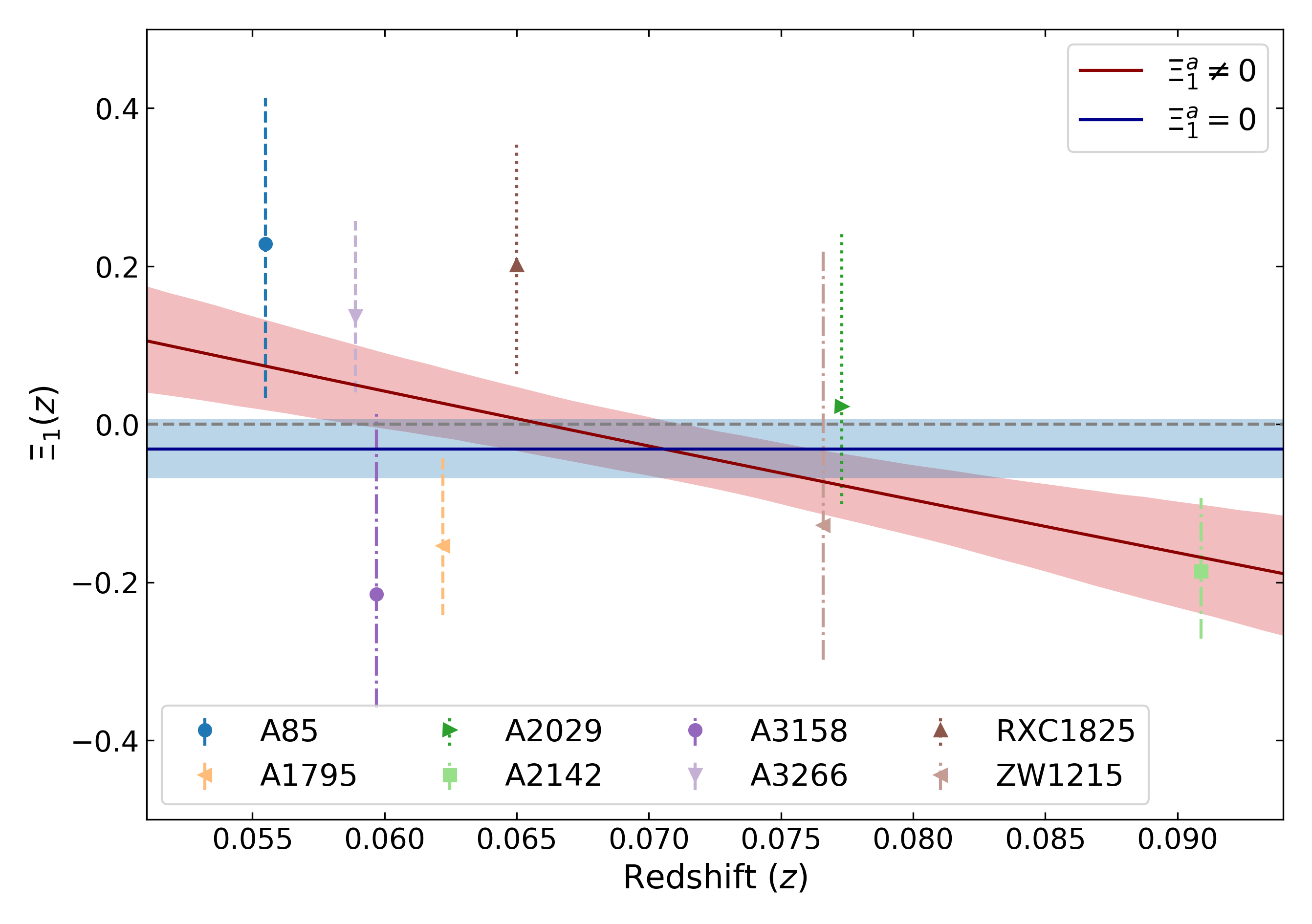}
 \caption{Taylor expansion fit to the posterior distributions of the $\Xi_1$ parameter. We show the 68$\%$ credible interval for the $\Xi_1^a\neq 0$ (red) and $\Xi_1^a =0$ (blue) cases. }
 \label{fig:XI_redshift}
\end{figure}

\subsection{Redshift evolution of $\Xi_1$}
\label{sec:XI_evolution}
While the analysis so far is done to constrain the DHOST parameters independently for each of the clusters, they are clearly redshift dependent. Alongside assessing the improvement of the fit to the pressure profile data when including the DHOST modifications, it is also important to assess their redshift behavior. We consider this fit to the posterior of $\Xi_1$ as the first-order proxy of a possible redshift dependent behavior, which however should be modeled simultaneously and is a computationally tedious analysis. In this context, having a larger dataset and an extended range of redshift would be of utmost importance to assess the DHOST gravity. Therefore, we fit the posterior distributions on the $\Xi_1$ parameters, using the 8 clusters, i.e., excluding the 4 non-NFW clusters to assess the redshift evolution of $\Xi_1$. We consider a simple Taylor expansion as,

\begin{equation}
    \Xi_1(z) = \Xi_{1}^{0} + \frac{z}{1+z} \Xi_{1}^{\rm a},
\end{equation}
where $\Xi_{1}^{0}$ is value of $\Xi_1(z)$ at redshift $z = 0$ and $\Xi_{1}^{\rm a}\neq 0$ provides the redshift dependence. As can be seen from \Cref{fig:XI_redshift}, there exists a redshift dependence, with $\Xi_1(z)$ deviating from GR expectation for larger redshifts ($z \gtrsim 0.08$). However, note that the redsfhit dependence is mostly dominated by cluster A2142, excluding which the assumed functional form is consistent with no evolution within the redshift range. In a similar approach fitting constant values through the 8 data points, assuming no redsfhit dependence provides a joint constrain\footnote{Performing a joint analysis of all the 8 clusters with a Gaussian prior on the $\Xi_1$ parameter would yield the same result. } on the $\Xi_1$ parameter from the combination of the 8 clusters. We find this value to be $\Xi_1 = -0.030 \pm 0.043$. This constraint is shown as the blue shaded region in the \Cref{fig:XI_redshift}. Note that the fit presented here is an equivalent approach as performing an importance sampling on the $\Xi_1$ parameter alone constrained from the individual clusters. Which is a valid approach owing to the fact that $\Xi_1$ is the only shared parameter among the clusters.

\section{Comparison with existing constraints}\label{sec:constraints}

{It is worth noticing that the parameter $\Xi_1$ numerically coincides with the $\Upsilon_1/4$ parameter in GLPV theories \cite{Sakstein:2016ggl}. The form of the gravitational force related to the potential $\Phi$ in \Cref{exp:grav_force:phi} is unchanged, in the GLPV theory while the term with $\Xi_3$ is an additional one. Therefore, for the observations where only the metric potential $\Phi$ is involved, one can compare $\Xi_1$ with the existing constraints in previous studies devoted to GLPV theory. For example, \cite{Koyama:2015oma, Saito:2015fza} study the effects of modified density profile for non-relativistic stars. On the other hand, astrophysical observables, such as weak lensing that also depend on the potential $\Psi$ (see e.g. \cite{Koyama:2015oma,Sakstein:2016ggl}), should be taken into account in order to  constrain  $\Xi_2$ and $\Xi_3$ parameters.}

Several works have earlier constrained the DHOST parameter $\Xi_1$, albeit with varied parametric form\footnote{For example, our parameter $\Xi_1$ here is equivalent to $\Upsilon_1/4$ used in \citep{Sakstein:2016ggl}.}, which however we can readily compare with our results. As mentioned earlier in \Cref{sec:Introduction}, in the current work we assess the modification to the hydrostatic equilibrium alone and do not consider the weak lensing counterpart, that can be utilized to place constraints on the $\Xi_2$ parameter. In this context, \citep{Sakstein:2016ggl}, is one of the earliest analyses to provide constraints on the $\Xi_1, \, \Xi_2$ parameters by contrasting the hydrostatic and weak lensing masses using a stacked dataset of 58 clusters in the redshift range of $0.1 \leq z\leq 1.2$. While the analysis in \citep{Sakstein:2016ggl}, improves the constraints by utilizing the stacked cluster profiles, losing the redshift dependence, here we have for the first time (as far as we are aware), tried to assess from individual unstacked clusters, which helps us also assess the redshift dependence of the constraints. Our constraints from individual clusters are however less stringent, and we find good consistency with $\Xi_1 \sim - 0.028^{+0.23}_{-0.17}$, reported there in\footnote{Converted from the $\Upsilon_1 = -0.11^{+0.98}_{-0.67}$ quoted in \citep{Sakstein:2016ggl}.}. The joint constraint of $\Xi_1 = -0.030\pm 0.043$, derived from our proxy-fit to the posteriors of $\Xi_1$ from individual clusters is in fact much more stringent, being $\sim 5$ times more precise. This in turn, reasserts the promise in utilizing the clusters with a combination of the X-ray data the SZ pressure profiles, which can improve upon the existing constraints by almost an order in precision. Note also that the redshift range ($z<0.1$) of the X-COP clusters used in our analysis is complementary to the redshift range utilized in \citep{Sakstein:2016ggl}, which is an added advantage when contrasting the constraints. Owing to the mild differences in the mass estimates obtained in our analysis to those quoted in the original analysis in \citetalias{Ettori:2018tus}, we also estimate a more conservative limit by utilizing only 4 clusters (A1795, A3158, RXC1825 and ZW1215), which agree very well. We find this limit to be $\Xi_1 = -0.061 \pm 0.074$, yet being at least twice tighter than the previous estimate. It is worth noting that these 4 clusters span a very small redshift range of $\Delta z \lesssim 0.02$ within $0.0597<z<0.0766$. For instance, the A1795 cluster provides an individual constrain of $\Xi_1 = -0.17^{+0.11}_{-0.10}$ using $\Px+\Psz$ data, which indicates a mild $\sim 1.55\, \sigma$ deviation from GR, however statistically disfavored in comparison with GR with a Bayesian evidence $\Delta \log(B) = -2.9$. As expected, we do not find such an evidence constraint when using only the $\Px$ data, as data from both the inner and outer regions of the cluster need to be weighed against the model to assess any variation in the gravitational potential along the radial direction, while accounting for the screening effects. 

Apart the cluster-scale constraints there exist lower limits from the non-relativistic stars of $\Xi_1>-1/6$, requiring a stable static solution \citep{Saito:2015fza} (see also \citep{Saltas:2018mxc}). In fact, one could reanalyze the current dataset by imposing the lower limit, which we expect to, however, be consistent, except for the 4 non-NFW clusters. And an upper limit of $\Xi_1< 7\times 10^{-3}$ from the consistency of the minimum mass for hydrogen burning in stars with the lowest mass red dwarf~\cite{Sakstein:2015zoa, Sakstein:2015aac}. Our constraints on the individual clusters (8 NFW) are very well in accordance with these limits obtained from the much smaller scale objects. Clusters A3266, RXC1825 and A85 have a positive mean value for $\Xi_1$ and extend beyond the upper limit, however, being consistent with the values lower than the limit as well. Taking the 4 non-NFW clusters at face value, one could argue for a strong disagreement with the existing lower limit from the non-relativistic stars, owing to i) non-NFW density profiles are not preferred as strongly as DHOST modification from the Bayesian evidence (see Fig. 2. of \citetalias{Ettori:2018tus}) and ii) they are not significantly affected by the non-thermal effects. {This would indicate invalidity of the aforementioned limits and scale-dependent behavior of the current screening mechanism. }However, we intend to interpret these results more carefully for the moment, and wait to perform additional analysis. 

More recently, \citep{Pizzuti:2020tdl} have performed a forecast analysis, showing the potential galaxy cluster constrains for beyond Horndeski models, scaling with the number of clusters. It has been shown that the posterior ( more precisely $\chi^2$ distribution ), could show large degeneracy for large values of $\Xi_1$ and the NFW parameters used to model the mass profile and that this should be taken into account when analyzing real observations. However, in our analysis, we find the $\Xi_1$ parameter not exceeding unity for any of the given 12 clusters. However, some confirming correlations are seen in one or two clusters especially with the $\Psz$ data alone.

\section{Summary}
We constrain deviation from GR on the cluster scales, modeled through a Vainshtein screening mechanism, utilizing the X-COP compilation which consists of 12 galaxy clusters. While we have performed our analysis on all the 12 clusters, we make a careful selection of subsets consisting of either 8 or 4 clusters to comment on the final inferences: i) we eliminate 4 clusters as they are not well represented by the NFW mass profile, ii) To be more conservative, we exclude 4 other clusters which have mild $\sim 1 \sigma$ variation from the masses quoted in the original analysis. 

Our main results are summarized as follows:

\begin{itemize}

    \item Performing a Bayesian analysis using the so-called backward method assuming the NFW mass profile, we find mild to moderate deviations from the GR scenario ($\Xi_1 \neq 0$) of the order $\sim 2 \sigma$ (see \Cref{tab:Constraints_PP}).
    
    \item Comparing the Bayesian evidence, both GR and DHOST scenarios perform equivalently, with mild to a moderate preference for GR at times. 

    \item As our main result, we report a constraint of $\Xi_1 = -0.030 \pm 0.043$ obtained using 8 clusters and a more conservative constraint of $\Xi_1 =-0.061\pm 0.074$ using only 4 clusters, for reasons quoted in the text. Our result shows no indication for a deviation from GR, while being stringent than the earlier constraints. 
    
    \item Assessing the redshift evolution we present the utility of datasets like the one here. In effect, we find possible evidence for redshift dependent behavior, however only dominated by constraints from A2142 cluster. 
    
\end{itemize}

While individual clusters do not immediately suggest a significant modification to GR, we find that a tentative redshift dependent behavior could be observed at a larger significance. This indeed makes it essential that we test the current setup against larger well-observed samples such as the NIKA2 SZ large program consisting of a sample size of 45 clusters, in a wider redshift range. We also intend to extend the discussion in terms of the physical parameters as outlined in \Cref{sec:physical_parameters}, which is of utmost importance to assess the feasibility of the models within the current formalism.

\section*{Acknowledgments}
Authors are grateful to Stefano Ettori and Dominique Eckert for useful comments and feedback on the use of data. BSH is supported by the INFN INDARK grant. PK, MDP, VC, RM acknowledge support from
INFN/Euclid Sezione di Roma. PK, MDP and RM also acknowledge support from Sapienza Universit\'a di Roma thanks to Progetti di Ricerca Medi 2018, prot. RM118164365E40D9 and 2019, prot. RM11916B7540DD8D. We acknowledge the use of publicly available python packages: \texttt{numpy, scipy, \& cmath}. 

\bibliography{bibliography} 

\begin{thebibliography}{83}%
\makeatletter
\providecommand \@ifxundefined [1]{%
 \@ifx{#1\undefined}
}%
\providecommand \@ifnum [1]{%
 \ifnum #1\expandafter \@firstoftwo
 \else \expandafter \@secondoftwo
 \fi
}%
\providecommand \@ifx [1]{%
 \ifx #1\expandafter \@firstoftwo
 \else \expandafter \@secondoftwo
 \fi
}%
\providecommand \natexlab [1]{#1}%
\providecommand \enquote  [1]{``#1''}%
\providecommand \bibnamefont  [1]{#1}%
\providecommand \bibfnamefont [1]{#1}%
\providecommand \citenamefont [1]{#1}%
\providecommand \href@noop [0]{\@secondoftwo}%
\providecommand \href [0]{\begingroup \@sanitize@url \@href}%
\providecommand \@href[1]{\@@startlink{#1}\@@href}%
\providecommand \@@href[1]{\endgroup#1\@@endlink}%
\providecommand \@sanitize@url [0]{\catcode `\\12\catcode `\$12\catcode
  `\&12\catcode `\#12\catcode `\^12\catcode `\_12\catcode `\%12\relax}%
\providecommand \@@startlink[1]{}%
\providecommand \@@endlink[0]{}%
\providecommand \url  [0]{\begingroup\@sanitize@url \@url }%
\providecommand \@url [1]{\endgroup\@href {#1}{\urlprefix }}%
\providecommand \urlprefix  [0]{URL }%
\providecommand \Eprint [0]{\href }%
\providecommand \doibase [0]{http://dx.doi.org/}%
\providecommand \selectlanguage [0]{\@gobble}%
\providecommand \bibinfo  [0]{\@secondoftwo}%
\providecommand \bibfield  [0]{\@secondoftwo}%
\providecommand \translation [1]{[#1]}%
\providecommand \BibitemOpen [0]{}%
\providecommand \bibitemStop [0]{}%
\providecommand \bibitemNoStop [0]{.\EOS\space}%
\providecommand \EOS [0]{\spacefactor3000\relax}%
\providecommand \BibitemShut  [1]{\csname bibitem#1\endcsname}%
\let\auto@bib@innerbib\@empty
\bibitem [{\citenamefont {Adam}\ \emph {et~al.}(2016)\citenamefont {Adam} \emph
  {et~al.}}]{Adam:2015rua}%
  \BibitemOpen
  \bibfield  {author} {\bibinfo {author} {\bibfnamefont {R.}~\bibnamefont
  {Adam}} \emph {et~al.} (\bibinfo {collaboration} {Planck}),\ }\href {\doibase
  10.1051/0004-6361/201527101} {\bibfield  {journal} {\bibinfo  {journal}
  {Astron. Astrophys.}\ }\textbf {\bibinfo {volume} {594}},\ \bibinfo {pages}
  {A1} (\bibinfo {year} {2016})},\ \Eprint {http://arxiv.org/abs/1502.01582}
  {arXiv:1502.01582 [astro-ph.CO]} \BibitemShut {NoStop}%
\bibitem [{\citenamefont {Ade}\ \emph {et~al.}(2016{\natexlab{a}})\citenamefont
  {Ade} \emph {et~al.}}]{Ade:2015xua}%
  \BibitemOpen
  \bibfield  {author} {\bibinfo {author} {\bibfnamefont {P.~A.~R.}\
  \bibnamefont {Ade}} \emph {et~al.} (\bibinfo {collaboration} {Planck}),\
  }\href {\doibase 10.1051/0004-6361/201525830} {\bibfield  {journal} {\bibinfo
   {journal} {Astron. Astrophys.}\ }\textbf {\bibinfo {volume} {594}},\
  \bibinfo {pages} {A13} (\bibinfo {year} {2016}{\natexlab{a}})},\ \Eprint
  {http://arxiv.org/abs/1502.01589} {arXiv:1502.01589 [astro-ph.CO]}
  \BibitemShut {NoStop}%
\bibitem [{\citenamefont {Riess}\ \emph {et~al.}(1998)\citenamefont {Riess}
  \emph {et~al.}}]{Riess:1998cb}%
  \BibitemOpen
  \bibfield  {author} {\bibinfo {author} {\bibfnamefont {A.~G.}\ \bibnamefont
  {Riess}} \emph {et~al.} (\bibinfo {collaboration} {Supernova Search Team}),\
  }\href {\doibase 10.1086/300499} {\bibfield  {journal} {\bibinfo  {journal}
  {Astron. J.}\ }\textbf {\bibinfo {volume} {116}},\ \bibinfo {pages} {1009}
  (\bibinfo {year} {1998})},\ \Eprint {http://arxiv.org/abs/astro-ph/9805201}
  {arXiv:astro-ph/9805201} \BibitemShut {NoStop}%
\bibitem [{\citenamefont {Perlmutter}\ \emph {et~al.}(1999)\citenamefont
  {Perlmutter} \emph {et~al.}}]{Perlmutter:1998np}%
  \BibitemOpen
  \bibfield  {author} {\bibinfo {author} {\bibfnamefont {S.}~\bibnamefont
  {Perlmutter}} \emph {et~al.} (\bibinfo {collaboration} {Supernova Cosmology
  Project}),\ }\href {\doibase 10.1086/307221} {\bibfield  {journal} {\bibinfo
  {journal} {Astrophys. J.}\ }\textbf {\bibinfo {volume} {517}},\ \bibinfo
  {pages} {565} (\bibinfo {year} {1999})},\ \Eprint
  {http://arxiv.org/abs/astro-ph/9812133} {arXiv:astro-ph/9812133} \BibitemShut
  {NoStop}%
\bibitem [{\citenamefont {Weinberg}(1989)}]{Weinberg:1988cp}%
  \BibitemOpen
  \bibfield  {author} {\bibinfo {author} {\bibfnamefont {S.}~\bibnamefont
  {Weinberg}},\ }\href {\doibase 10.1103/RevModPhys.61.1} {\bibfield  {journal}
  {\bibinfo  {journal} {Rev. Mod. Phys.}\ }\textbf {\bibinfo {volume} {61}},\
  \bibinfo {pages} {1} (\bibinfo {year} {1989})}\BibitemShut {NoStop}%
\bibitem [{\citenamefont {Bull}\ \emph {et~al.}(2016)\citenamefont {Bull} \emph
  {et~al.}}]{Bull:2015stt}%
  \BibitemOpen
  \bibfield  {author} {\bibinfo {author} {\bibfnamefont {P.}~\bibnamefont
  {Bull}} \emph {et~al.},\ }\href {\doibase 10.1016/j.dark.2016.02.001}
  {\bibfield  {journal} {\bibinfo  {journal} {Phys. Dark Univ.}\ }\textbf
  {\bibinfo {volume} {12}},\ \bibinfo {pages} {56} (\bibinfo {year} {2016})},\
  \Eprint {http://arxiv.org/abs/1512.05356} {arXiv:1512.05356 [astro-ph.CO]}
  \BibitemShut {NoStop}%
\bibitem [{\citenamefont {Caldwell}\ \emph {et~al.}(1998)\citenamefont
  {Caldwell}, \citenamefont {Dave},\ and\ \citenamefont
  {Steinhardt}}]{Caldwell:1997ii}%
  \BibitemOpen
  \bibfield  {author} {\bibinfo {author} {\bibfnamefont {R.~R.}\ \bibnamefont
  {Caldwell}}, \bibinfo {author} {\bibfnamefont {R.}~\bibnamefont {Dave}}, \
  and\ \bibinfo {author} {\bibfnamefont {P.~J.}\ \bibnamefont {Steinhardt}},\
  }\href {\doibase 10.1103/PhysRevLett.80.1582} {\bibfield  {journal} {\bibinfo
   {journal} {Phys. Rev. Lett.}\ }\textbf {\bibinfo {volume} {80}},\ \bibinfo
  {pages} {1582} (\bibinfo {year} {1998})},\ \Eprint
  {http://arxiv.org/abs/astro-ph/9708069} {arXiv:astro-ph/9708069} \BibitemShut
  {NoStop}%
\bibitem [{\citenamefont {Chiba}\ \emph {et~al.}(2000)\citenamefont {Chiba},
  \citenamefont {Okabe},\ and\ \citenamefont {Yamaguchi}}]{Chiba:1999ka}%
  \BibitemOpen
  \bibfield  {author} {\bibinfo {author} {\bibfnamefont {T.}~\bibnamefont
  {Chiba}}, \bibinfo {author} {\bibfnamefont {T.}~\bibnamefont {Okabe}}, \ and\
  \bibinfo {author} {\bibfnamefont {M.}~\bibnamefont {Yamaguchi}},\ }\href
  {\doibase 10.1103/PhysRevD.62.023511} {\bibfield  {journal} {\bibinfo
  {journal} {Phys. Rev. D}\ }\textbf {\bibinfo {volume} {62}},\ \bibinfo
  {pages} {023511} (\bibinfo {year} {2000})},\ \Eprint
  {http://arxiv.org/abs/astro-ph/9912463} {arXiv:astro-ph/9912463} \BibitemShut
  {NoStop}%
\bibitem [{\citenamefont {Armendariz-Picon}\ \emph {et~al.}(2001)\citenamefont
  {Armendariz-Picon}, \citenamefont {Mukhanov},\ and\ \citenamefont
  {Steinhardt}}]{ArmendarizPicon:2000ah}%
  \BibitemOpen
  \bibfield  {author} {\bibinfo {author} {\bibfnamefont {C.}~\bibnamefont
  {Armendariz-Picon}}, \bibinfo {author} {\bibfnamefont {V.~F.}\ \bibnamefont
  {Mukhanov}}, \ and\ \bibinfo {author} {\bibfnamefont {P.~J.}\ \bibnamefont
  {Steinhardt}},\ }\href {\doibase 10.1103/PhysRevD.63.103510} {\bibfield
  {journal} {\bibinfo  {journal} {Phys. Rev. D}\ }\textbf {\bibinfo {volume}
  {63}},\ \bibinfo {pages} {103510} (\bibinfo {year} {2001})},\ \Eprint
  {http://arxiv.org/abs/astro-ph/0006373} {arXiv:astro-ph/0006373} \BibitemShut
  {NoStop}%
\bibitem [{\citenamefont {Tsujikawa}(2010)}]{Tsujikawa:2010zza}%
  \BibitemOpen
  \bibfield  {author} {\bibinfo {author} {\bibfnamefont {S.}~\bibnamefont
  {Tsujikawa}},\ }\href {\doibase 10.1007/978-3-642-10598-2_3} {\bibfield
  {journal} {\bibinfo  {journal} {Lect. Notes Phys.}\ }\textbf {\bibinfo
  {volume} {800}},\ \bibinfo {pages} {99} (\bibinfo {year} {2010})},\ \Eprint
  {http://arxiv.org/abs/1101.0191} {arXiv:1101.0191 [gr-qc]} \BibitemShut
  {NoStop}%
\bibitem [{\citenamefont {Nojiri}\ and\ \citenamefont
  {Odintsov}(2011)}]{Nojiri:2010wj}%
  \BibitemOpen
  \bibfield  {author} {\bibinfo {author} {\bibfnamefont {S.}~\bibnamefont
  {Nojiri}}\ and\ \bibinfo {author} {\bibfnamefont {S.~D.}\ \bibnamefont
  {Odintsov}},\ }\href {\doibase 10.1016/j.physrep.2011.04.001} {\bibfield
  {journal} {\bibinfo  {journal} {Phys. Rept.}\ }\textbf {\bibinfo {volume}
  {505}},\ \bibinfo {pages} {59} (\bibinfo {year} {2011})},\ \Eprint
  {http://arxiv.org/abs/1011.0544} {arXiv:1011.0544 [gr-qc]} \BibitemShut
  {NoStop}%
\bibitem [{\citenamefont {Clifton}\ \emph {et~al.}(2012)\citenamefont
  {Clifton}, \citenamefont {Ferreira}, \citenamefont {Padilla},\ and\
  \citenamefont {Skordis}}]{Clifton:2011jh}%
  \BibitemOpen
  \bibfield  {author} {\bibinfo {author} {\bibfnamefont {T.}~\bibnamefont
  {Clifton}}, \bibinfo {author} {\bibfnamefont {P.~G.}\ \bibnamefont
  {Ferreira}}, \bibinfo {author} {\bibfnamefont {A.}~\bibnamefont {Padilla}}, \
  and\ \bibinfo {author} {\bibfnamefont {C.}~\bibnamefont {Skordis}},\ }\href
  {\doibase 10.1016/j.physrep.2012.01.001} {\bibfield  {journal} {\bibinfo
  {journal} {Phys. Rept.}\ }\textbf {\bibinfo {volume} {513}},\ \bibinfo
  {pages} {1} (\bibinfo {year} {2012})},\ \Eprint
  {http://arxiv.org/abs/1106.2476} {arXiv:1106.2476 [astro-ph.CO]} \BibitemShut
  {NoStop}%
\bibitem [{\citenamefont {Joyce}\ \emph {et~al.}(2016)\citenamefont {Joyce},
  \citenamefont {Lombriser},\ and\ \citenamefont {Schmidt}}]{Joyce:2016vqv}%
  \BibitemOpen
  \bibfield  {author} {\bibinfo {author} {\bibfnamefont {A.}~\bibnamefont
  {Joyce}}, \bibinfo {author} {\bibfnamefont {L.}~\bibnamefont {Lombriser}}, \
  and\ \bibinfo {author} {\bibfnamefont {F.}~\bibnamefont {Schmidt}},\ }\href
  {\doibase 10.1146/annurev-nucl-102115-044553} {\bibfield  {journal} {\bibinfo
   {journal} {Ann. Rev. Nucl. Part. Sci.}\ }\textbf {\bibinfo {volume} {66}},\
  \bibinfo {pages} {95} (\bibinfo {year} {2016})},\ \Eprint
  {http://arxiv.org/abs/1601.06133} {arXiv:1601.06133 [astro-ph.CO]}
  \BibitemShut {NoStop}%
\bibitem [{\citenamefont {Langlois}\ and\ \citenamefont
  {Noui}(2016)}]{Langlois:2015cwa}%
  \BibitemOpen
  \bibfield  {author} {\bibinfo {author} {\bibfnamefont {D.}~\bibnamefont
  {Langlois}}\ and\ \bibinfo {author} {\bibfnamefont {K.}~\bibnamefont
  {Noui}},\ }\href {\doibase 10.1088/1475-7516/2016/02/034} {\bibfield
  {journal} {\bibinfo  {journal} {JCAP}\ }\textbf {\bibinfo {volume} {1602}},\
  \bibinfo {pages} {034} (\bibinfo {year} {2016})},\ \Eprint
  {http://arxiv.org/abs/1510.06930} {arXiv:1510.06930 [gr-qc]} \BibitemShut
  {NoStop}%
\bibitem [{\citenamefont {Crisostomi}\ \emph {et~al.}(2016)\citenamefont
  {Crisostomi}, \citenamefont {Koyama},\ and\ \citenamefont
  {Tasinato}}]{Crisostomi:2016czh}%
  \BibitemOpen
  \bibfield  {author} {\bibinfo {author} {\bibfnamefont {M.}~\bibnamefont
  {Crisostomi}}, \bibinfo {author} {\bibfnamefont {K.}~\bibnamefont {Koyama}},
  \ and\ \bibinfo {author} {\bibfnamefont {G.}~\bibnamefont {Tasinato}},\
  }\href {\doibase 10.1088/1475-7516/2016/04/044} {\bibfield  {journal}
  {\bibinfo  {journal} {JCAP}\ }\textbf {\bibinfo {volume} {1604}},\ \bibinfo
  {pages} {044} (\bibinfo {year} {2016})},\ \Eprint
  {http://arxiv.org/abs/1602.03119} {arXiv:1602.03119 [hep-th]} \BibitemShut
  {NoStop}%
\bibitem [{\citenamefont {Ben~Achour}\ \emph
  {et~al.}(2016{\natexlab{a}})\citenamefont {Ben~Achour}, \citenamefont
  {Langlois},\ and\ \citenamefont {Noui}}]{Achour:2016rkg}%
  \BibitemOpen
  \bibfield  {author} {\bibinfo {author} {\bibfnamefont {J.}~\bibnamefont
  {Ben~Achour}}, \bibinfo {author} {\bibfnamefont {D.}~\bibnamefont
  {Langlois}}, \ and\ \bibinfo {author} {\bibfnamefont {K.}~\bibnamefont
  {Noui}},\ }\href {\doibase 10.1103/PhysRevD.93.124005} {\bibfield  {journal}
  {\bibinfo  {journal} {Phys. Rev.}\ }\textbf {\bibinfo {volume} {D93}},\
  \bibinfo {pages} {124005} (\bibinfo {year} {2016}{\natexlab{a}})},\ \Eprint
  {http://arxiv.org/abs/1602.08398} {arXiv:1602.08398 [gr-qc]} \BibitemShut
  {NoStop}%
\bibitem [{\citenamefont {Motohashi}\ \emph {et~al.}(2016)\citenamefont
  {Motohashi}, \citenamefont {Noui}, \citenamefont {Suyama}, \citenamefont
  {Yamaguchi},\ and\ \citenamefont {Langlois}}]{Motohashi:2016ftl}%
  \BibitemOpen
  \bibfield  {author} {\bibinfo {author} {\bibfnamefont {H.}~\bibnamefont
  {Motohashi}}, \bibinfo {author} {\bibfnamefont {K.}~\bibnamefont {Noui}},
  \bibinfo {author} {\bibfnamefont {T.}~\bibnamefont {Suyama}}, \bibinfo
  {author} {\bibfnamefont {M.}~\bibnamefont {Yamaguchi}}, \ and\ \bibinfo
  {author} {\bibfnamefont {D.}~\bibnamefont {Langlois}},\ }\href {\doibase
  10.1088/1475-7516/2016/07/033} {\bibfield  {journal} {\bibinfo  {journal}
  {JCAP}\ }\textbf {\bibinfo {volume} {1607}},\ \bibinfo {pages} {033}
  (\bibinfo {year} {2016})},\ \Eprint {http://arxiv.org/abs/1603.09355}
  {arXiv:1603.09355 [hep-th]} \BibitemShut {NoStop}%
\bibitem [{\citenamefont {Ben~Achour}\ \emph
  {et~al.}(2016{\natexlab{b}})\citenamefont {Ben~Achour}, \citenamefont
  {Crisostomi}, \citenamefont {Koyama}, \citenamefont {Langlois}, \citenamefont
  {Noui},\ and\ \citenamefont {Tasinato}}]{BenAchour:2016fzp}%
  \BibitemOpen
  \bibfield  {author} {\bibinfo {author} {\bibfnamefont {J.}~\bibnamefont
  {Ben~Achour}}, \bibinfo {author} {\bibfnamefont {M.}~\bibnamefont
  {Crisostomi}}, \bibinfo {author} {\bibfnamefont {K.}~\bibnamefont {Koyama}},
  \bibinfo {author} {\bibfnamefont {D.}~\bibnamefont {Langlois}}, \bibinfo
  {author} {\bibfnamefont {K.}~\bibnamefont {Noui}}, \ and\ \bibinfo {author}
  {\bibfnamefont {G.}~\bibnamefont {Tasinato}},\ }\href {\doibase
  10.1007/JHEP12(2016)100} {\bibfield  {journal} {\bibinfo  {journal} {JHEP}\
  }\textbf {\bibinfo {volume} {12}},\ \bibinfo {pages} {100} (\bibinfo {year}
  {2016}{\natexlab{b}})},\ \Eprint {http://arxiv.org/abs/1608.08135}
  {arXiv:1608.08135 [hep-th]} \BibitemShut {NoStop}%
\bibitem [{\citenamefont {Kobayashi}(2019)}]{Kobayashi:2019hrl}%
  \BibitemOpen
  \bibfield  {author} {\bibinfo {author} {\bibfnamefont {T.}~\bibnamefont
  {Kobayashi}},\ }\href {\doibase 10.1088/1361-6633/ab2429} {\bibfield
  {journal} {\bibinfo  {journal} {Rept. Prog. Phys.}\ }\textbf {\bibinfo
  {volume} {82}},\ \bibinfo {pages} {086901} (\bibinfo {year} {2019})},\
  \Eprint {http://arxiv.org/abs/1901.07183} {arXiv:1901.07183 [gr-qc]}
  \BibitemShut {NoStop}%
\bibitem [{\citenamefont {Brans}\ and\ \citenamefont
  {Dicke}(1961)}]{Brans:1961sx}%
  \BibitemOpen
  \bibfield  {author} {\bibinfo {author} {\bibfnamefont {C.}~\bibnamefont
  {Brans}}\ and\ \bibinfo {author} {\bibfnamefont {R.~H.}\ \bibnamefont
  {Dicke}},\ }\href {\doibase 10.1103/PhysRev.124.925} {\bibfield  {journal}
  {\bibinfo  {journal} {Phys. Rev.}\ }\textbf {\bibinfo {volume} {124}},\
  \bibinfo {pages} {925} (\bibinfo {year} {1961})}\BibitemShut {NoStop}%
\bibitem [{\citenamefont {De~Felice}\ and\ \citenamefont
  {Tsujikawa}(2010)}]{DeFelice:2010aj}%
  \BibitemOpen
  \bibfield  {author} {\bibinfo {author} {\bibfnamefont {A.}~\bibnamefont
  {De~Felice}}\ and\ \bibinfo {author} {\bibfnamefont {S.}~\bibnamefont
  {Tsujikawa}},\ }\href {\doibase 10.12942/lrr-2010-3} {\bibfield  {journal}
  {\bibinfo  {journal} {Living Rev. Rel.}\ }\textbf {\bibinfo {volume} {13}},\
  \bibinfo {pages} {3} (\bibinfo {year} {2010})},\ \Eprint
  {http://arxiv.org/abs/1002.4928} {arXiv:1002.4928 [gr-qc]} \BibitemShut
  {NoStop}%
\bibitem [{\citenamefont {Deffayet}\ \emph {et~al.}(2009)\citenamefont
  {Deffayet}, \citenamefont {Esposito-Farese},\ and\ \citenamefont
  {Vikman}}]{Deffayet:2009wt}%
  \BibitemOpen
  \bibfield  {author} {\bibinfo {author} {\bibfnamefont {C.}~\bibnamefont
  {Deffayet}}, \bibinfo {author} {\bibfnamefont {G.}~\bibnamefont
  {Esposito-Farese}}, \ and\ \bibinfo {author} {\bibfnamefont {A.}~\bibnamefont
  {Vikman}},\ }\href {\doibase 10.1103/PhysRevD.79.084003} {\bibfield
  {journal} {\bibinfo  {journal} {Phys. Rev. D}\ }\textbf {\bibinfo {volume}
  {79}},\ \bibinfo {pages} {084003} (\bibinfo {year} {2009})},\ \Eprint
  {http://arxiv.org/abs/0901.1314} {arXiv:0901.1314 [hep-th]} \BibitemShut
  {NoStop}%
\bibitem [{\citenamefont {Neveu}\ \emph {et~al.}(2013)\citenamefont {Neveu},
  \citenamefont {Ruhlmann-Kleider}, \citenamefont {Conley}, \citenamefont
  {Palanque-Delabrouille}, \citenamefont {Astier}, \citenamefont {Guy},\ and\
  \citenamefont {Babichev}}]{Neveu:2013mfa}%
  \BibitemOpen
  \bibfield  {author} {\bibinfo {author} {\bibfnamefont {J.}~\bibnamefont
  {Neveu}}, \bibinfo {author} {\bibfnamefont {V.}~\bibnamefont
  {Ruhlmann-Kleider}}, \bibinfo {author} {\bibfnamefont {A.}~\bibnamefont
  {Conley}}, \bibinfo {author} {\bibfnamefont {N.}~\bibnamefont
  {Palanque-Delabrouille}}, \bibinfo {author} {\bibfnamefont {P.}~\bibnamefont
  {Astier}}, \bibinfo {author} {\bibfnamefont {J.}~\bibnamefont {Guy}}, \ and\
  \bibinfo {author} {\bibfnamefont {E.}~\bibnamefont {Babichev}},\ }\href
  {\doibase 10.1051/0004-6361/201321256} {\bibfield  {journal} {\bibinfo
  {journal} {Astron. Astrophys.}\ }\textbf {\bibinfo {volume} {555}},\ \bibinfo
  {pages} {A53} (\bibinfo {year} {2013})},\ \Eprint
  {http://arxiv.org/abs/1302.2786} {arXiv:1302.2786 [gr-qc]} \BibitemShut
  {NoStop}%
\bibitem [{\citenamefont {Neveu}\ \emph {et~al.}(2017)\citenamefont {Neveu},
  \citenamefont {Ruhlmann-Kleider}, \citenamefont {Astier}, \citenamefont
  {Besan\c{c}on}, \citenamefont {Guy}, \citenamefont {M\"oller},\ and\
  \citenamefont {Babichev}}]{Neveu:2016gxp}%
  \BibitemOpen
  \bibfield  {author} {\bibinfo {author} {\bibfnamefont {J.}~\bibnamefont
  {Neveu}}, \bibinfo {author} {\bibfnamefont {V.}~\bibnamefont
  {Ruhlmann-Kleider}}, \bibinfo {author} {\bibfnamefont {P.}~\bibnamefont
  {Astier}}, \bibinfo {author} {\bibfnamefont {M.}~\bibnamefont
  {Besan\c{c}on}}, \bibinfo {author} {\bibfnamefont {J.}~\bibnamefont {Guy}},
  \bibinfo {author} {\bibfnamefont {A.}~\bibnamefont {M\"oller}}, \ and\
  \bibinfo {author} {\bibfnamefont {E.}~\bibnamefont {Babichev}},\ }\href
  {\doibase 10.1051/0004-6361/201628878} {\bibfield  {journal} {\bibinfo
  {journal} {Astron. Astrophys.}\ }\textbf {\bibinfo {volume} {600}},\ \bibinfo
  {pages} {A40} (\bibinfo {year} {2017})},\ \Eprint
  {http://arxiv.org/abs/1605.02627} {arXiv:1605.02627 [gr-qc]} \BibitemShut
  {NoStop}%
\bibitem [{\citenamefont {Horndeski}(1974)}]{Horndeski:1974wa}%
  \BibitemOpen
  \bibfield  {author} {\bibinfo {author} {\bibfnamefont {G.~W.}\ \bibnamefont
  {Horndeski}},\ }\href {\doibase 10.1007/BF01807638} {\bibfield  {journal}
  {\bibinfo  {journal} {Int. J. Theor. Phys.}\ }\textbf {\bibinfo {volume}
  {10}},\ \bibinfo {pages} {363} (\bibinfo {year} {1974})}\BibitemShut
  {NoStop}%
\bibitem [{\citenamefont {Kobayashi}\ \emph {et~al.}(2011)\citenamefont
  {Kobayashi}, \citenamefont {Yamaguchi},\ and\ \citenamefont
  {Yokoyama}}]{Kobayashi:2011nu}%
  \BibitemOpen
  \bibfield  {author} {\bibinfo {author} {\bibfnamefont {T.}~\bibnamefont
  {Kobayashi}}, \bibinfo {author} {\bibfnamefont {M.}~\bibnamefont
  {Yamaguchi}}, \ and\ \bibinfo {author} {\bibfnamefont {J.}~\bibnamefont
  {Yokoyama}},\ }\href {\doibase 10.1143/PTP.126.511} {\bibfield  {journal}
  {\bibinfo  {journal} {Prog. Theor. Phys.}\ }\textbf {\bibinfo {volume}
  {126}},\ \bibinfo {pages} {511} (\bibinfo {year} {2011})},\ \Eprint
  {http://arxiv.org/abs/1105.5723} {arXiv:1105.5723 [hep-th]} \BibitemShut
  {NoStop}%
\bibitem [{\citenamefont {Zumalac\'arregui}\ and\ \citenamefont
  {Garc\'\i{}a-Bellido}(2014)}]{Zumalacarregui:2013pma}%
  \BibitemOpen
  \bibfield  {author} {\bibinfo {author} {\bibfnamefont {M.}~\bibnamefont
  {Zumalac\'arregui}}\ and\ \bibinfo {author} {\bibfnamefont {J.}~\bibnamefont
  {Garc\'\i{}a-Bellido}},\ }\href {\doibase 10.1103/PhysRevD.89.064046}
  {\bibfield  {journal} {\bibinfo  {journal} {Phys. Rev. D}\ }\textbf {\bibinfo
  {volume} {89}},\ \bibinfo {pages} {064046} (\bibinfo {year} {2014})},\
  \Eprint {http://arxiv.org/abs/1308.4685} {arXiv:1308.4685 [gr-qc]}
  \BibitemShut {NoStop}%
\bibitem [{\citenamefont {Gleyzes}\ \emph {et~al.}(2015)\citenamefont
  {Gleyzes}, \citenamefont {Langlois}, \citenamefont {Piazza},\ and\
  \citenamefont {Vernizzi}}]{Gleyzes:2014dya}%
  \BibitemOpen
  \bibfield  {author} {\bibinfo {author} {\bibfnamefont {J.}~\bibnamefont
  {Gleyzes}}, \bibinfo {author} {\bibfnamefont {D.}~\bibnamefont {Langlois}},
  \bibinfo {author} {\bibfnamefont {F.}~\bibnamefont {Piazza}}, \ and\ \bibinfo
  {author} {\bibfnamefont {F.}~\bibnamefont {Vernizzi}},\ }\href {\doibase
  10.1103/PhysRevLett.114.211101} {\bibfield  {journal} {\bibinfo  {journal}
  {Phys. Rev. Lett.}\ }\textbf {\bibinfo {volume} {114}},\ \bibinfo {pages}
  {211101} (\bibinfo {year} {2015})},\ \Eprint {http://arxiv.org/abs/1404.6495}
  {arXiv:1404.6495 [hep-th]} \BibitemShut {NoStop}%
\bibitem [{\citenamefont {Vainshtein}(1972)}]{Vainshtein:1972sx}%
  \BibitemOpen
  \bibfield  {author} {\bibinfo {author} {\bibfnamefont {A.~I.}\ \bibnamefont
  {Vainshtein}},\ }\href {\doibase 10.1016/0370-2693(72)90147-5} {\bibfield
  {journal} {\bibinfo  {journal} {Phys. Lett. B}\ }\textbf {\bibinfo {volume}
  {39}},\ \bibinfo {pages} {393} (\bibinfo {year} {1972})}\BibitemShut
  {NoStop}%
\bibitem [{\citenamefont {Babichev}\ and\ \citenamefont
  {Deffayet}(2013)}]{Babichev:2013usa}%
  \BibitemOpen
  \bibfield  {author} {\bibinfo {author} {\bibfnamefont {E.}~\bibnamefont
  {Babichev}}\ and\ \bibinfo {author} {\bibfnamefont {C.}~\bibnamefont
  {Deffayet}},\ }\href {\doibase 10.1088/0264-9381/30/18/184001} {\bibfield
  {journal} {\bibinfo  {journal} {Class. Quant. Grav.}\ }\textbf {\bibinfo
  {volume} {30}},\ \bibinfo {pages} {184001} (\bibinfo {year} {2013})},\
  \Eprint {http://arxiv.org/abs/1304.7240} {arXiv:1304.7240 [gr-qc]}
  \BibitemShut {NoStop}%
\bibitem [{\citenamefont {Brax}\ \emph {et~al.}(2004)\citenamefont {Brax},
  \citenamefont {van~de Bruck}, \citenamefont {Davis}, \citenamefont {Khoury},\
  and\ \citenamefont {Weltman}}]{Brax:2004qh}%
  \BibitemOpen
  \bibfield  {author} {\bibinfo {author} {\bibfnamefont {P.}~\bibnamefont
  {Brax}}, \bibinfo {author} {\bibfnamefont {C.}~\bibnamefont {van~de Bruck}},
  \bibinfo {author} {\bibfnamefont {A.-C.}\ \bibnamefont {Davis}}, \bibinfo
  {author} {\bibfnamefont {J.}~\bibnamefont {Khoury}}, \ and\ \bibinfo {author}
  {\bibfnamefont {A.}~\bibnamefont {Weltman}},\ }\href {\doibase
  10.1103/PhysRevD.70.123518} {\bibfield  {journal} {\bibinfo  {journal} {Phys.
  Rev. D}\ }\textbf {\bibinfo {volume} {70}},\ \bibinfo {pages} {123518}
  (\bibinfo {year} {2004})},\ \Eprint {http://arxiv.org/abs/astro-ph/0408415}
  {arXiv:astro-ph/0408415} \BibitemShut {NoStop}%
\bibitem [{\citenamefont {Crisostomi}\ and\ \citenamefont
  {Koyama}(2018)}]{Crisostomi:2017lbg}%
  \BibitemOpen
  \bibfield  {author} {\bibinfo {author} {\bibfnamefont {M.}~\bibnamefont
  {Crisostomi}}\ and\ \bibinfo {author} {\bibfnamefont {K.}~\bibnamefont
  {Koyama}},\ }\href {\doibase 10.1103/PhysRevD.97.021301} {\bibfield
  {journal} {\bibinfo  {journal} {Phys. Rev.}\ }\textbf {\bibinfo {volume}
  {D97}},\ \bibinfo {pages} {021301} (\bibinfo {year} {2018})},\ \Eprint
  {http://arxiv.org/abs/1711.06661} {arXiv:1711.06661 [astro-ph.CO]}
  \BibitemShut {NoStop}%
\bibitem [{\citenamefont {Langlois}\ \emph
  {et~al.}(2018{\natexlab{a}})\citenamefont {Langlois}, \citenamefont {Saito},
  \citenamefont {Yamauchi},\ and\ \citenamefont {Noui}}]{Langlois:2017dyl}%
  \BibitemOpen
  \bibfield  {author} {\bibinfo {author} {\bibfnamefont {D.}~\bibnamefont
  {Langlois}}, \bibinfo {author} {\bibfnamefont {R.}~\bibnamefont {Saito}},
  \bibinfo {author} {\bibfnamefont {D.}~\bibnamefont {Yamauchi}}, \ and\
  \bibinfo {author} {\bibfnamefont {K.}~\bibnamefont {Noui}},\ }\href {\doibase
  10.1103/PhysRevD.97.061501} {\bibfield  {journal} {\bibinfo  {journal} {Phys.
  Rev. D}\ }\textbf {\bibinfo {volume} {97}},\ \bibinfo {pages} {061501}
  (\bibinfo {year} {2018}{\natexlab{a}})},\ \Eprint
  {http://arxiv.org/abs/1711.07403} {arXiv:1711.07403 [gr-qc]} \BibitemShut
  {NoStop}%
\bibitem [{\citenamefont {Bartolo}\ \emph {et~al.}(2018)\citenamefont
  {Bartolo}, \citenamefont {Karmakar}, \citenamefont {Matarrese},\ and\
  \citenamefont {Scomparin}}]{Bartolo:2017ibw}%
  \BibitemOpen
  \bibfield  {author} {\bibinfo {author} {\bibfnamefont {N.}~\bibnamefont
  {Bartolo}}, \bibinfo {author} {\bibfnamefont {P.}~\bibnamefont {Karmakar}},
  \bibinfo {author} {\bibfnamefont {S.}~\bibnamefont {Matarrese}}, \ and\
  \bibinfo {author} {\bibfnamefont {M.}~\bibnamefont {Scomparin}},\ }\href
  {\doibase 10.1088/1475-7516/2018/05/048} {\bibfield  {journal} {\bibinfo
  {journal} {JCAP}\ }\textbf {\bibinfo {volume} {1805}},\ \bibinfo {pages}
  {048} (\bibinfo {year} {2018})},\ \Eprint {http://arxiv.org/abs/1712.04002}
  {arXiv:1712.04002 [gr-qc]} \BibitemShut {NoStop}%
\bibitem [{\citenamefont {Dima}\ and\ \citenamefont
  {Vernizzi}(2018)}]{Dima:2017pwp}%
  \BibitemOpen
  \bibfield  {author} {\bibinfo {author} {\bibfnamefont {A.}~\bibnamefont
  {Dima}}\ and\ \bibinfo {author} {\bibfnamefont {F.}~\bibnamefont
  {Vernizzi}},\ }\href {\doibase 10.1103/PhysRevD.97.101302} {\bibfield
  {journal} {\bibinfo  {journal} {Phys. Rev.}\ }\textbf {\bibinfo {volume}
  {D97}},\ \bibinfo {pages} {101302} (\bibinfo {year} {2018})},\ \Eprint
  {http://arxiv.org/abs/1712.04731} {arXiv:1712.04731 [gr-qc]} \BibitemShut
  {NoStop}%
\bibitem [{\citenamefont {Hirano}\ \emph {et~al.}(2019)\citenamefont {Hirano},
  \citenamefont {Kobayashi},\ and\ \citenamefont {Yamauchi}}]{Hirano:2019scf}%
  \BibitemOpen
  \bibfield  {author} {\bibinfo {author} {\bibfnamefont {S.}~\bibnamefont
  {Hirano}}, \bibinfo {author} {\bibfnamefont {T.}~\bibnamefont {Kobayashi}}, \
  and\ \bibinfo {author} {\bibfnamefont {D.}~\bibnamefont {Yamauchi}},\ }\href
  {\doibase 10.1103/PhysRevD.99.104073} {\bibfield  {journal} {\bibinfo
  {journal} {Phys. Rev. D}\ }\textbf {\bibinfo {volume} {99}},\ \bibinfo
  {pages} {104073} (\bibinfo {year} {2019})},\ \Eprint
  {http://arxiv.org/abs/1903.08399} {arXiv:1903.08399 [gr-qc]} \BibitemShut
  {NoStop}%
\bibitem [{\citenamefont {Crisostomi}\ \emph {et~al.}(2019)\citenamefont
  {Crisostomi}, \citenamefont {Lewandowski},\ and\ \citenamefont
  {Vernizzi}}]{Crisostomi:2019yfo}%
  \BibitemOpen
  \bibfield  {author} {\bibinfo {author} {\bibfnamefont {M.}~\bibnamefont
  {Crisostomi}}, \bibinfo {author} {\bibfnamefont {M.}~\bibnamefont
  {Lewandowski}}, \ and\ \bibinfo {author} {\bibfnamefont {F.}~\bibnamefont
  {Vernizzi}},\ }\href {\doibase 10.1103/PhysRevD.100.024025} {\bibfield
  {journal} {\bibinfo  {journal} {Phys. Rev. D}\ }\textbf {\bibinfo {volume}
  {100}},\ \bibinfo {pages} {024025} (\bibinfo {year} {2019})},\ \Eprint
  {http://arxiv.org/abs/1903.11591} {arXiv:1903.11591 [gr-qc]} \BibitemShut
  {NoStop}%
\bibitem [{\citenamefont {Kimura}\ \emph {et~al.}(2012)\citenamefont {Kimura},
  \citenamefont {Kobayashi},\ and\ \citenamefont {Yamamoto}}]{Kimura:2011dc}%
  \BibitemOpen
  \bibfield  {author} {\bibinfo {author} {\bibfnamefont {R.}~\bibnamefont
  {Kimura}}, \bibinfo {author} {\bibfnamefont {T.}~\bibnamefont {Kobayashi}}, \
  and\ \bibinfo {author} {\bibfnamefont {K.}~\bibnamefont {Yamamoto}},\ }\href
  {\doibase 10.1103/PhysRevD.85.024023} {\bibfield  {journal} {\bibinfo
  {journal} {Phys. Rev. D}\ }\textbf {\bibinfo {volume} {85}},\ \bibinfo
  {pages} {024023} (\bibinfo {year} {2012})},\ \Eprint
  {http://arxiv.org/abs/1111.6749} {arXiv:1111.6749 [astro-ph.CO]} \BibitemShut
  {NoStop}%
\bibitem [{\citenamefont {Narikawa}\ \emph {et~al.}(2013)\citenamefont
  {Narikawa}, \citenamefont {Kobayashi}, \citenamefont {Yamauchi},\ and\
  \citenamefont {Saito}}]{Narikawa:2013pjr}%
  \BibitemOpen
  \bibfield  {author} {\bibinfo {author} {\bibfnamefont {T.}~\bibnamefont
  {Narikawa}}, \bibinfo {author} {\bibfnamefont {T.}~\bibnamefont {Kobayashi}},
  \bibinfo {author} {\bibfnamefont {D.}~\bibnamefont {Yamauchi}}, \ and\
  \bibinfo {author} {\bibfnamefont {R.}~\bibnamefont {Saito}},\ }\href
  {\doibase 10.1103/PhysRevD.87.124006} {\bibfield  {journal} {\bibinfo
  {journal} {Phys. Rev. D}\ }\textbf {\bibinfo {volume} {87}},\ \bibinfo
  {pages} {124006} (\bibinfo {year} {2013})},\ \Eprint
  {http://arxiv.org/abs/1302.2311} {arXiv:1302.2311 [astro-ph.CO]} \BibitemShut
  {NoStop}%
\bibitem [{\citenamefont {Koyama}\ \emph {et~al.}(2013)\citenamefont {Koyama},
  \citenamefont {Niz},\ and\ \citenamefont {Tasinato}}]{Koyama:2013paa}%
  \BibitemOpen
  \bibfield  {author} {\bibinfo {author} {\bibfnamefont {K.}~\bibnamefont
  {Koyama}}, \bibinfo {author} {\bibfnamefont {G.}~\bibnamefont {Niz}}, \ and\
  \bibinfo {author} {\bibfnamefont {G.}~\bibnamefont {Tasinato}},\ }\href
  {\doibase 10.1103/PhysRevD.88.021502} {\bibfield  {journal} {\bibinfo
  {journal} {Phys. Rev. D}\ }\textbf {\bibinfo {volume} {88}},\ \bibinfo
  {pages} {021502} (\bibinfo {year} {2013})},\ \Eprint
  {http://arxiv.org/abs/1305.0279} {arXiv:1305.0279 [hep-th]} \BibitemShut
  {NoStop}%
\bibitem [{\citenamefont {Kase}\ and\ \citenamefont
  {Tsujikawa}(2013)}]{Kase:2013uja}%
  \BibitemOpen
  \bibfield  {author} {\bibinfo {author} {\bibfnamefont {R.}~\bibnamefont
  {Kase}}\ and\ \bibinfo {author} {\bibfnamefont {S.}~\bibnamefont
  {Tsujikawa}},\ }\href {\doibase 10.1088/1475-7516/2013/08/054} {\bibfield
  {journal} {\bibinfo  {journal} {JCAP}\ }\textbf {\bibinfo {volume} {08}},\
  \bibinfo {pages} {054} (\bibinfo {year} {2013})},\ \Eprint
  {http://arxiv.org/abs/1306.6401} {arXiv:1306.6401 [gr-qc]} \BibitemShut
  {NoStop}%
\bibitem [{\citenamefont {Kobayashi}\ \emph {et~al.}(2015)\citenamefont
  {Kobayashi}, \citenamefont {Watanabe},\ and\ \citenamefont
  {Yamauchi}}]{Kobayashi:2014ida}%
  \BibitemOpen
  \bibfield  {author} {\bibinfo {author} {\bibfnamefont {T.}~\bibnamefont
  {Kobayashi}}, \bibinfo {author} {\bibfnamefont {Y.}~\bibnamefont {Watanabe}},
  \ and\ \bibinfo {author} {\bibfnamefont {D.}~\bibnamefont {Yamauchi}},\
  }\href {\doibase 10.1103/PhysRevD.91.064013} {\bibfield  {journal} {\bibinfo
  {journal} {Phys. Rev. D}\ }\textbf {\bibinfo {volume} {91}},\ \bibinfo
  {pages} {064013} (\bibinfo {year} {2015})},\ \Eprint
  {http://arxiv.org/abs/1411.4130} {arXiv:1411.4130 [gr-qc]} \BibitemShut
  {NoStop}%
\bibitem [{\citenamefont {Koyama}\ and\ \citenamefont
  {Sakstein}(2015)}]{Koyama:2015oma}%
  \BibitemOpen
  \bibfield  {author} {\bibinfo {author} {\bibfnamefont {K.}~\bibnamefont
  {Koyama}}\ and\ \bibinfo {author} {\bibfnamefont {J.}~\bibnamefont
  {Sakstein}},\ }\href {\doibase 10.1103/PhysRevD.91.124066} {\bibfield
  {journal} {\bibinfo  {journal} {Phys. Rev.}\ }\textbf {\bibinfo {volume}
  {D91}},\ \bibinfo {pages} {124066} (\bibinfo {year} {2015})},\ \Eprint
  {http://arxiv.org/abs/1502.06872} {arXiv:1502.06872 [astro-ph.CO]}
  \BibitemShut {NoStop}%
\bibitem [{\citenamefont {Saito}\ \emph {et~al.}(2015)\citenamefont {Saito},
  \citenamefont {Yamauchi}, \citenamefont {Mizuno}, \citenamefont {Gleyzes},\
  and\ \citenamefont {Langlois}}]{Saito:2015fza}%
  \BibitemOpen
  \bibfield  {author} {\bibinfo {author} {\bibfnamefont {R.}~\bibnamefont
  {Saito}}, \bibinfo {author} {\bibfnamefont {D.}~\bibnamefont {Yamauchi}},
  \bibinfo {author} {\bibfnamefont {S.}~\bibnamefont {Mizuno}}, \bibinfo
  {author} {\bibfnamefont {J.}~\bibnamefont {Gleyzes}}, \ and\ \bibinfo
  {author} {\bibfnamefont {D.}~\bibnamefont {Langlois}},\ }\href {\doibase
  10.1088/1475-7516/2015/06/008} {\bibfield  {journal} {\bibinfo  {journal}
  {JCAP}\ }\textbf {\bibinfo {volume} {1506}},\ \bibinfo {pages} {008}
  (\bibinfo {year} {2015})},\ \Eprint {http://arxiv.org/abs/1503.01448}
  {arXiv:1503.01448 [gr-qc]} \BibitemShut {NoStop}%
\bibitem [{\citenamefont {Sakstein}(2015{\natexlab{a}})}]{Sakstein:2015zoa}%
  \BibitemOpen
  \bibfield  {author} {\bibinfo {author} {\bibfnamefont {J.}~\bibnamefont
  {Sakstein}},\ }\href {\doibase 10.1103/PhysRevLett.115.201101} {\bibfield
  {journal} {\bibinfo  {journal} {Phys. Rev. Lett.}\ }\textbf {\bibinfo
  {volume} {115}},\ \bibinfo {pages} {201101} (\bibinfo {year}
  {2015}{\natexlab{a}})},\ \Eprint {http://arxiv.org/abs/1510.05964}
  {arXiv:1510.05964 [astro-ph.CO]} \BibitemShut {NoStop}%
\bibitem [{\citenamefont {Sakstein}(2015{\natexlab{b}})}]{Sakstein:2015aac}%
  \BibitemOpen
  \bibfield  {author} {\bibinfo {author} {\bibfnamefont {J.}~\bibnamefont
  {Sakstein}},\ }\href {\doibase 10.1103/PhysRevD.92.124045} {\bibfield
  {journal} {\bibinfo  {journal} {Phys. Rev.}\ }\textbf {\bibinfo {volume}
  {D92}},\ \bibinfo {pages} {124045} (\bibinfo {year} {2015}{\natexlab{b}})},\
  \Eprint {http://arxiv.org/abs/1511.01685} {arXiv:1511.01685 [astro-ph.CO]}
  \BibitemShut {NoStop}%
\bibitem [{\citenamefont {Jain}\ \emph {et~al.}(2016)\citenamefont {Jain},
  \citenamefont {Kouvaris},\ and\ \citenamefont {Nielsen}}]{Jain:2015edg}%
  \BibitemOpen
  \bibfield  {author} {\bibinfo {author} {\bibfnamefont {R.~K.}\ \bibnamefont
  {Jain}}, \bibinfo {author} {\bibfnamefont {C.}~\bibnamefont {Kouvaris}}, \
  and\ \bibinfo {author} {\bibfnamefont {N.~G.}\ \bibnamefont {Nielsen}},\
  }\href {\doibase 10.1103/PhysRevLett.116.151103} {\bibfield  {journal}
  {\bibinfo  {journal} {Phys. Rev. Lett.}\ }\textbf {\bibinfo {volume} {116}},\
  \bibinfo {pages} {151103} (\bibinfo {year} {2016})},\ \Eprint
  {http://arxiv.org/abs/1512.05946} {arXiv:1512.05946 [astro-ph.CO]}
  \BibitemShut {NoStop}%
\bibitem [{\citenamefont {Babichev}\ \emph {et~al.}(2016)\citenamefont
  {Babichev}, \citenamefont {Koyama}, \citenamefont {Langlois}, \citenamefont
  {Saito},\ and\ \citenamefont {Sakstein}}]{Babichev:2016jom}%
  \BibitemOpen
  \bibfield  {author} {\bibinfo {author} {\bibfnamefont {E.}~\bibnamefont
  {Babichev}}, \bibinfo {author} {\bibfnamefont {K.}~\bibnamefont {Koyama}},
  \bibinfo {author} {\bibfnamefont {D.}~\bibnamefont {Langlois}}, \bibinfo
  {author} {\bibfnamefont {R.}~\bibnamefont {Saito}}, \ and\ \bibinfo {author}
  {\bibfnamefont {J.}~\bibnamefont {Sakstein}},\ }\href {\doibase
  10.1088/0264-9381/33/23/235014} {\bibfield  {journal} {\bibinfo  {journal}
  {Class. Quant. Grav.}\ }\textbf {\bibinfo {volume} {33}},\ \bibinfo {pages}
  {235014} (\bibinfo {year} {2016})},\ \Eprint
  {http://arxiv.org/abs/1606.06627} {arXiv:1606.06627 [gr-qc]} \BibitemShut
  {NoStop}%
\bibitem [{\citenamefont {Sakstein}\ \emph {et~al.}(2016)\citenamefont
  {Sakstein}, \citenamefont {Wilcox}, \citenamefont {Bacon}, \citenamefont
  {Koyama},\ and\ \citenamefont {Nichol}}]{Sakstein:2016ggl}%
  \BibitemOpen
  \bibfield  {author} {\bibinfo {author} {\bibfnamefont {J.}~\bibnamefont
  {Sakstein}}, \bibinfo {author} {\bibfnamefont {H.}~\bibnamefont {Wilcox}},
  \bibinfo {author} {\bibfnamefont {D.}~\bibnamefont {Bacon}}, \bibinfo
  {author} {\bibfnamefont {K.}~\bibnamefont {Koyama}}, \ and\ \bibinfo {author}
  {\bibfnamefont {R.~C.}\ \bibnamefont {Nichol}},\ }\href {\doibase
  10.1088/1475-7516/2016/07/019} {\bibfield  {journal} {\bibinfo  {journal}
  {JCAP}\ }\textbf {\bibinfo {volume} {1607}},\ \bibinfo {pages} {019}
  (\bibinfo {year} {2016})},\ \Eprint {http://arxiv.org/abs/1603.06368}
  {arXiv:1603.06368 [astro-ph.CO]} \BibitemShut {NoStop}%
\bibitem [{\citenamefont {Salzano}\ \emph {et~al.}(2017)\citenamefont
  {Salzano}, \citenamefont {Mota}, \citenamefont {Capozziello},\ and\
  \citenamefont {Donahue}}]{Salzano:2017qac}%
  \BibitemOpen
  \bibfield  {author} {\bibinfo {author} {\bibfnamefont {V.}~\bibnamefont
  {Salzano}}, \bibinfo {author} {\bibfnamefont {D.~F.}\ \bibnamefont {Mota}},
  \bibinfo {author} {\bibfnamefont {S.}~\bibnamefont {Capozziello}}, \ and\
  \bibinfo {author} {\bibfnamefont {M.}~\bibnamefont {Donahue}},\ }\href
  {\doibase 10.1103/PhysRevD.95.044038} {\bibfield  {journal} {\bibinfo
  {journal} {Phys. Rev. D}\ }\textbf {\bibinfo {volume} {95}},\ \bibinfo
  {pages} {044038} (\bibinfo {year} {2017})},\ \Eprint
  {http://arxiv.org/abs/1701.03517} {arXiv:1701.03517 [astro-ph.CO]}
  \BibitemShut {NoStop}%
\bibitem [{\citenamefont {Eckert}\ \emph {et~al.}(2017)\citenamefont {Eckert},
  \citenamefont {Ettori}, \citenamefont {Pointecouteau}, \citenamefont
  {Molendi}, \citenamefont {Paltani},\ and\ \citenamefont
  {Tchernin}}]{Eckert:2016bfe}%
  \BibitemOpen
  \bibfield  {author} {\bibinfo {author} {\bibfnamefont {D.}~\bibnamefont
  {Eckert}}, \bibinfo {author} {\bibfnamefont {S.}~\bibnamefont {Ettori}},
  \bibinfo {author} {\bibfnamefont {E.}~\bibnamefont {Pointecouteau}}, \bibinfo
  {author} {\bibfnamefont {S.}~\bibnamefont {Molendi}}, \bibinfo {author}
  {\bibfnamefont {S.}~\bibnamefont {Paltani}}, \ and\ \bibinfo {author}
  {\bibfnamefont {C.}~\bibnamefont {Tchernin}},\ }\href {\doibase
  10.1002/asna.201713345} {\bibfield  {journal} {\bibinfo  {journal} {Astron.
  Nachr.}\ }\textbf {\bibinfo {volume} {338}},\ \bibinfo {pages} {293}
  (\bibinfo {year} {2017})},\ \Eprint {http://arxiv.org/abs/1611.05051}
  {arXiv:1611.05051 [astro-ph.CO]} \BibitemShut {NoStop}%
\bibitem [{\citenamefont {Ettori}\ \emph {et~al.}(2019)\citenamefont {Ettori},
  \citenamefont {Ghirardini}, \citenamefont {Eckert}, \citenamefont
  {Pointecouteau}, \citenamefont {Gastaldello}, \citenamefont {Sereno},
  \citenamefont {Gaspari}, \citenamefont {Ghizzardi}, \citenamefont
  {Roncarelli},\ and\ \citenamefont {Rossetti}}]{Ettori:2018tus}%
  \BibitemOpen
  \bibfield  {author} {\bibinfo {author} {\bibfnamefont {S.}~\bibnamefont
  {Ettori}}, \bibinfo {author} {\bibfnamefont {V.}~\bibnamefont {Ghirardini}},
  \bibinfo {author} {\bibfnamefont {D.}~\bibnamefont {Eckert}}, \bibinfo
  {author} {\bibfnamefont {E.}~\bibnamefont {Pointecouteau}}, \bibinfo {author}
  {\bibfnamefont {F.}~\bibnamefont {Gastaldello}}, \bibinfo {author}
  {\bibfnamefont {M.}~\bibnamefont {Sereno}}, \bibinfo {author} {\bibfnamefont
  {M.}~\bibnamefont {Gaspari}}, \bibinfo {author} {\bibfnamefont
  {S.}~\bibnamefont {Ghizzardi}}, \bibinfo {author} {\bibfnamefont
  {M.}~\bibnamefont {Roncarelli}}, \ and\ \bibinfo {author} {\bibfnamefont
  {M.}~\bibnamefont {Rossetti}},\ }\href {\doibase 10.1051/0004-6361/201833323}
  {\bibfield  {journal} {\bibinfo  {journal} {Astron. Astrophys.}\ }\textbf
  {\bibinfo {volume} {621}},\ \bibinfo {pages} {A39} (\bibinfo {year}
  {2019})},\ \Eprint {http://arxiv.org/abs/1805.00035} {arXiv:1805.00035
  [astro-ph.CO]} \BibitemShut {NoStop}%
\bibitem [{\citenamefont {Eckert}\ \emph {et~al.}(2019)\citenamefont {Eckert}
  \emph {et~al.}}]{Eckert:2018mlz}%
  \BibitemOpen
  \bibfield  {author} {\bibinfo {author} {\bibfnamefont {D.}~\bibnamefont
  {Eckert}} \emph {et~al.},\ }\href {\doibase 10.1051/0004-6361/201833324}
  {\bibfield  {journal} {\bibinfo  {journal} {Astron. Astrophys.}\ }\textbf
  {\bibinfo {volume} {621}},\ \bibinfo {pages} {A40} (\bibinfo {year}
  {2019})},\ \Eprint {http://arxiv.org/abs/1805.00034} {arXiv:1805.00034
  [astro-ph.CO]} \BibitemShut {NoStop}%
\bibitem [{\citenamefont {Ghirardini}\ \emph {et~al.}(2019)\citenamefont
  {Ghirardini} \emph {et~al.}}]{Ghirardini:2018byi}%
  \BibitemOpen
  \bibfield  {author} {\bibinfo {author} {\bibfnamefont {V.}~\bibnamefont
  {Ghirardini}} \emph {et~al.},\ }\href {\doibase 10.1051/0004-6361/201833325}
  {\bibfield  {journal} {\bibinfo  {journal} {Astron. Astrophys.}\ }\textbf
  {\bibinfo {volume} {621}},\ \bibinfo {pages} {A41} (\bibinfo {year}
  {2019})},\ \Eprint {http://arxiv.org/abs/1805.00042} {arXiv:1805.00042
  [astro-ph.CO]} \BibitemShut {NoStop}%
\bibitem [{\citenamefont {Ade}\ \emph {et~al.}(2016{\natexlab{b}})\citenamefont
  {Ade} \emph {et~al.}}]{Planck:2015lwi}%
  \BibitemOpen
  \bibfield  {author} {\bibinfo {author} {\bibfnamefont {P.~A.~R.}\
  \bibnamefont {Ade}} \emph {et~al.} (\bibinfo {collaboration} {Planck}),\
  }\href {\doibase 10.1051/0004-6361/201525833} {\bibfield  {journal} {\bibinfo
   {journal} {Astron. Astrophys.}\ }\textbf {\bibinfo {volume} {594}},\
  \bibinfo {pages} {A24} (\bibinfo {year} {2016}{\natexlab{b}})},\ \Eprint
  {http://arxiv.org/abs/1502.01597} {arXiv:1502.01597 [astro-ph.CO]}
  \BibitemShut {NoStop}%
\bibitem [{\citenamefont {Terukina}\ \emph {et~al.}(2014)\citenamefont
  {Terukina}, \citenamefont {Lombriser}, \citenamefont {Yamamoto},
  \citenamefont {Bacon}, \citenamefont {Koyama},\ and\ \citenamefont
  {Nichol}}]{Terukina:2013eqa}%
  \BibitemOpen
  \bibfield  {author} {\bibinfo {author} {\bibfnamefont {A.}~\bibnamefont
  {Terukina}}, \bibinfo {author} {\bibfnamefont {L.}~\bibnamefont {Lombriser}},
  \bibinfo {author} {\bibfnamefont {K.}~\bibnamefont {Yamamoto}}, \bibinfo
  {author} {\bibfnamefont {D.}~\bibnamefont {Bacon}}, \bibinfo {author}
  {\bibfnamefont {K.}~\bibnamefont {Koyama}}, \ and\ \bibinfo {author}
  {\bibfnamefont {R.~C.}\ \bibnamefont {Nichol}},\ }\href {\doibase
  10.1088/1475-7516/2014/04/013} {\bibfield  {journal} {\bibinfo  {journal}
  {JCAP}\ }\textbf {\bibinfo {volume} {04}},\ \bibinfo {pages} {013} (\bibinfo
  {year} {2014})},\ \Eprint {http://arxiv.org/abs/1312.5083} {arXiv:1312.5083
  [astro-ph.CO]} \BibitemShut {NoStop}%
\bibitem [{\citenamefont {Wilcox}\ \emph {et~al.}(2015)\citenamefont {Wilcox}
  \emph {et~al.}}]{Wilcox:2015kna}%
  \BibitemOpen
  \bibfield  {author} {\bibinfo {author} {\bibfnamefont {H.}~\bibnamefont
  {Wilcox}} \emph {et~al.},\ }\href {\doibase 10.1093/mnras/stv1366} {\bibfield
   {journal} {\bibinfo  {journal} {Mon. Not. Roy. Astron. Soc.}\ }\textbf
  {\bibinfo {volume} {452}},\ \bibinfo {pages} {1171} (\bibinfo {year}
  {2015})},\ \Eprint {http://arxiv.org/abs/1504.03937} {arXiv:1504.03937
  [astro-ph.CO]} \BibitemShut {NoStop}%
\bibitem [{\citenamefont {Pizzuti}\ \emph {et~al.}(2020)\citenamefont
  {Pizzuti}, \citenamefont {Saltas},\ and\ \citenamefont
  {Amendola}}]{Pizzuti:2020tdl}%
  \BibitemOpen
  \bibfield  {author} {\bibinfo {author} {\bibfnamefont {L.}~\bibnamefont
  {Pizzuti}}, \bibinfo {author} {\bibfnamefont {I.~D.}\ \bibnamefont {Saltas}},
  \ and\ \bibinfo {author} {\bibfnamefont {L.}~\bibnamefont {Amendola}},\
  }\href@noop {} {\  (\bibinfo {year} {2020})},\ \Eprint
  {http://arxiv.org/abs/2011.15089} {arXiv:2011.15089 [astro-ph.CO]}
  \BibitemShut {NoStop}%
\bibitem [{\citenamefont {Langlois}\ \emph
  {et~al.}(2018{\natexlab{b}})\citenamefont {Langlois}, \citenamefont {Saito},
  \citenamefont {Yamauchi},\ and\ \citenamefont {Noui}}]{1711.07403}%
  \BibitemOpen
  \bibfield  {author} {\bibinfo {author} {\bibfnamefont {D.}~\bibnamefont
  {Langlois}}, \bibinfo {author} {\bibfnamefont {R.}~\bibnamefont {Saito}},
  \bibinfo {author} {\bibfnamefont {D.}~\bibnamefont {Yamauchi}}, \ and\
  \bibinfo {author} {\bibfnamefont {K.}~\bibnamefont {Noui}},\ }\href {\doibase
  10.1103/PhysRevD.97.061501} {\bibfield  {journal} {\bibinfo  {journal} {Phys.
  Rev.}\ }\textbf {\bibinfo {volume} {D97}},\ \bibinfo {pages} {061501}
  (\bibinfo {year} {2018}{\natexlab{b}})},\ \Eprint
  {http://arxiv.org/abs/1711.07403} {arXiv:1711.07403 [gr-qc]} \BibitemShut
  {NoStop}%
\bibitem [{\citenamefont {Langlois}\ \emph {et~al.}(2017)\citenamefont
  {Langlois}, \citenamefont {Mancarella}, \citenamefont {Noui},\ and\
  \citenamefont {Vernizzi}}]{Langlois:2017mxy}%
  \BibitemOpen
  \bibfield  {author} {\bibinfo {author} {\bibfnamefont {D.}~\bibnamefont
  {Langlois}}, \bibinfo {author} {\bibfnamefont {M.}~\bibnamefont
  {Mancarella}}, \bibinfo {author} {\bibfnamefont {K.}~\bibnamefont {Noui}}, \
  and\ \bibinfo {author} {\bibfnamefont {F.}~\bibnamefont {Vernizzi}},\ }\href
  {\doibase 10.1088/1475-7516/2017/05/033} {\bibfield  {journal} {\bibinfo
  {journal} {JCAP}\ }\textbf {\bibinfo {volume} {1705}},\ \bibinfo {pages}
  {033} (\bibinfo {year} {2017})},\ \Eprint {http://arxiv.org/abs/1703.03797}
  {arXiv:1703.03797 [hep-th]} \BibitemShut {NoStop}%
\bibitem [{\citenamefont {{Kravtsov}}\ and\ \citenamefont
  {{Borgani}}(2012)}]{Kravtsov12}%
  \BibitemOpen
  \bibfield  {author} {\bibinfo {author} {\bibfnamefont {A.~V.}\ \bibnamefont
  {{Kravtsov}}}\ and\ \bibinfo {author} {\bibfnamefont {S.}~\bibnamefont
  {{Borgani}}},\ }\href {\doibase 10.1146/annurev-astro-081811-125502}
  {\bibfield  {journal} {\bibinfo  {journal} {\araa}\ }\textbf {\bibinfo
  {volume} {50}},\ \bibinfo {pages} {353} (\bibinfo {year} {2012})},\ \Eprint
  {http://arxiv.org/abs/1205.5556} {arXiv:1205.5556 [astro-ph.CO]} \BibitemShut
  {NoStop}%
\bibitem [{\citenamefont {Vikhlinin}\ \emph {et~al.}(2006)\citenamefont
  {Vikhlinin}, \citenamefont {Kravtsov}, \citenamefont {Forman}, \citenamefont
  {Jones}, \citenamefont {Markevitch}, \citenamefont {Murray},\ and\
  \citenamefont {Van~Speybroeck}}]{Vikhlinin:2005mp}%
  \BibitemOpen
  \bibfield  {author} {\bibinfo {author} {\bibfnamefont {A.}~\bibnamefont
  {Vikhlinin}}, \bibinfo {author} {\bibfnamefont {A.}~\bibnamefont {Kravtsov}},
  \bibinfo {author} {\bibfnamefont {W.}~\bibnamefont {Forman}}, \bibinfo
  {author} {\bibfnamefont {C.}~\bibnamefont {Jones}}, \bibinfo {author}
  {\bibfnamefont {M.}~\bibnamefont {Markevitch}}, \bibinfo {author}
  {\bibfnamefont {S.~S.}\ \bibnamefont {Murray}}, \ and\ \bibinfo {author}
  {\bibfnamefont {L.}~\bibnamefont {Van~Speybroeck}},\ }\href {\doibase
  10.1086/500288} {\bibfield  {journal} {\bibinfo  {journal} {Astrophys. J.}\
  }\textbf {\bibinfo {volume} {640}},\ \bibinfo {pages} {691} (\bibinfo {year}
  {2006})},\ \Eprint {http://arxiv.org/abs/astro-ph/0507092}
  {arXiv:astro-ph/0507092 [astro-ph]} \BibitemShut {NoStop}%
\bibitem [{\citenamefont {Vikhlinin}(2006)}]{astro-ph/0504098}%
  \BibitemOpen
  \bibfield  {author} {\bibinfo {author} {\bibfnamefont {A.}~\bibnamefont
  {Vikhlinin}},\ }\href {\doibase 10.1086/500121} {\bibfield  {journal}
  {\bibinfo  {journal} {Astrophys. J.}\ }\textbf {\bibinfo {volume} {640}},\
  \bibinfo {pages} {710} (\bibinfo {year} {2006})},\ \Eprint
  {http://arxiv.org/abs/astro-ph/0504098} {arXiv:astro-ph/0504098 [astro-ph]}
  \BibitemShut {NoStop}%
\bibitem [{\citenamefont {Navarro}\ \emph {et~al.}(1996)\citenamefont
  {Navarro}, \citenamefont {Frenk},\ and\ \citenamefont
  {White}}]{Navarro:1995iw}%
  \BibitemOpen
  \bibfield  {author} {\bibinfo {author} {\bibfnamefont {J.~F.}\ \bibnamefont
  {Navarro}}, \bibinfo {author} {\bibfnamefont {C.~S.}\ \bibnamefont {Frenk}},
  \ and\ \bibinfo {author} {\bibfnamefont {S.~D.~M.}\ \bibnamefont {White}},\
  }\href {\doibase 10.1086/177173} {\bibfield  {journal} {\bibinfo  {journal}
  {Astrophys. J.}\ }\textbf {\bibinfo {volume} {462}},\ \bibinfo {pages} {563}
  (\bibinfo {year} {1996})},\ \Eprint {http://arxiv.org/abs/astro-ph/9508025}
  {arXiv:astro-ph/9508025 [astro-ph]} \BibitemShut {NoStop}%
\bibitem [{\citenamefont {Ade}\ \emph {et~al.}(2014)\citenamefont {Ade} \emph
  {et~al.}}]{Ade:2013skr}%
  \BibitemOpen
  \bibfield  {author} {\bibinfo {author} {\bibfnamefont {P.}~\bibnamefont
  {Ade}} \emph {et~al.} (\bibinfo {collaboration} {Planck}),\ }\href {\doibase
  10.1051/0004-6361/201321523} {\bibfield  {journal} {\bibinfo  {journal}
  {Astron. Astrophys.}\ }\textbf {\bibinfo {volume} {571}},\ \bibinfo {pages}
  {A29} (\bibinfo {year} {2014})},\ \Eprint {http://arxiv.org/abs/1303.5089}
  {arXiv:1303.5089 [astro-ph.CO]} \BibitemShut {NoStop}%
\bibitem [{\citenamefont {Ghirardini}\ \emph {et~al.}(2018)\citenamefont
  {Ghirardini}, \citenamefont {Ettori}, \citenamefont {Eckert}, \citenamefont
  {Molendi}, \citenamefont {Gastaldello}, \citenamefont {Pointecouteau},
  \citenamefont {Hurier},\ and\ \citenamefont {Bourdin}}]{Ghirardini:2017apw}%
  \BibitemOpen
  \bibfield  {author} {\bibinfo {author} {\bibfnamefont {V.}~\bibnamefont
  {Ghirardini}}, \bibinfo {author} {\bibfnamefont {S.}~\bibnamefont {Ettori}},
  \bibinfo {author} {\bibfnamefont {D.}~\bibnamefont {Eckert}}, \bibinfo
  {author} {\bibfnamefont {S.}~\bibnamefont {Molendi}}, \bibinfo {author}
  {\bibfnamefont {F.}~\bibnamefont {Gastaldello}}, \bibinfo {author}
  {\bibfnamefont {E.}~\bibnamefont {Pointecouteau}}, \bibinfo {author}
  {\bibfnamefont {G.}~\bibnamefont {Hurier}}, \ and\ \bibinfo {author}
  {\bibfnamefont {H.}~\bibnamefont {Bourdin}},\ }\href {\doibase
  10.1051/0004-6361/201731748} {\bibfield  {journal} {\bibinfo  {journal}
  {Astron. Astrophys.}\ }\textbf {\bibinfo {volume} {614}},\ \bibinfo {pages}
  {A7} (\bibinfo {year} {2018})},\ \Eprint {http://arxiv.org/abs/1708.02954}
  {arXiv:1708.02954 [astro-ph.CO]} \BibitemShut {NoStop}%
\bibitem [{\citenamefont {Eckert}\ \emph {et~al.}(2016)\citenamefont {Eckert}
  \emph {et~al.}}]{Eckert:2015rlr}%
  \BibitemOpen
  \bibfield  {author} {\bibinfo {author} {\bibfnamefont {D.}~\bibnamefont
  {Eckert}} \emph {et~al.},\ }\href {\doibase 10.1051/0004-6361/201527293}
  {\bibfield  {journal} {\bibinfo  {journal} {Astron. Astrophys.}\ }\textbf
  {\bibinfo {volume} {592}},\ \bibinfo {pages} {A12} (\bibinfo {year}
  {2016})},\ \Eprint {http://arxiv.org/abs/1512.03814} {arXiv:1512.03814
  [astro-ph.CO]} \BibitemShut {NoStop}%
\bibitem [{\citenamefont {Eckert}\ \emph {et~al.}(2015)\citenamefont {Eckert},
  \citenamefont {Ettori}, \citenamefont {Molendi}, \citenamefont {Vazza},
  \citenamefont {Roncarelli}, \citenamefont {Gastaldello},\ and\ \citenamefont
  {Rossetti}}]{Eckert:2013faa}%
  \BibitemOpen
  \bibfield  {author} {\bibinfo {author} {\bibfnamefont {D.}~\bibnamefont
  {Eckert}}, \bibinfo {author} {\bibfnamefont {S.}~\bibnamefont {Ettori}},
  \bibinfo {author} {\bibfnamefont {S.}~\bibnamefont {Molendi}}, \bibinfo
  {author} {\bibfnamefont {F.}~\bibnamefont {Vazza}}, \bibinfo {author}
  {\bibfnamefont {M.}~\bibnamefont {Roncarelli}}, \bibinfo {author}
  {\bibfnamefont {F.}~\bibnamefont {Gastaldello}}, \ and\ \bibinfo {author}
  {\bibfnamefont {M.}~\bibnamefont {Rossetti}},\ }\href {\doibase
  10.1093/mnras/stu2590} {\bibfield  {journal} {\bibinfo  {journal} {Mon. Not.
  Roy. Astron. Soc.}\ }\textbf {\bibinfo {volume} {447}},\ \bibinfo {pages}
  {2198} (\bibinfo {year} {2015})},\ \Eprint {http://arxiv.org/abs/1310.8389}
  {arXiv:1310.8389 [astro-ph.CO]} \BibitemShut {NoStop}%
\bibitem [{\citenamefont {Hogg}\ and\ \citenamefont
  {Foreman-Mackey}(2018)}]{Hogg:2017akh}%
  \BibitemOpen
  \bibfield  {author} {\bibinfo {author} {\bibfnamefont {D.~W.}\ \bibnamefont
  {Hogg}}\ and\ \bibinfo {author} {\bibfnamefont {D.}~\bibnamefont
  {Foreman-Mackey}},\ }\href {\doibase 10.3847/1538-4365/aab76e} {\bibfield
  {journal} {\bibinfo  {journal} {Astrophys. J. Suppl.}\ }\textbf {\bibinfo
  {volume} {236}},\ \bibinfo {pages} {11} (\bibinfo {year} {2018})},\ \Eprint
  {http://arxiv.org/abs/1710.06068} {arXiv:1710.06068 [astro-ph.IM]}
  \BibitemShut {NoStop}%
\bibitem [{\citenamefont {{Foreman-Mackey}}\ \emph {et~al.}(2013)\citenamefont
  {{Foreman-Mackey}}, \citenamefont {{Hogg}}, \citenamefont {{Lang}},\ and\
  \citenamefont {{Goodman}}}]{Foreman-Mackey13}%
  \BibitemOpen
  \bibfield  {author} {\bibinfo {author} {\bibfnamefont {D.}~\bibnamefont
  {{Foreman-Mackey}}}, \bibinfo {author} {\bibfnamefont {D.~W.}\ \bibnamefont
  {{Hogg}}}, \bibinfo {author} {\bibfnamefont {D.}~\bibnamefont {{Lang}}}, \
  and\ \bibinfo {author} {\bibfnamefont {J.}~\bibnamefont {{Goodman}}},\ }\href
  {\doibase 10.1086/670067} {\bibfield  {journal} {\bibinfo  {journal}
  {Publications of the Astronomical Society of the Pacific}\ }\textbf {\bibinfo
  {volume} {125}},\ \bibinfo {pages} {306} (\bibinfo {year} {2013})},\ \Eprint
  {http://arxiv.org/abs/1202.3665} {arXiv:1202.3665 [astro-ph.IM]} \BibitemShut
  {NoStop}%
\bibitem [{\citenamefont {{Hinton}}(2016)}]{Hinton16}%
  \BibitemOpen
  \bibfield  {author} {\bibinfo {author} {\bibfnamefont {S.~R.}\ \bibnamefont
  {{Hinton}}},\ }\href {\doibase 10.21105/joss.00045} {\bibfield  {journal}
  {\bibinfo  {journal} {The Journal of Open Source Software}\ }\textbf
  {\bibinfo {volume} {1}},\ \bibinfo {eid} {00045} (\bibinfo {year}
  {2016})}\BibitemShut {NoStop}%
\bibitem [{\citenamefont {Trotta}(2008)}]{Trotta:2008qt}%
  \BibitemOpen
  \bibfield  {author} {\bibinfo {author} {\bibfnamefont {R.}~\bibnamefont
  {Trotta}},\ }\href {\doibase 10.1080/00107510802066753} {\bibfield  {journal}
  {\bibinfo  {journal} {Contemp. Phys.}\ }\textbf {\bibinfo {volume} {49}},\
  \bibinfo {pages} {71} (\bibinfo {year} {2008})},\ \Eprint
  {http://arxiv.org/abs/0803.4089} {arXiv:0803.4089 [astro-ph]} \BibitemShut
  {NoStop}%
\bibitem [{\citenamefont {Trotta}(2017)}]{Trotta:2017wnx}%
  \BibitemOpen
  \bibfield  {author} {\bibinfo {author} {\bibfnamefont {R.}~\bibnamefont
  {Trotta}}\ }(\bibinfo {year} {2017})\ \Eprint
  {http://arxiv.org/abs/1701.01467} {arXiv:1701.01467 [astro-ph.CO]}
  \BibitemShut {NoStop}%
\bibitem [{\citenamefont {Heavens}\ \emph
  {et~al.}(2017{\natexlab{a}})\citenamefont {Heavens}, \citenamefont {Fantaye},
  \citenamefont {Sellentin}, \citenamefont {Eggers}, \citenamefont {Hosenie},
  \citenamefont {Kroon},\ and\ \citenamefont {Mootoovaloo}}]{Heavens:2017hkr}%
  \BibitemOpen
  \bibfield  {author} {\bibinfo {author} {\bibfnamefont {A.}~\bibnamefont
  {Heavens}}, \bibinfo {author} {\bibfnamefont {Y.}~\bibnamefont {Fantaye}},
  \bibinfo {author} {\bibfnamefont {E.}~\bibnamefont {Sellentin}}, \bibinfo
  {author} {\bibfnamefont {H.}~\bibnamefont {Eggers}}, \bibinfo {author}
  {\bibfnamefont {Z.}~\bibnamefont {Hosenie}}, \bibinfo {author} {\bibfnamefont
  {S.}~\bibnamefont {Kroon}}, \ and\ \bibinfo {author} {\bibfnamefont
  {A.}~\bibnamefont {Mootoovaloo}},\ }\href {\doibase
  10.1103/PhysRevLett.119.101301} {\bibfield  {journal} {\bibinfo  {journal}
  {Phys. Rev. Lett.}\ }\textbf {\bibinfo {volume} {119}},\ \bibinfo {pages}
  {101301} (\bibinfo {year} {2017}{\natexlab{a}})},\ \Eprint
  {http://arxiv.org/abs/1704.03467} {arXiv:1704.03467 [astro-ph.CO]}
  \BibitemShut {NoStop}%
\bibitem [{\citenamefont {Heavens}\ \emph
  {et~al.}(2017{\natexlab{b}})\citenamefont {Heavens}, \citenamefont {Fantaye},
  \citenamefont {Mootoovaloo}, \citenamefont {Eggers}, \citenamefont {Hosenie},
  \citenamefont {Kroon},\ and\ \citenamefont {Sellentin}}]{Heavens:2017afc}%
  \BibitemOpen
  \bibfield  {author} {\bibinfo {author} {\bibfnamefont {A.}~\bibnamefont
  {Heavens}}, \bibinfo {author} {\bibfnamefont {Y.}~\bibnamefont {Fantaye}},
  \bibinfo {author} {\bibfnamefont {A.}~\bibnamefont {Mootoovaloo}}, \bibinfo
  {author} {\bibfnamefont {H.}~\bibnamefont {Eggers}}, \bibinfo {author}
  {\bibfnamefont {Z.}~\bibnamefont {Hosenie}}, \bibinfo {author} {\bibfnamefont
  {S.}~\bibnamefont {Kroon}}, \ and\ \bibinfo {author} {\bibfnamefont
  {E.}~\bibnamefont {Sellentin}},\ }\href@noop {} {\  (\bibinfo {year}
  {2017}{\natexlab{b}})},\ \Eprint {http://arxiv.org/abs/1704.03472}
  {arXiv:1704.03472 [stat.CO]} \BibitemShut {NoStop}%
\bibitem [{\citenamefont {Jeffreys}(1961)}]{Jeffreys:1939xee}%
  \BibitemOpen
  \bibfield  {author} {\bibinfo {author} {\bibfnamefont {H.}~\bibnamefont
  {Jeffreys}},\ }\href@noop {} {\emph {\bibinfo {title} {{The Theory of
  Probability}}}},\ Oxford Classic Texts in the Physical Sciences\ (\bibinfo
  {year} {1961})\BibitemShut {NoStop}%
\bibitem [{\citenamefont {Ameglio}\ \emph {et~al.}(2009)\citenamefont
  {Ameglio}, \citenamefont {Borgani}, \citenamefont {Pierpaoli}, \citenamefont
  {Dolag}, \citenamefont {Ettori},\ and\ \citenamefont
  {Morandi}}]{Ameglio:2008ip}%
  \BibitemOpen
  \bibfield  {author} {\bibinfo {author} {\bibfnamefont {S.}~\bibnamefont
  {Ameglio}}, \bibinfo {author} {\bibfnamefont {S.}~\bibnamefont {Borgani}},
  \bibinfo {author} {\bibfnamefont {E.}~\bibnamefont {Pierpaoli}}, \bibinfo
  {author} {\bibfnamefont {K.}~\bibnamefont {Dolag}}, \bibinfo {author}
  {\bibfnamefont {S.}~\bibnamefont {Ettori}}, \ and\ \bibinfo {author}
  {\bibfnamefont {A.}~\bibnamefont {Morandi}},\ }\href {\doibase
  10.1111/j.1365-2966.2008.14324.x} {\bibfield  {journal} {\bibinfo  {journal}
  {Mon. Not. Roy. Astron. Soc.}\ }\textbf {\bibinfo {volume} {394}},\ \bibinfo
  {pages} {479} (\bibinfo {year} {2009})},\ \Eprint
  {http://arxiv.org/abs/0811.2199} {arXiv:0811.2199 [astro-ph]} \BibitemShut
  {NoStop}%
\bibitem [{\citenamefont {Croston}\ \emph {et~al.}(2006)\citenamefont
  {Croston}, \citenamefont {Arnaud}, \citenamefont {Pointecouteau},\ and\
  \citenamefont {Pratt}}]{Croston:2006nq}%
  \BibitemOpen
  \bibfield  {author} {\bibinfo {author} {\bibfnamefont {J.~H.}\ \bibnamefont
  {Croston}}, \bibinfo {author} {\bibfnamefont {M.}~\bibnamefont {Arnaud}},
  \bibinfo {author} {\bibfnamefont {E.}~\bibnamefont {Pointecouteau}}, \ and\
  \bibinfo {author} {\bibfnamefont {G.~W.}\ \bibnamefont {Pratt}},\ }\href
  {\doibase 10.1051/0004-6361:20065795} {\bibfield  {journal} {\bibinfo
  {journal} {Astron. Astrophys.}\ }\textbf {\bibinfo {volume} {459}},\ \bibinfo
  {pages} {1007} (\bibinfo {year} {2006})},\ \Eprint
  {http://arxiv.org/abs/astro-ph/0608700} {arXiv:astro-ph/0608700} \BibitemShut
  {NoStop}%
\bibitem [{\citenamefont {Mathiesen}\ \emph {et~al.}(1999)\citenamefont
  {Mathiesen}, \citenamefont {Evrard},\ and\ \citenamefont
  {Mohr}}]{Mathiesen:1999pn}%
  \BibitemOpen
  \bibfield  {author} {\bibinfo {author} {\bibfnamefont {B.}~\bibnamefont
  {Mathiesen}}, \bibinfo {author} {\bibfnamefont {A.~E.}\ \bibnamefont
  {Evrard}}, \ and\ \bibinfo {author} {\bibfnamefont {J.~J.}\ \bibnamefont
  {Mohr}},\ }\href {\doibase 10.1086/312138} {\bibfield  {journal} {\bibinfo
  {journal} {Astrophys. J. Lett.}\ }\textbf {\bibinfo {volume} {520}},\
  \bibinfo {pages} {L21} (\bibinfo {year} {1999})},\ \Eprint
  {http://arxiv.org/abs/astro-ph/9904429} {arXiv:astro-ph/9904429} \BibitemShut
  {NoStop}%
\bibitem [{\citenamefont {Ettori}\ \emph {et~al.}(2013)\citenamefont {Ettori},
  \citenamefont {Donnarumma}, \citenamefont {Pointecouteau}, \citenamefont
  {Reiprich}, \citenamefont {Giodini}, \citenamefont {Lovisari},\ and\
  \citenamefont {Schmidt}}]{Ettori:2013tka}%
  \BibitemOpen
  \bibfield  {author} {\bibinfo {author} {\bibfnamefont {S.}~\bibnamefont
  {Ettori}}, \bibinfo {author} {\bibfnamefont {A.}~\bibnamefont {Donnarumma}},
  \bibinfo {author} {\bibfnamefont {E.}~\bibnamefont {Pointecouteau}}, \bibinfo
  {author} {\bibfnamefont {T.}~\bibnamefont {Reiprich}}, \bibinfo {author}
  {\bibfnamefont {S.}~\bibnamefont {Giodini}}, \bibinfo {author} {\bibfnamefont
  {L.}~\bibnamefont {Lovisari}}, \ and\ \bibinfo {author} {\bibfnamefont
  {R.}~\bibnamefont {Schmidt}},\ }\href {\doibase 10.1007/s11214-013-9976-7}
  {\bibfield  {journal} {\bibinfo  {journal} {Space Sci. Rev.}\ }\textbf
  {\bibinfo {volume} {177}},\ \bibinfo {pages} {119} (\bibinfo {year}
  {2013})},\ \Eprint {http://arxiv.org/abs/1303.3530} {arXiv:1303.3530
  [astro-ph.CO]} \BibitemShut {NoStop}%
\bibitem [{\citenamefont {Ettori}\ \emph {et~al.}(2017)\citenamefont {Ettori},
  \citenamefont {Ghirardini}, \citenamefont {Eckert}, \citenamefont {Dubath},\
  and\ \citenamefont {Pointecouteau}}]{Ettori:2016kll}%
  \BibitemOpen
  \bibfield  {author} {\bibinfo {author} {\bibfnamefont {S.}~\bibnamefont
  {Ettori}}, \bibinfo {author} {\bibfnamefont {V.}~\bibnamefont {Ghirardini}},
  \bibinfo {author} {\bibfnamefont {D.}~\bibnamefont {Eckert}}, \bibinfo
  {author} {\bibfnamefont {F.}~\bibnamefont {Dubath}}, \ and\ \bibinfo {author}
  {\bibfnamefont {E.}~\bibnamefont {Pointecouteau}},\ }\href {\doibase
  10.1093/mnrasl/slx074} {\bibfield  {journal} {\bibinfo  {journal} {Mon. Not.
  Roy. Astron. Soc.}\ }\textbf {\bibinfo {volume} {470}},\ \bibinfo {pages}
  {L29} (\bibinfo {year} {2017})},\ \Eprint {http://arxiv.org/abs/1612.07288}
  {arXiv:1612.07288 [astro-ph.CO]} \BibitemShut {NoStop}%
\bibitem [{\citenamefont {Saltas}\ \emph {et~al.}(2018)\citenamefont {Saltas},
  \citenamefont {Sawicki},\ and\ \citenamefont {Lopes}}]{Saltas:2018mxc}%
  \BibitemOpen
  \bibfield  {author} {\bibinfo {author} {\bibfnamefont {I.~D.}\ \bibnamefont
  {Saltas}}, \bibinfo {author} {\bibfnamefont {I.}~\bibnamefont {Sawicki}}, \
  and\ \bibinfo {author} {\bibfnamefont {I.}~\bibnamefont {Lopes}},\ }\href
  {\doibase 10.1088/1475-7516/2018/05/028} {\bibfield  {journal} {\bibinfo
  {journal} {JCAP}\ }\textbf {\bibinfo {volume} {05}},\ \bibinfo {pages} {028}
  (\bibinfo {year} {2018})},\ \Eprint {http://arxiv.org/abs/1803.00541}
  {arXiv:1803.00541 [astro-ph.CO]} \BibitemShut {NoStop}%
\bibitem [{\citenamefont {Arai}\ \emph {et~al.}(2020)\citenamefont {Arai},
  \citenamefont {Karmakar},\ and\ \citenamefont {Nishizawa}}]{Arai:2019zul}%
  \BibitemOpen
  \bibfield  {author} {\bibinfo {author} {\bibfnamefont {S.}~\bibnamefont
  {Arai}}, \bibinfo {author} {\bibfnamefont {P.}~\bibnamefont {Karmakar}}, \
  and\ \bibinfo {author} {\bibfnamefont {A.}~\bibnamefont {Nishizawa}},\ }\href
  {\doibase 10.1103/PhysRevD.102.024003} {\bibfield  {journal} {\bibinfo
  {journal} {Phys. Rev. D}\ }\textbf {\bibinfo {volume} {102}},\ \bibinfo
  {pages} {024003} (\bibinfo {year} {2020})},\ \Eprint
  {http://arxiv.org/abs/1912.01768} {arXiv:1912.01768 [gr-qc]} \BibitemShut
  {NoStop}%
\end{thebibliography}%

\appendix

\section{DHOST action}\label{app:dhost}
The action for the viable Class Ia DHOST theory after GW170817 event ($c_g^2=c^2$) is 
\begin{eqnarray}\label{ac:host}
 S&=& \int d^4x\,  \sqrt{-g}\,  \mathcal{L}\,, \label{ac:total}
\end{eqnarray}
where
\begin{equation} \label{ac:dhost:gw}
\begin{split}
& L^{_{\rm DHOST}}_{c_g^2=c^2}=  P + Q\,  \Box\phi +  F \,  R +  A_3\phi^\mu \phi^\nu \phi_{\mu \nu} \Box \phi  \\
&+\frac{1}{8F} \bigg(48 {F_X}^2 -8(F-X{F_X}) A_3-X^2 A_3^2 \bigg) \phi^\mu \phi_{\mu \nu} \phi_\lambda \phi^{\lambda \nu} \\
&+\frac{1}{2 F}\left(4{F_X}+X A_3\right) A_3(\phi_\mu \phi^{\mu \nu } \phi_\nu)^2 \;. 
\end{split}
\end{equation}
$P, Q, F, A_3$ are the arbitrary functions of the scalar field $\phi$ and its kinetic energy, $X$, which reduces to the GR case when we set $F= 1 / 2\kappa$ with $\kappa =8\pi \GN c^{-4}$, and $P, Q, A_3=0$.

In effective field theory (EFT) formalism, the action is expressed in terms of time dependent linear operators. $c_g=c$ tightly constrains the tensor speed alteration parameter, $\alpha_{\rm{T}}$. The viable DHOST theory ($c_g=c$) is expressed in terms of five time-dependent linear effective field theory (EFT) parameters, i.e., $\alpha_{\rm{M,B,K,H}}$ and $\bone$, which are the measures to the deviation from the $\Lambda$CDM model \cite{Langlois:2017mxy}. The coefficient of the linear perturbations only depends on the background value, hence only on the time. Subsequently, $\alpha_{\rm{M,B,K,H}}$ and $\bone$ are solely functions of time. 

\section{A note on change of basis}
\label{sec:physical_parameters}

\begin{figure}[!t]
\centering
\includegraphics[scale=0.37]{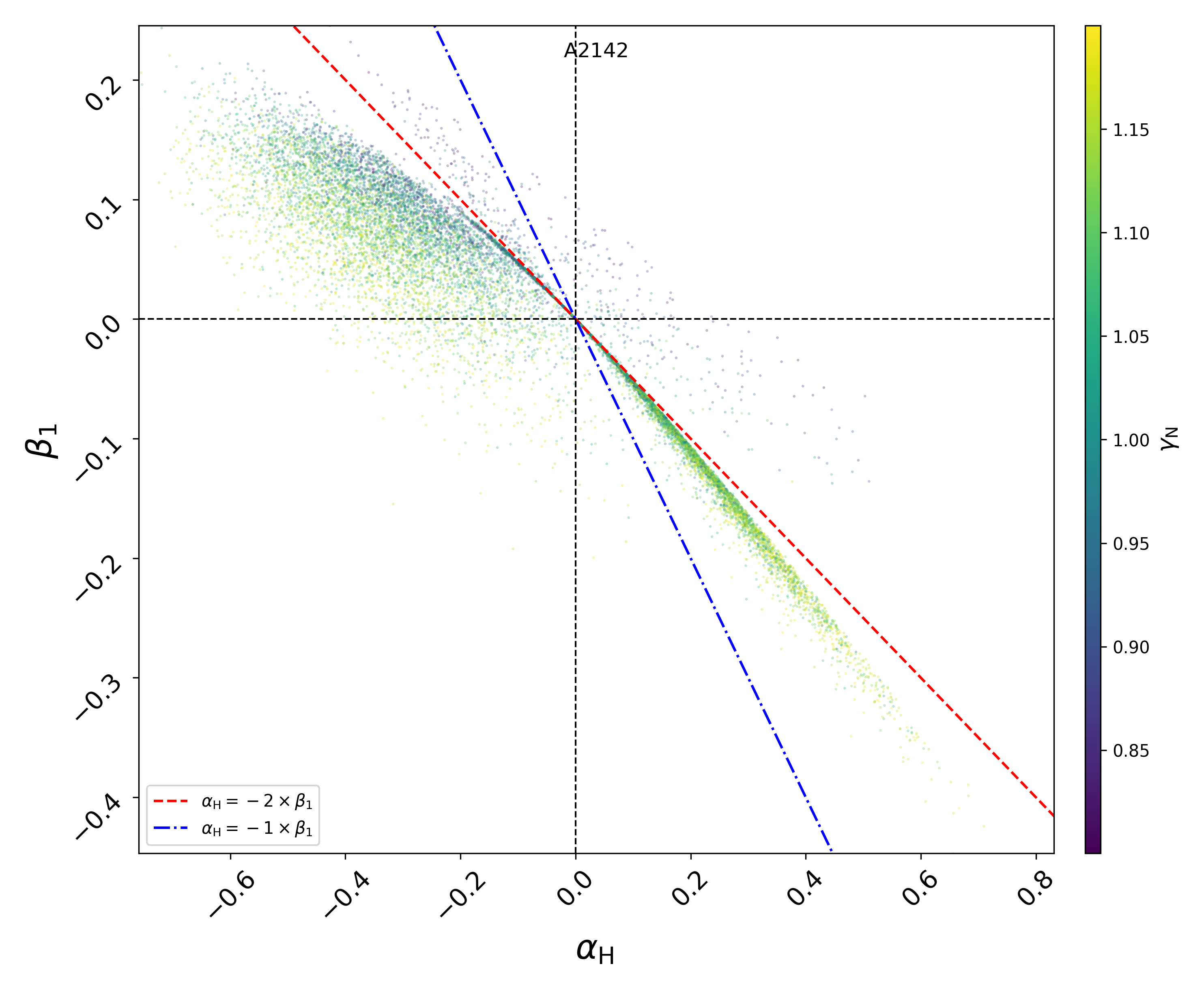}
 \caption{We show the scatter for the physical parameters of the DHOST theory, $\{\aH,\, \bone\}$. See \Cref{sec:physical_parameters} for discussion.}
 \label{fig:aH_beta1}
\end{figure}

While we have performed the analysis using the effective parameters $\{\Xi_1, \gNtilde\}$, they are both functions of the physical parameters $\{\aH,\, \bone,\, \gN \}$, which in turn are functions of time, and hence redshift. As mentioned earlier, $\gNtilde$ is unconstrained when sampled over as a free parameter and the posterior for $\Xi_1$ remains unchanged when fixing $\gNtilde = 1$. Therefore one could advantage of the analysis performed assuming $\gNtilde = 1$, and estimate, 
\begin{equation}
\label{eq:gNtilde}
\gNtilde \equiv \frac{\gNtilde\times \M^{\gNtilde \neq 1}}{\M^{\gNtilde = 1}}.
\end{equation}

Doing so, we find strictly a mean value of $\gNtilde \sim 1$ as expected, however with the dispersion that varies from cluster to cluster. If one were to assume a fixed but $\gNtilde \neq 1$ value, the mean obtained from the above expression would be expected to be same as the assumption. Therefore, we infer this dispersion as the tentative uncertainty on the GR expectation of $\gNtilde = 1$, in contrast to having either a fixed value or a completely unconstrained quantity. This validates the fact that the $\Xi_1$ and $\gNtilde$ are uncorrelated parameters and hence reasserts the method followed using two different MCMC samples to estimate this dispersion. In other words, it is an equivalent approach assuming that the mass of the cluster is accurately known in the DHOST gravity and that the uncertainty is present only on the $\gNtilde$ parameter. Needless to say, having probes such as weak lensing or any background expansion history, could break the degeneracy and help correctly constrain this quantity. Here we only anticipate an analysis assigning the uncertainty on the parameter $\gNtilde\times \M$ completely to $\gNtilde$ alone, however to obtain an expectation for the allowed parameter space of the physical quantities ($\aH,\,\bone$) within the DHOST theory. 

Through a simple change of basis we can represent the $\aH$ and $\bone$ in terms of effective parameters $\{\Xi_1, \gNtilde\}$ as,  

\begin{eqnarray} \label{eq:alphaH_beta1}
\begin{aligned}
 \aH &= \frac{1}{2}\left[\left(-1 +\frac{\gN}{\gNtilde} + \frac{3 \Xi_1}{2}\right) \pm \frac{3}{2}\sqrt{\xi}\right] \nonumber \\
 \bone &= \frac{1}{2}\left[\left(1 -\frac{\gN}{\gNtilde} - \frac{\Xi_1}{2} \right) \mp \frac{1}{2}\sqrt{\xi}\right] \nonumber \\
 \textrm{where,} \\ 
 \xi &= \Xi_1^{2}+ 4\Xi_1 \left(1 - \frac{\gN}{\gNtilde}\right).
 \end{aligned}
\end{eqnarray}

Note that the above solutions for $\aH$ and $\bone$ are valid under the conditions, $\left[\Xi_1>0,\, \,  \gNtilde > \gN \times 4/(4+\Xi_1) \right]$ or $\left[\Xi_1 < 0,\,\, \gNtilde < \gN \times 4/(4+\Xi_1) \right]$. While we have the distribution on $\gNtilde$ and $\Xi_1$, the parameter $\gN$ itself is free quantity and therefore certain values for the same have to be assumed before obtaining the distributions of $\{\aH, \bone\}$. As in, the scatter for values of $\{\aH,\, \bone \}$, for viable physical models can obtained using a flat distribution of $0.8< \gN <1.2$. Note that the phenomenology of DHOST gravity can be equivalent to that of GR, even if $\aH,\, \bone \neq 0$, when $\aH = - \bone$ and $\gN = 1-2\bone \equiv 1+2\aH$. The distributions of DHOST parameters have been estimated by computing the approximated time evolution of a scalar field through numerical simulations in \cite{Arai:2019zul}. A comparison can be done with simulations and data from future observations. We leave the possibility of a detailed analysis of the physical parameters along these lines for a future consideration, also with the inclusion of complementary weak lensing probes, which would allow to break additional degeneracies.

\begin{figure}[!ht]
\centering
\includegraphics[scale=0.38]{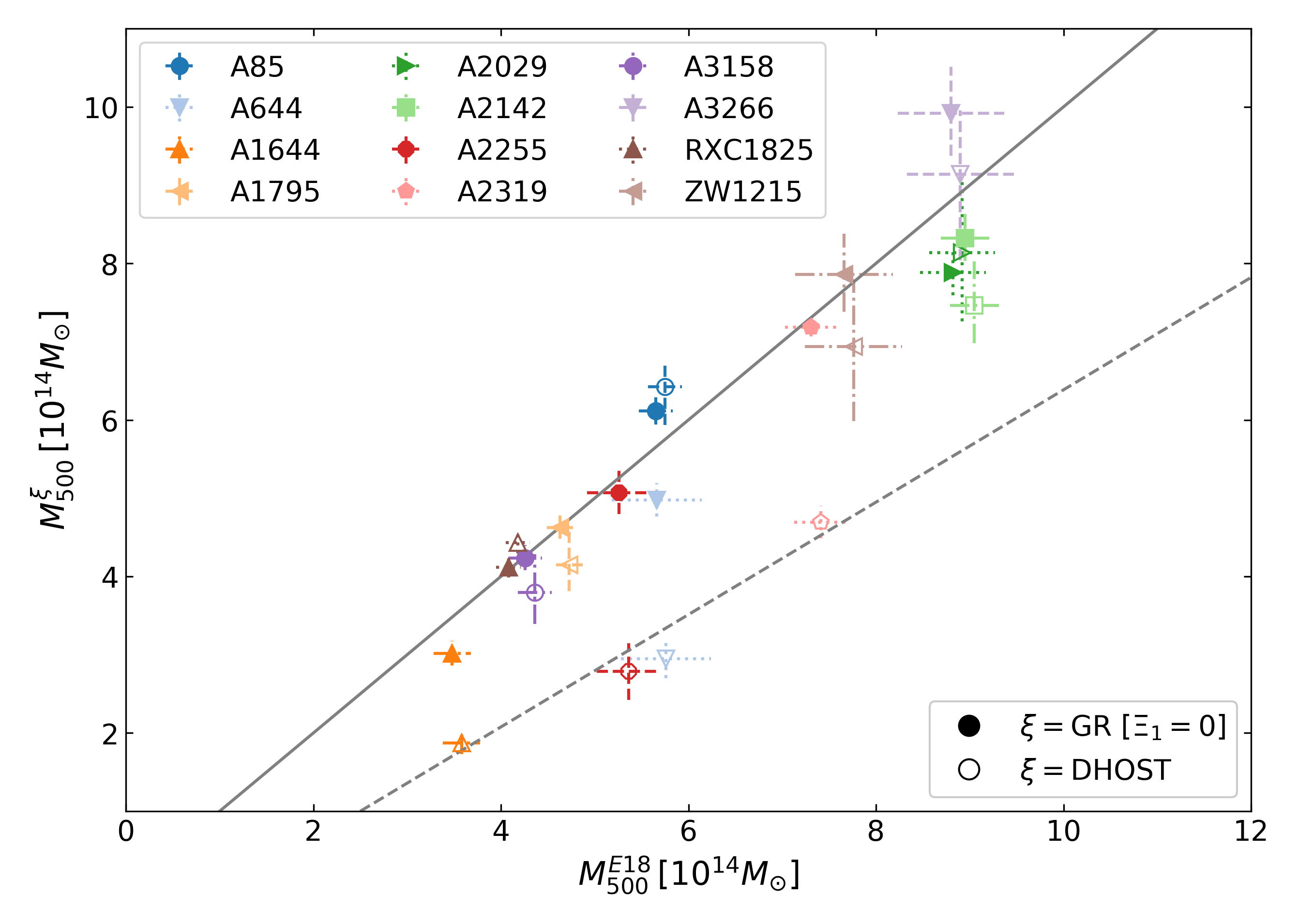}
 \caption{Here we show the comparison  of the mass estimates ($M_{500}$), obtained in our analysis for both GR and DHOST cases, against the GR estimates in \citetalias{Ettori:2018tus}. }
 \label{fig:mass_comp}
\end{figure}

\section{Mass comparison}
\label{sec:Mass_comparison}

In this section we provide a comparison the $M_{500}$ estimates obtained within our analysis and the original analysis presented in \citetalias{Ettori:2018tus}, as in \Cref{fig:mass_comp}. We also compare constrains in the DHOST scenario, which are shown as open markers. Firstly we notice an overall agreement for all the clusters in the GR case. However interestingly, we find that in our analysis, the uncertainties estimated for the non-NFW clusters are lower, while the mean values are in better agreement. Also the DHOST mass estimates for these 4 clusters tend to follow a constant scaling w.r.t the GR masses, of $M_{500}^{\rm DHOST} \sim 0.6 \times M_{500}^{\rm GR}$. The effects of mass variation are also represented equivalently in the lower panel of \Cref{fig:Xi_gamma_redshift} as a ratio. This is indeed an interesting feature, which requires more attention and we intend to explore this in a future study. 

The upper panels in \Cref{fig:Fit_Profiles}, show the mass reconstructions both in the GR (blue) and the DHOST (red) scenarios. In here one can notice that mass reconstructions in the DHOST scenario for the non-NFW clusters show larger variation from the GR case, and this illustrates the large values of the Bayesian evidence in favor of the DHOST modeling, reported in \Cref{tab:Constraints_PP} (see also \Cref{fig:Bayesian_evidence}). Th lower panels, also illustrate how the constraining ability of the $\Psz$ and $\Px$ data independently contribute to the joint likelihood (see Equation \ref{eq:chisqeff}), through the variation of uncertainty in the radial distance. 

As mentioned earlier we have utilized the Vikhlinin profile with only 6 parameters neglecting a second component (see eq.3 of \citep{Vikhlinin:2005mp}). Firstly, we find that the agreement between our mass estimates here and the mass estimates reported in \citetalias{Ettori:2018tus} worsens when we utilize the full parametric form with 9 free parameters. Alongside this, we also obtain much larger uncertainties on the mass estimates, which deters us from further using the full parametric form in the main analysis.

\begin{figure}[!ht]
\centering
\includegraphics[scale=0.5]{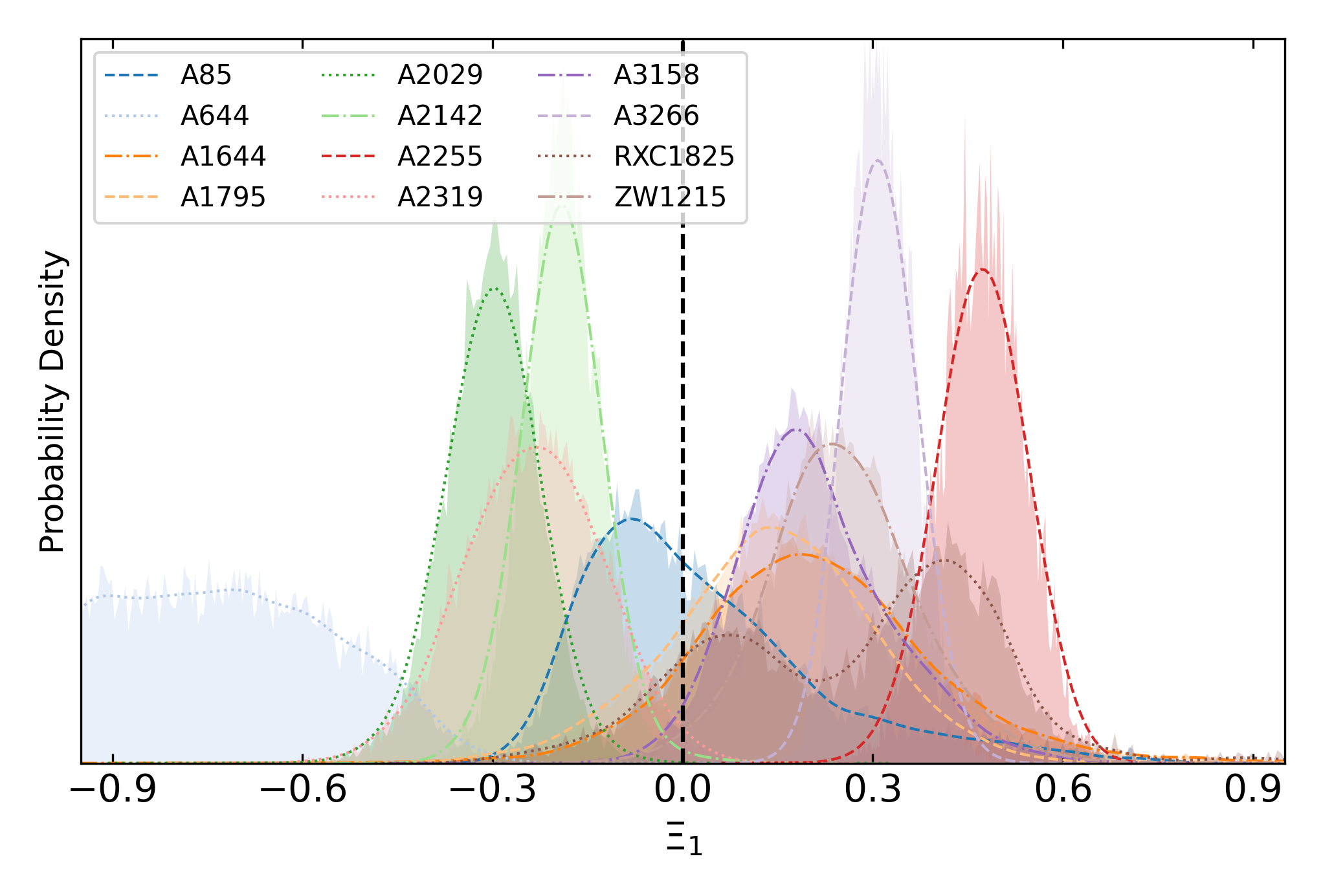}
 \caption{Same as \Cref{fig:XI}, using only the $\Psz$ data. Note here that the limits on the x-axis are different from those in \Cref{fig:XI}. Also, we have not excluded the three inner most data points as done in the main analysis.}
 \label{fig:XI_PSZ}
\end{figure}

\section{Constraints using $P_{\rm SZ}$ data alone}
\label{sec:SZ_alone}
Alongside our main analysis, we also assess the constraints on the DHOST parameter $\Xi_1$ when utilizing the $\Psz$ data alone. In \Cref{fig:XI_PSZ}, we show the marginalized posterior distributions, which can be compared with the main results in \Cref{fig:XI}. In here, however, we include also the 3 internal points of the $\Psz$ data, which were excluded in the main analysis. Notice that the constraints on the parameter $\Xi_1$, vary significantly and we find good consistency between our main analysis and the $\Psz$ analysis only for the clusters, A2142, RXC1825. For the rest of the clusters, however, the constraints vary, signifying the need for the formalism followed in the main analysis, i.e., exclusion of 3 innermost points of $\Psz$ data and joint analysis of $\Psz$ and $\Px$. We no longer notice the significant clustering of the constraints on $\Xi_1 \ll 0$ obtained for the non-NFW clusters and in general the constraints from individual clusters are more dispersed.

\begin{figure*}
    \centering
    \includegraphics[scale=0.3]{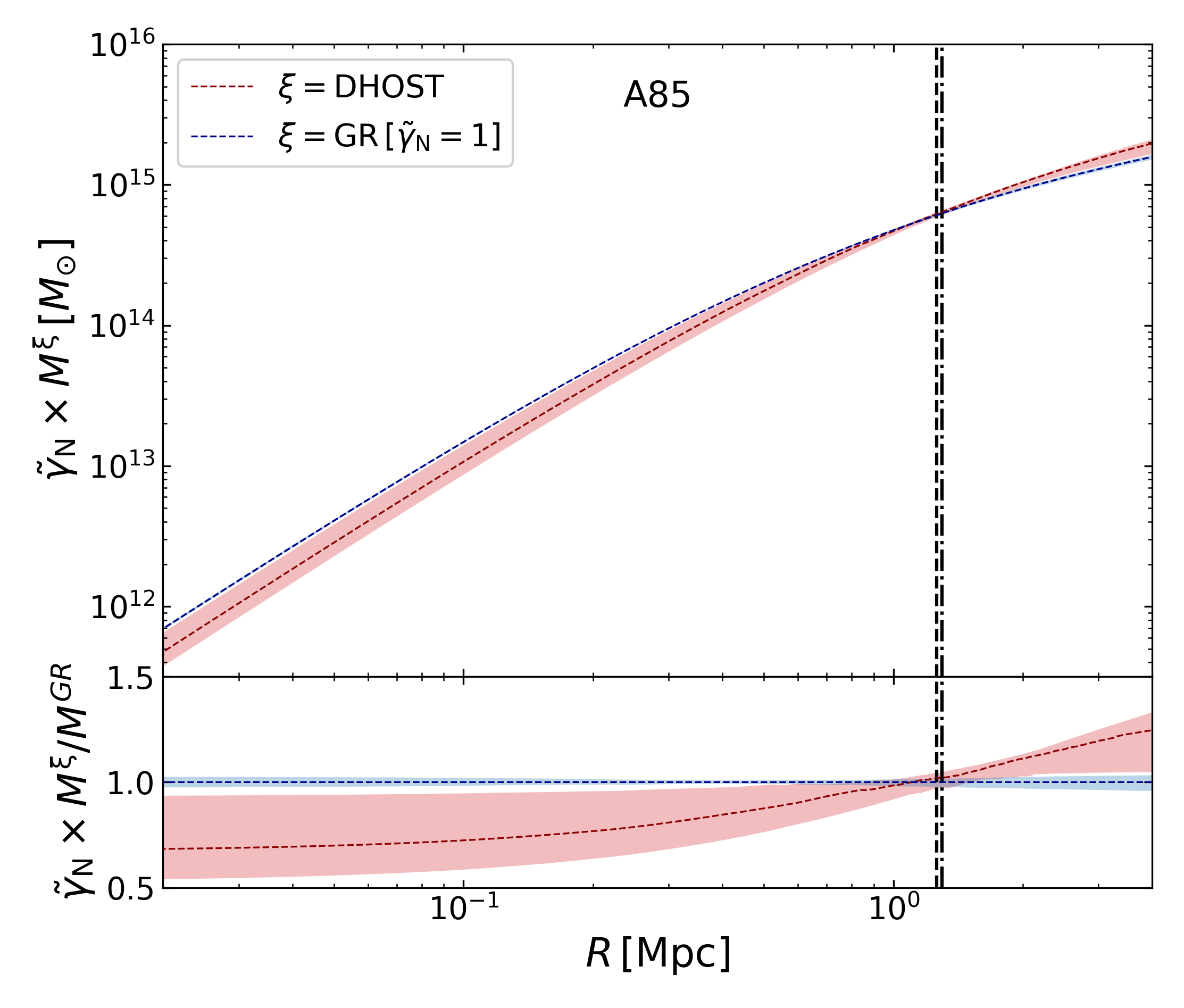}
    \includegraphics[scale=0.3]{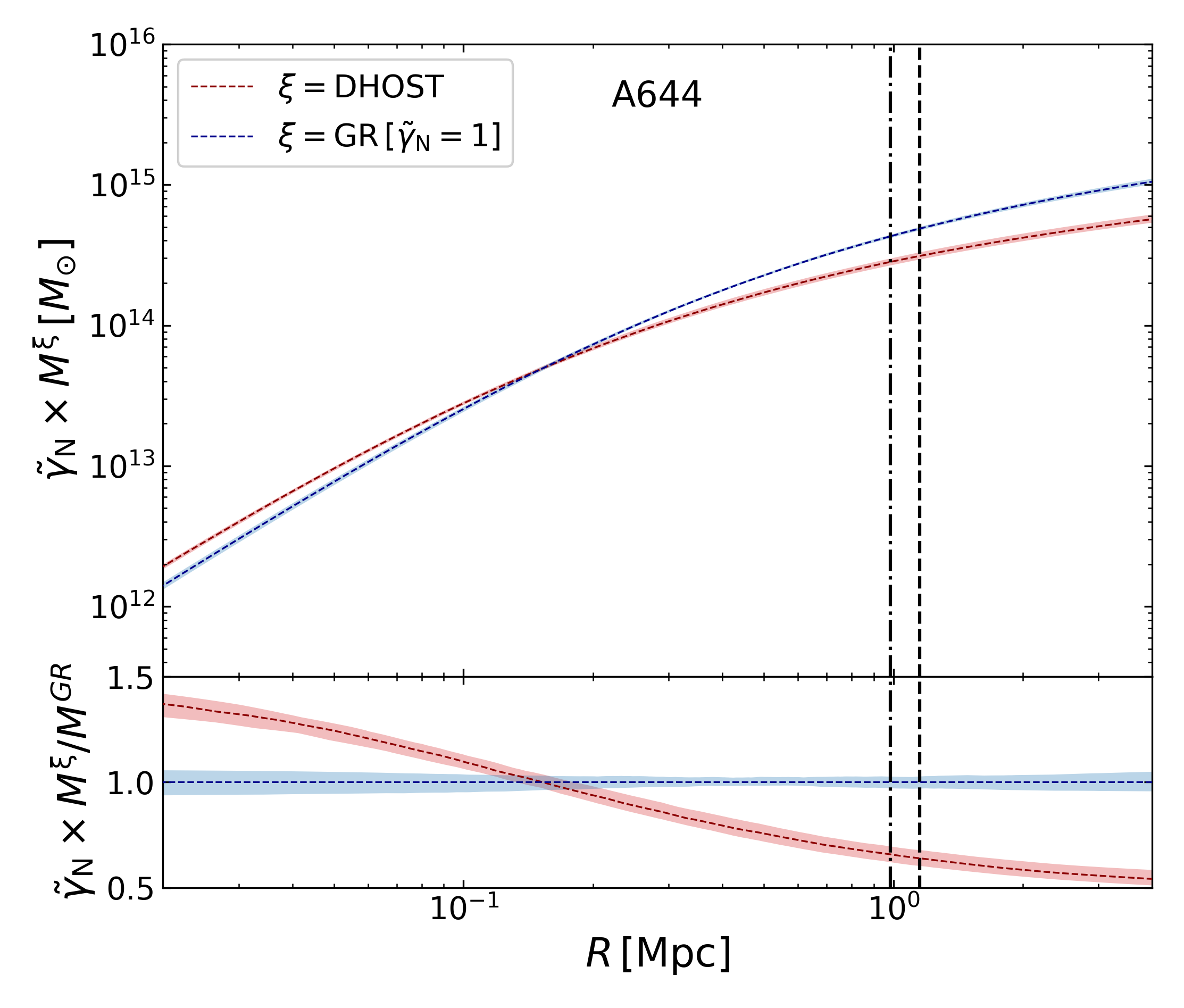}
    \includegraphics[scale=0.3]{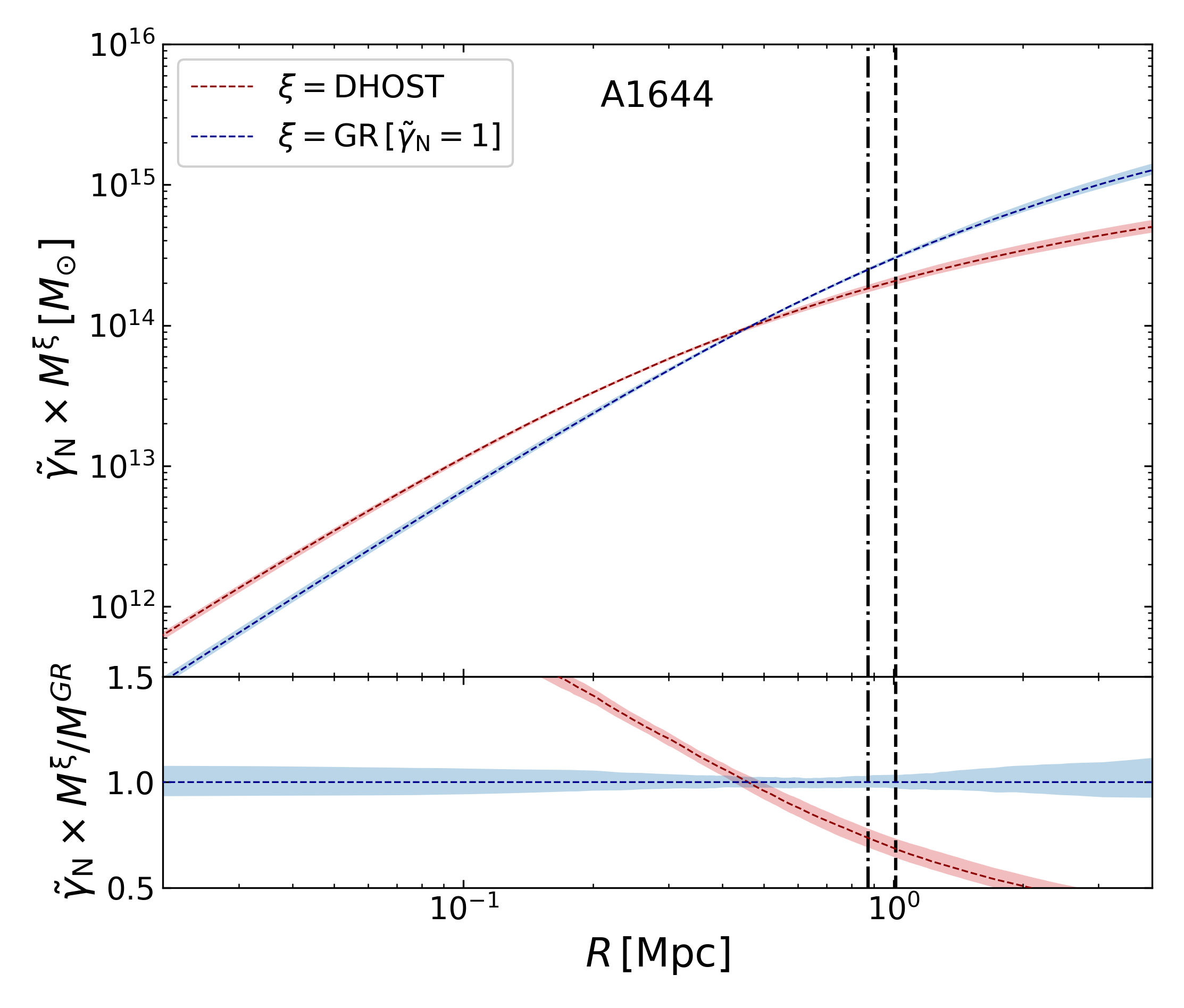}
    
    \includegraphics[scale=0.3]{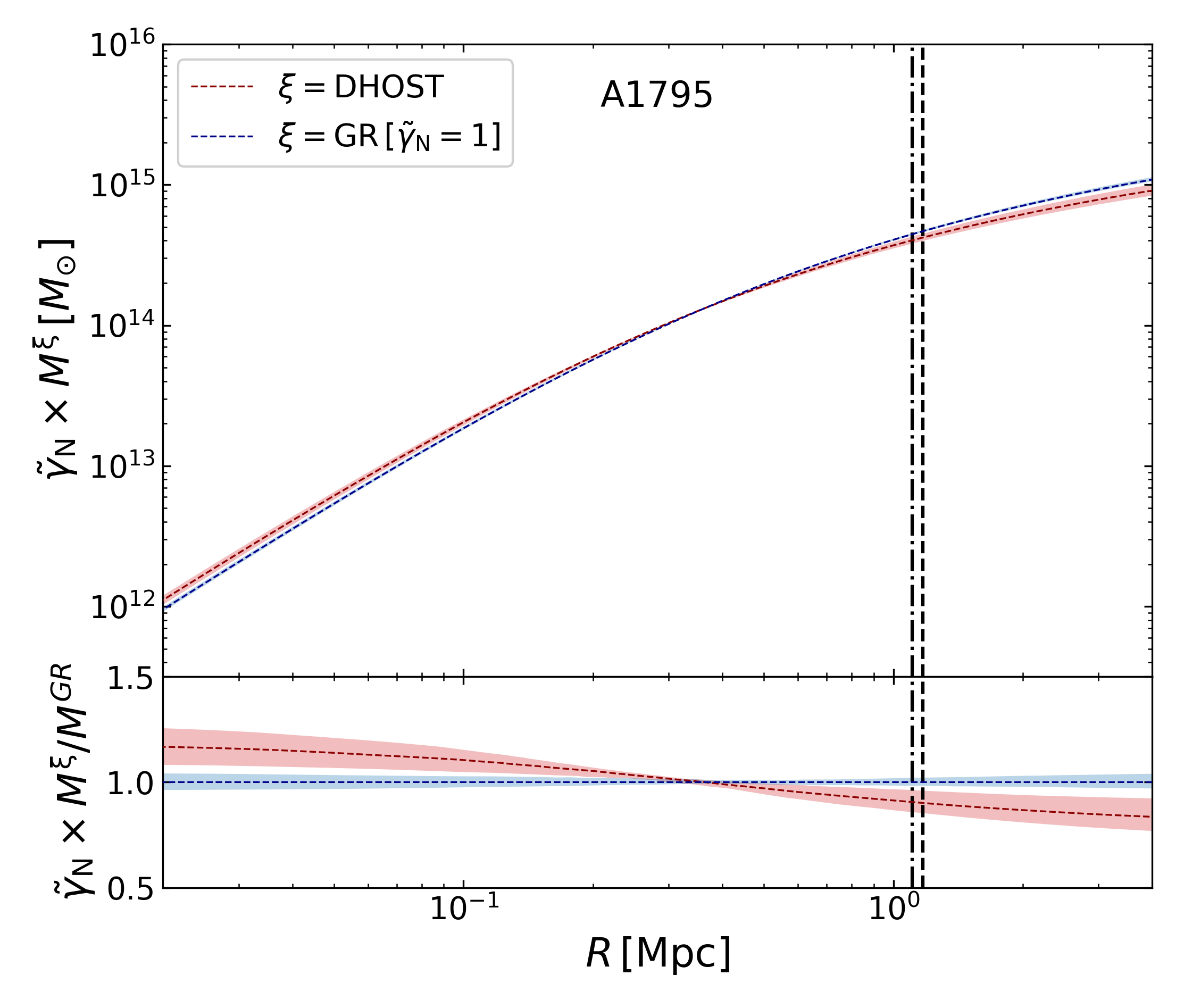}
    \includegraphics[scale=0.3]{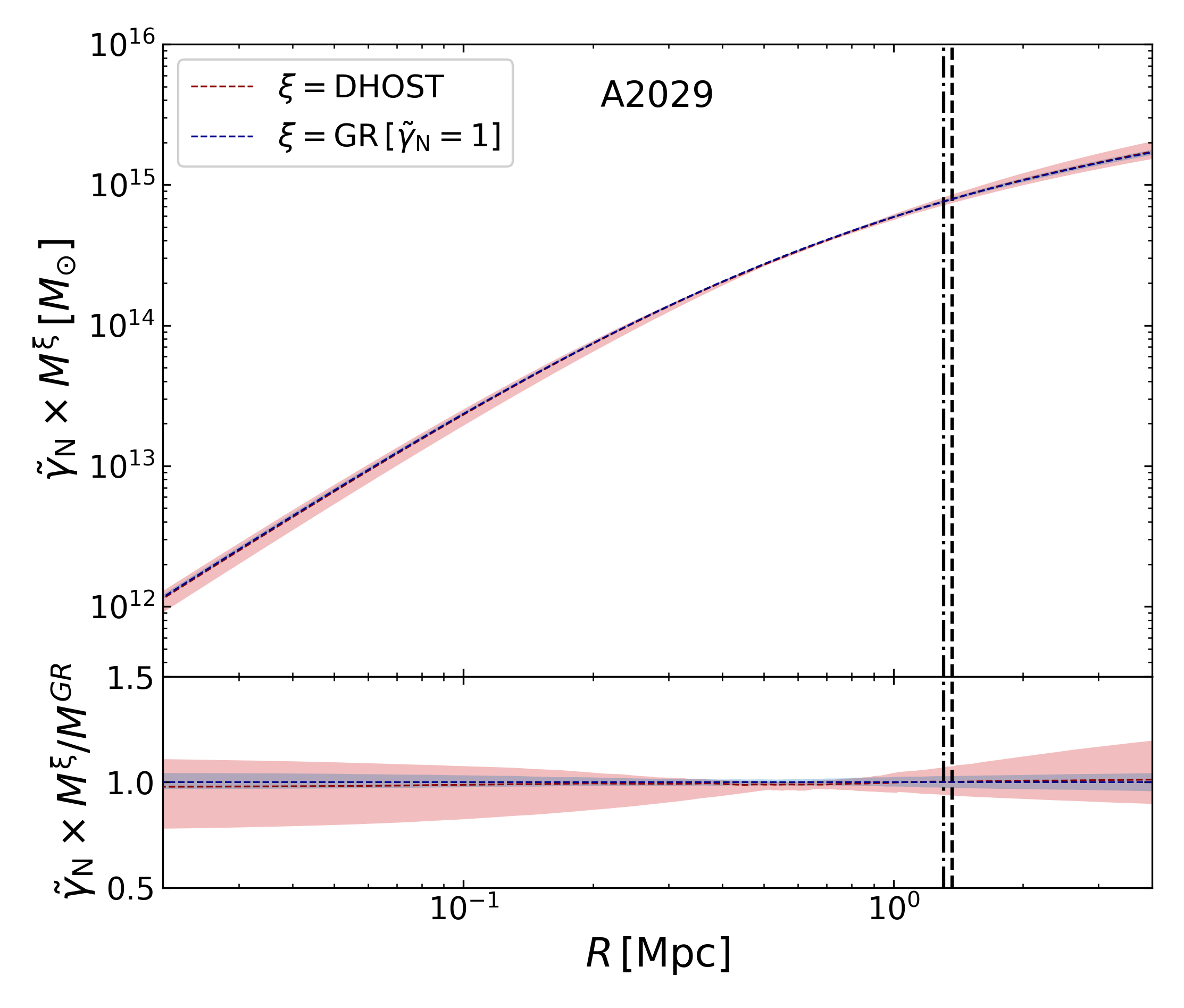}
    \includegraphics[scale=0.3]{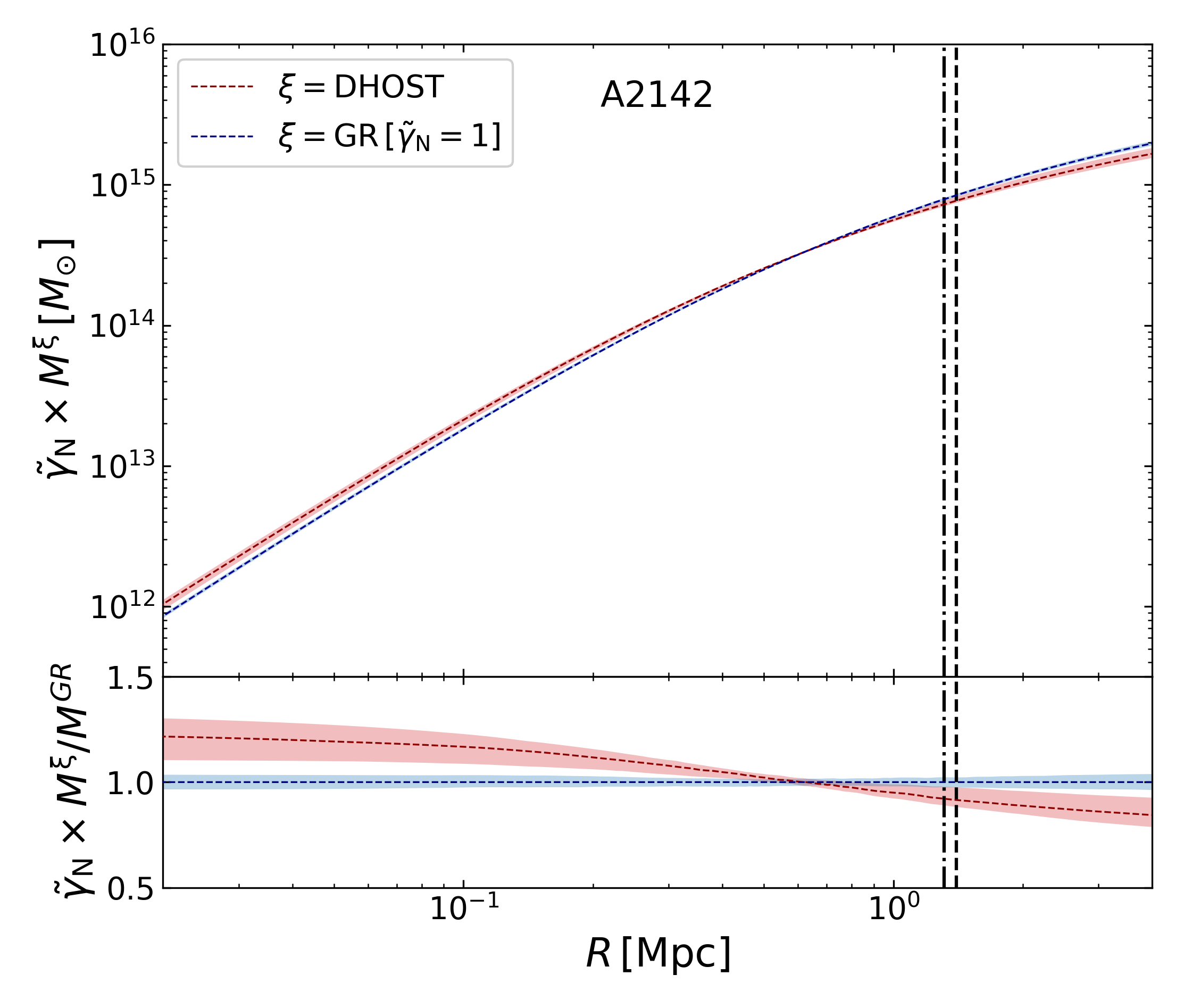}
    
    \includegraphics[scale=0.3]{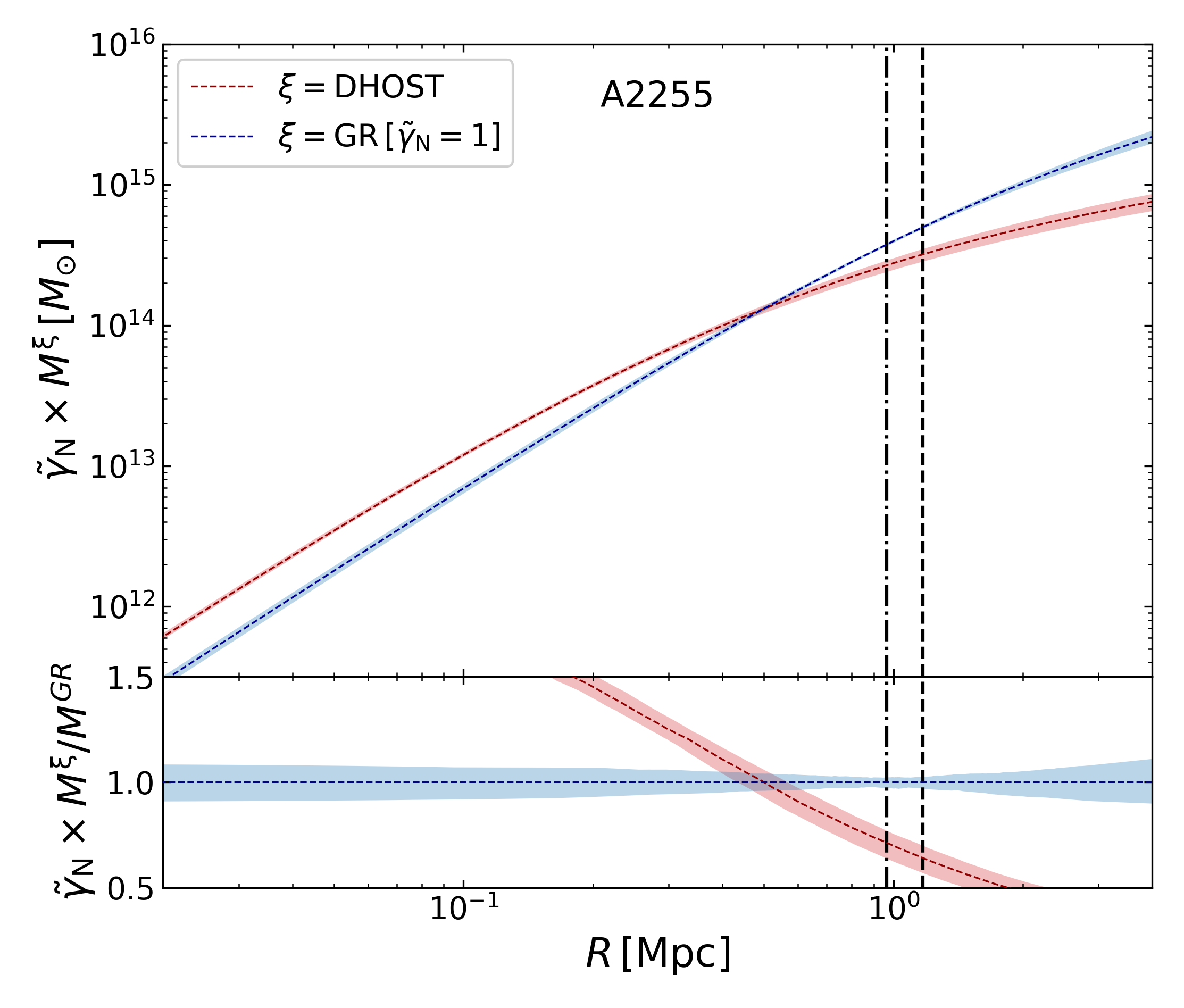}
    \includegraphics[scale=0.3]{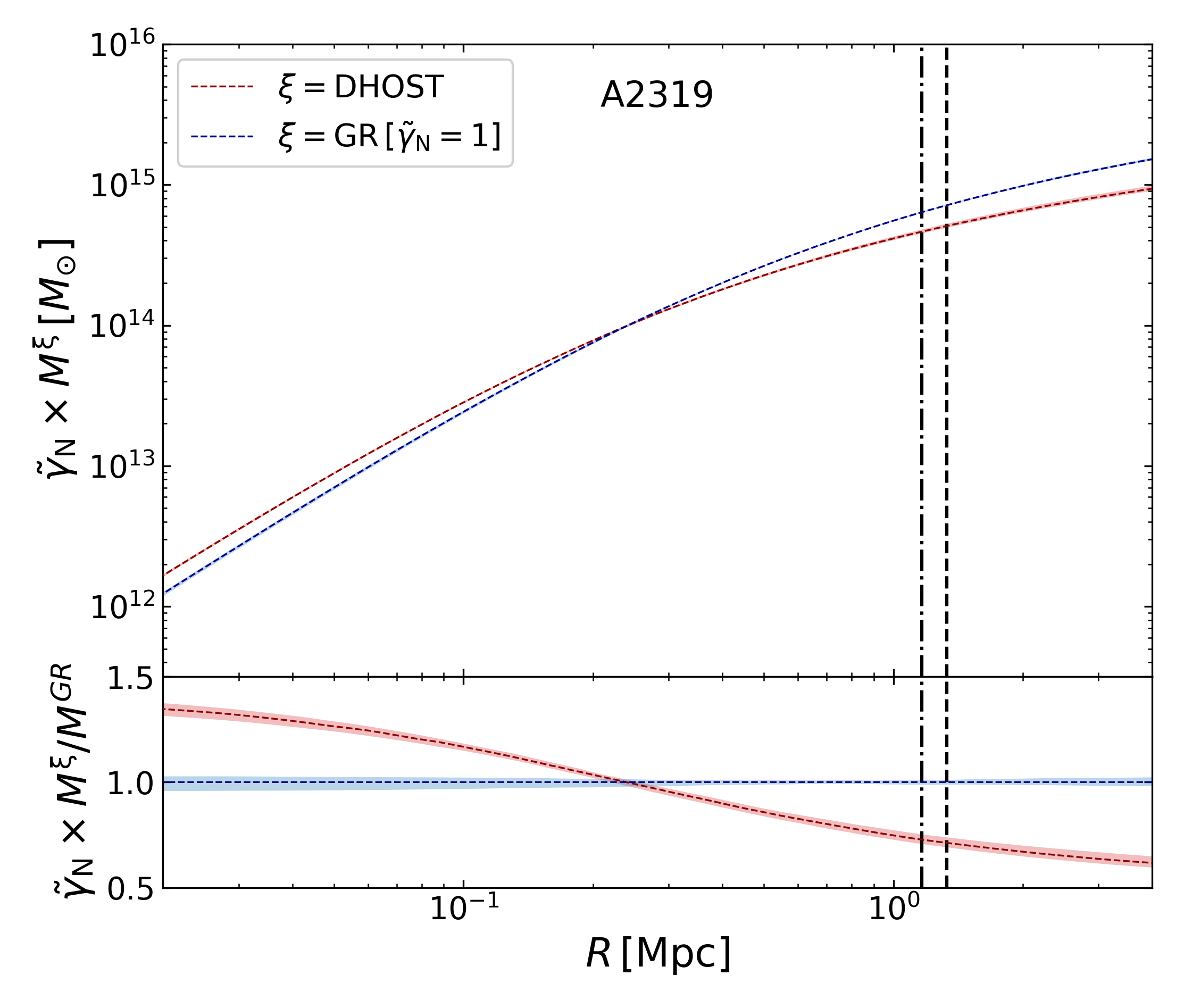}
    \includegraphics[scale=0.3]{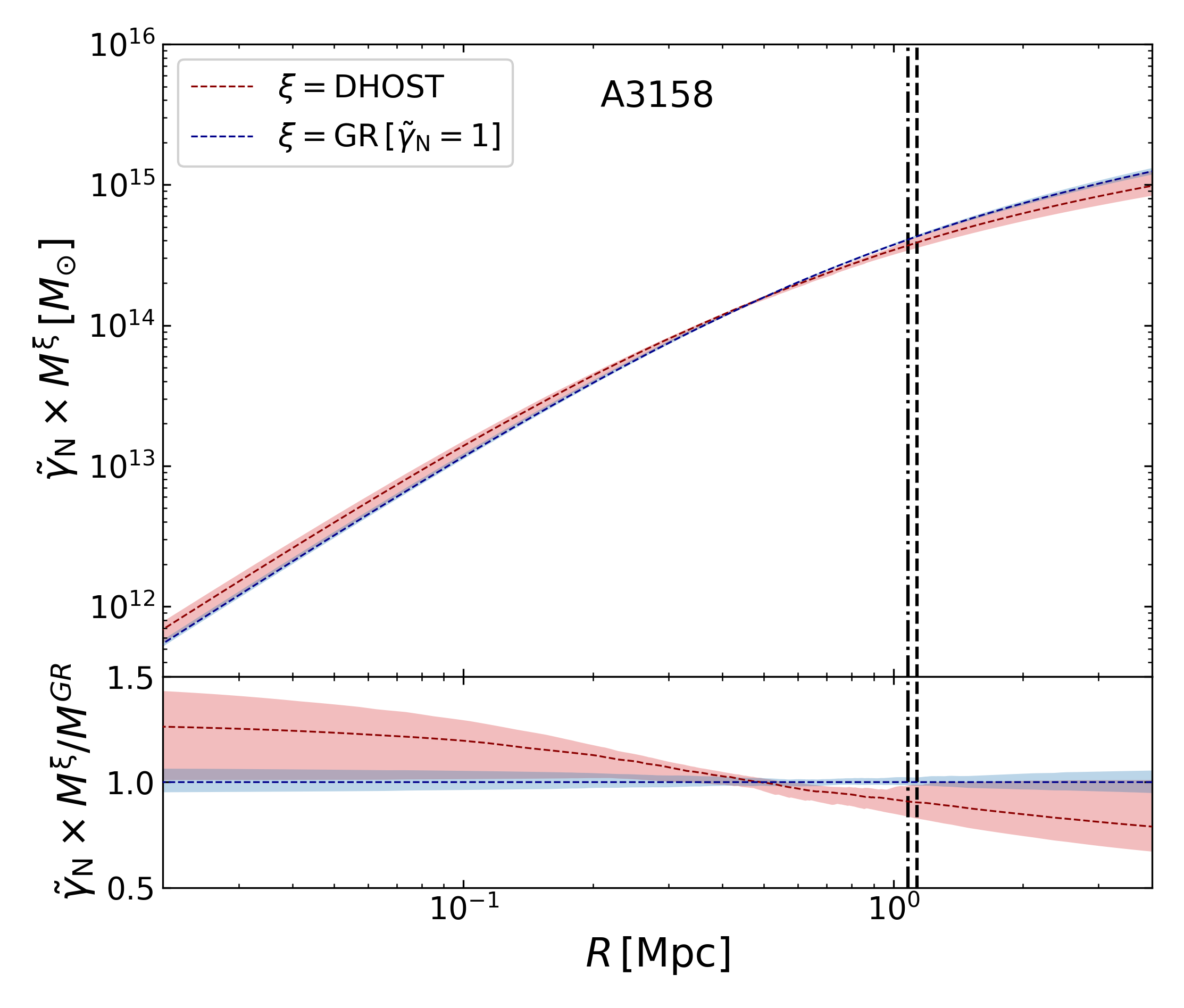}
    
    \includegraphics[scale=0.3]{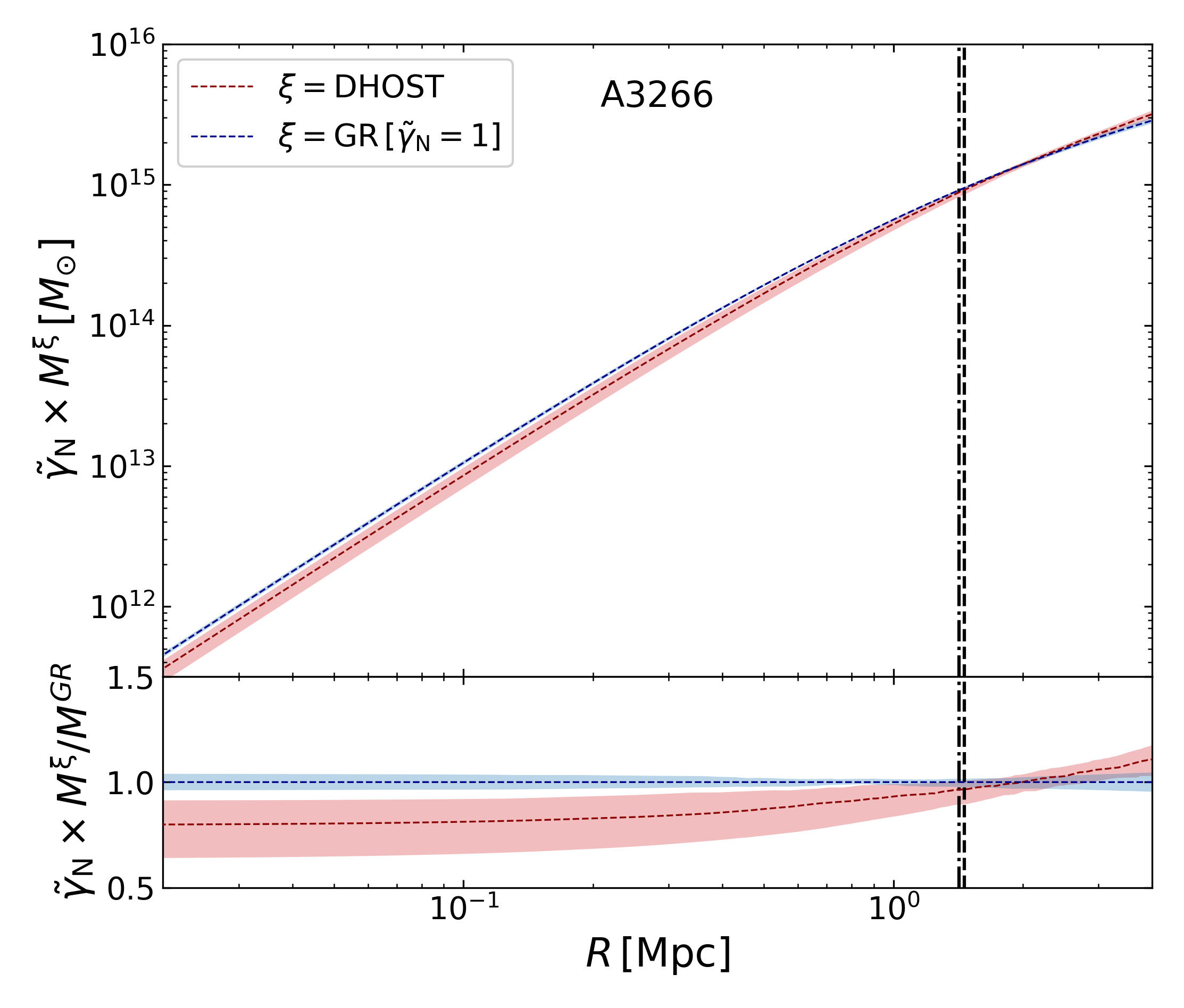}
    \includegraphics[scale=0.3]{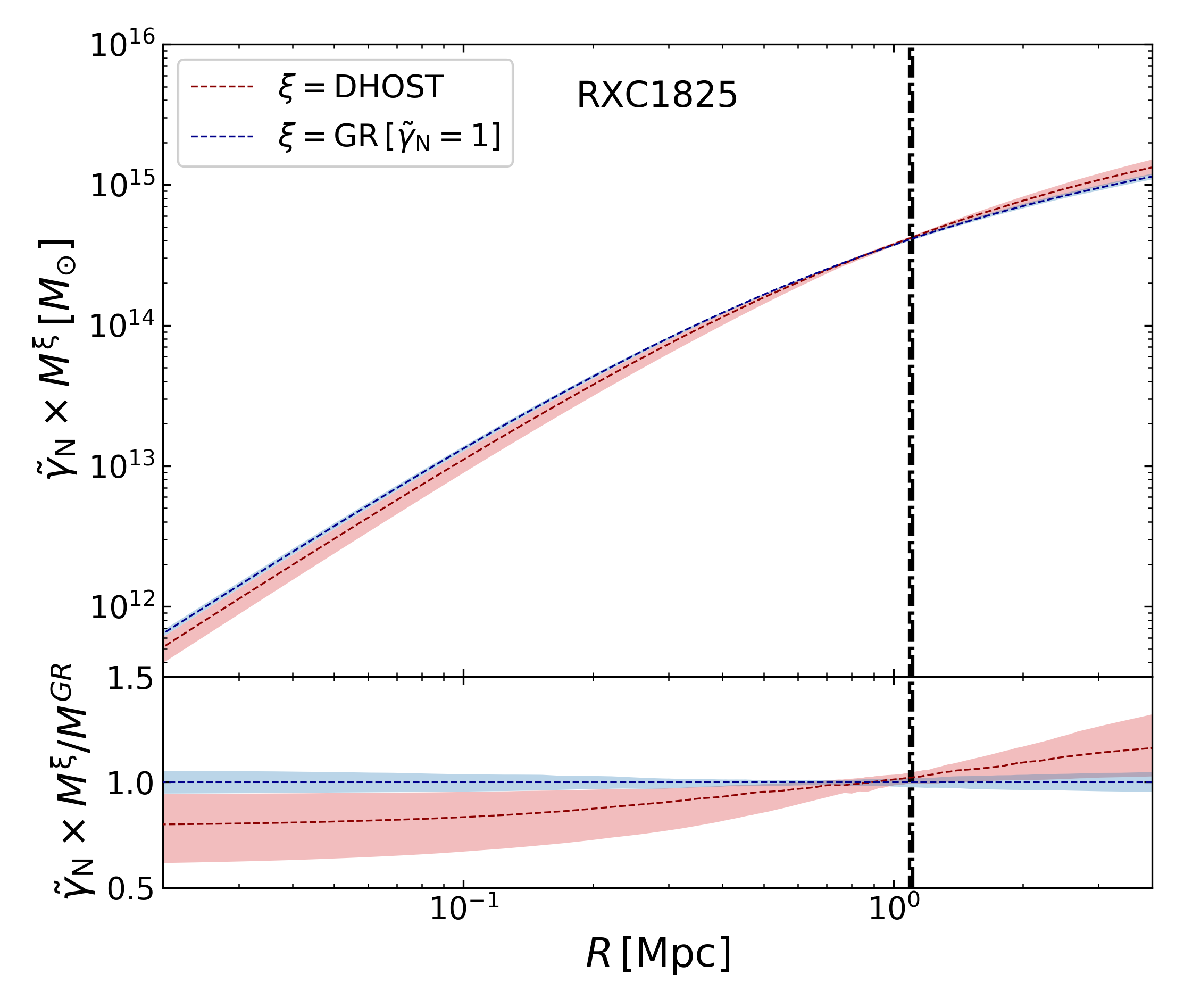}
    \includegraphics[scale=0.3]{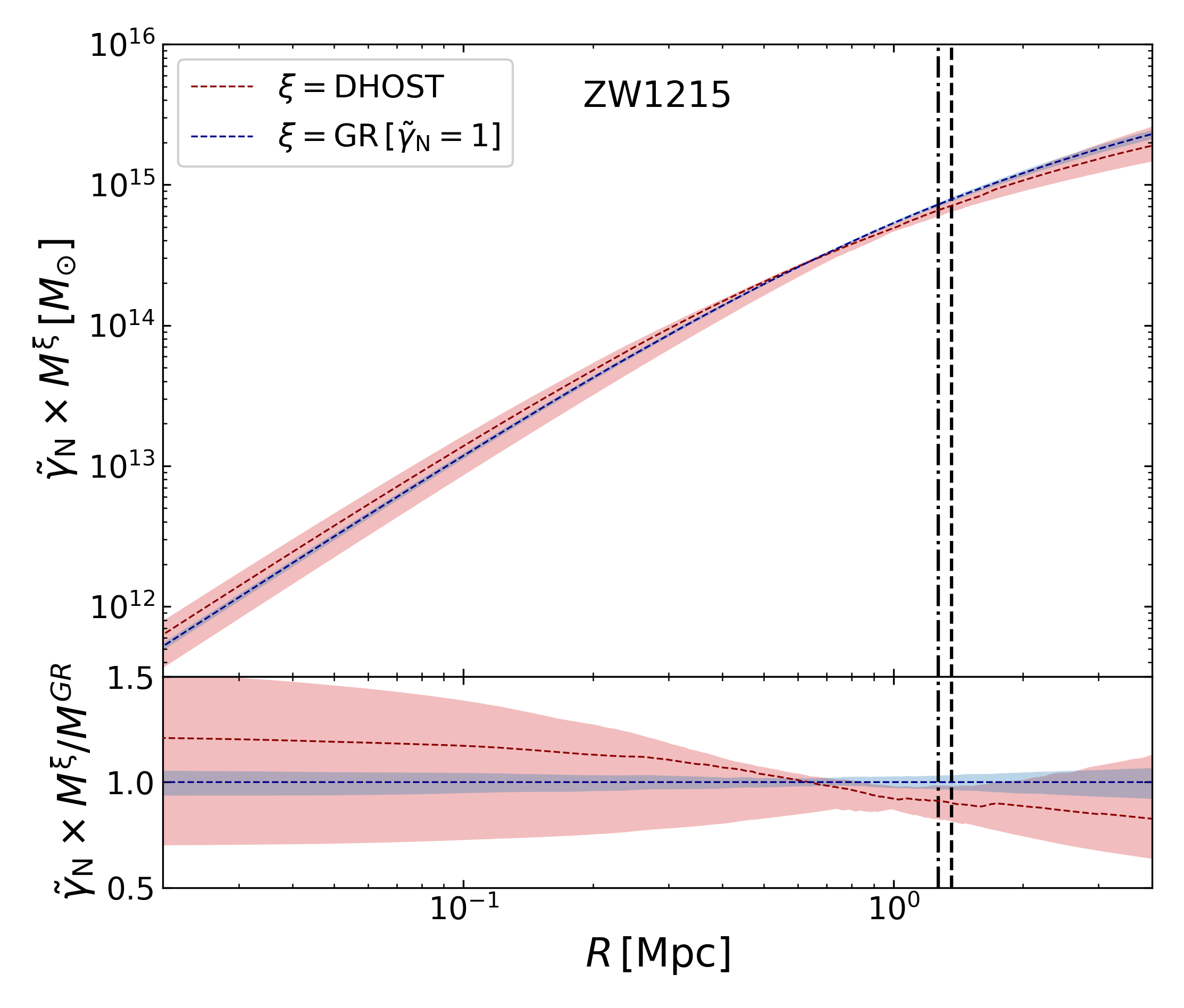}
    
    \caption{We show the comparison of mass profiles (\textit{Top}) and the relative difference  (\textit{Bottom}), between the GR and the DHOST modification. The dashed and dash-dotted vertical lines show the $R_{500}$ in the GR and DHOST cases, respectively. The shaded region corresponds to a $68\%$ C.L. limits on the mass profiles. The corresponding mass ($M_{500}$) and concentration ($c_{500}$) parameters are shown in \Cref{tab:Constraints_PP}.}
    \label{fig:Fit_Profiles}
\end{figure*}

%
\end{document}